\documentclass[11pt,a4paper]{article}
\pdfoutput=1 

\usepackage{amsmath,amssymb}
\usepackage{appendix}
\usepackage{color}
\usepackage{cite}
\usepackage{feynmp}
\usepackage[urlcolor=blue]{hyperref}
\usepackage[capitalise]{cleveref}
\usepackage{epsfig,graphicx}
\usepackage{graphicx}
\usepackage{subcaption}
\usepackage{chngpage}
\usepackage{calc}
\usepackage{lipsum}

\renewcommand{\Re}{\mathrm{Re}}

\newcommand{\zp}{{Z^\prime}}

\renewcommand\({\left(}
\renewcommand\){\right)}
\renewcommand\[{\left[}
\renewcommand\]{\right]}

\newcommand{\exclude}[1]{}

\def\bra{\langle}
\def\ket{\rangle}
\def\beq{\begin{equation}}
\def\eeq{\end{equation}}
\newcommand{\C}[1]{\mathcal{#1}}

\begin{document}
\numberwithin{equation}{section}
\title{
\vspace{2.5cm} 
\Large{\textbf{Purely flavor-changing $\boldsymbol Z^{\boldsymbol\prime}$ bosons and where they might 
hide
\vspace{0.5cm}}}}

\author{Patrick Foldenauer and Joerg Jaeckel\\[2ex]
\small{\em Institut f\"ur Theoretische Physik, Universit\"at 
Heidelberg,} \\
\small{\em Philosophenweg 16, 69120 Heidelberg, Germany}\\[0.5ex]}

\date{}
\maketitle

\begin{abstract}
\noindent
A plethora of ultraviolet completions of the Standard Model have extra U(1) gauge symmetries. In general, the associated massive $Z^\prime$ gauge boson can mediate flavor-changing neutral current processes at tree level. We consider a situation where the $Z^\prime$ boson couples solely via flavor-changing interactions to quarks and leptons. In this scenario the model parameter space is, in general, quite well constrained by existing flavor bounds. However, we argue that cancellation effects shelter islands in parameter space from strong flavor constraints and that these can be probed by multipurpose collider experiments like ATLAS or CMS as well as LHCb in upcoming runs at the LHC. In still allowed regions of parameter space these scenarios may help to explain the current tension between theory and experiment of $(g-2)_\mu$ as well as a small anomaly in $\tau$ decays.
\end{abstract}

\newpage

\tableofcontents

\section{Introduction}

In searching for new physics it is prudent to explore the limits of 
applicability of standard tests and probe for corners in parameter space where 
they can be evaded. We can then turn around and check if in these regions other tests become more powerful. It is in this spirit that in 
this paper we want to examine flavor-changing $\zp$ bosons coupling to quarks 
and leptons. In this case severe constraints arise from precision tests of 
flavor-changing neutral currents (FCNCs), in particular on mesons. Yet we will 
see that there are still interesting areas of parameter space that can be probed 
with direct production at the LHC and ``non-flavored'' measurements such as 
$(g-2)$.

Flavor-changing $\zp$ bosons could be a remnant of a solution to the still 
unsolved question
of the origin of the flavor structure of the Standard Model (SM). Indeed, one 
of the earliest approaches towards an explanation of Yukawa coupling patterns 
and the family structure of the SM fermions was the 
introduction of so-called horizontal or gauged flavor symmetries 
\cite{Barr:1978rv, Wilczek:1978xi, Davidson:1979wr, Ong:1978tq, 
Yanagida:1979gs, Mohapatra:1974wk}. For example, the 
different copies of up- and down-type quarks, charged leptons and neutrinos can
transform as a multiplet under a new horizontal SU(2) symmetry group. 
Likewise, one can assign charges under a new local U(1) gauge symmetry. The breaking of such symmetries generally leads 
to the emergence of new massive gauge bosons mediating FCNCs. In such scenarios, care has to be taken in that the magnitude 
of such an effect does not violate experimental constraints 
\cite{Langacker:2000ju}. 
Nevertheless, gauged flavor models are enjoying a
renewed popularity~\cite{Guadagnoli:2011id,Grinstein:2010ve}. Especially, 
the case of a new U(1) gauge symmetry has been studied extensively in the 
past (see Refs.~\cite{Leike:1998wr, Langacker:2009im} for reviews). 

In this paper we take a phenomenological approach of extending the SM by a 
neutral massive $\zp$ boson, which is possibly the remnant of a broken 
gauge symmetry, with the simplest possibility being a U(1)\footnote{In the 
following we do not take care of anomalies. For our simple phenomenological 
considerations we implicitly assume that anomalies will be canceled in a more 
complete model.}. 
Specifically, we consider models with exclusively flavor-changing couplings, one in the quark and one in 
the lepton sector
\begin{equation}
\mathcal{L}_{\zp} =  \bar{q}\, \gamma^\mu\ [g^L_{qq'}\, P_L + g^R_{qq'}\, P_R]\ 
q^\prime \, \zp_\mu 
  +   \bar{\ell}\, \gamma^\mu\ [g^L_{\ell\ell'}\, P_L + g^R_{\ell\ell'}\, P_R]\ 
\ell' \, \zp_\mu + h.c. \,.
\label{eq_gen_lag}
\end{equation}
Purely flavor-changing interactions provide a simple but interesting test case.
On the one hand they provide a maximally flavor-changing effect. On the other hand they are often more difficult to detect. For example, if the quark part of the interaction involves a $b$- and an $s$-quark, production at proton colliders like the LHC requires reliance on the sea-quarks in the protons which are less abundant\footnote{This is also the reason why we do not consider interactions involving $t$-quarks. The corresponding limits are much weaker.}. 
Similarly at LEP simple $s$-channel production of $\zp$-bosons via the lepton couplings is not possible as the initial state is not flavored.

\bigskip
The paper is structured as follows. In \cref{sec_coll} we will discuss  collider constraints on our 
model from reinterpreting an ATLAS search for neutral resonances in  
$e\mu$, $e\tau$ and $\mu\tau$ final states \cite{Aad:2015pfa}. In 
\cref{sec_constraints} we review relevant existing constraints on our 
model. In this context we will discuss meson mixing, meson decays into charged 
leptons and neutrinos, lepton decays, muonium-antimuonium oscillations, LEP searches as well as electron and muon $(g-2)$ measurements. In view of the $(g-2)_\mu$ anomaly, we will also discuss a possible explanation of the observed shift $\Delta a_\mu$ within our model together with a small anomaly in $\tau$-decays (cf. also~\cite{Altmannshofer:2014cfa,Altmannshofer:2016brv}). Finally, we will interpret and wrap up our results in 
\cref{sec_summary}. The summary plots of our findings of collider and existing 
flavor constraints on our model can be found in \cref{sec_coll} in 
\cref{fig_bsem,fig_bsmt,fig_bdem} and in the Appendix~\ref{sec_app_plots} in 
\cref{fig_plots1,fig_plots2,fig_plots3}. An example interpretation of the $(g-2)_{\mu}$ and $\tau$-decay anomalies is depicted in \cref{fig_propaganda}. We focus on the situation where the $\zp$ bosons are heavy $M_{\zp}\gg M_{Z}$.
Unless otherwise stated we take the lepton sector couplings to be purely right-handed. Most plots, however, are also applicable to the purely left-handed case. The additional limits present in this situation are given as dotted and dash-dotted lines.

\section{Flavor violation from a collider point of view}
\label{sec_coll}

One main goal of this paper is to reinterpret existing neutral resonance 
searches at the LHC in the context of a flavor-changing $\zp$ boson. This 
provides us with new constraints on the induced FCNCs, complementary to the 
usual bounds coming from flavor and electroweak precision experiments (see 
\cref{sec_constraints}).
Previously, Davidson et al. \cite{Davidson:2013fxa} have investigated 
flavor-changing four-quark 
contact interactions coming from new physics from a scale $M\gg M_W$ in an 
effective field theory (EFT) approach. They consider the 
various four-quark operators the LHC is sensitive to, also including quark 
flavor violating operators, which are of the type
\begin{equation}
 \mathcal{O}_{minj}^{XY} = (\bar{q}_m \, \gamma^\mu\, P_X \, q_i)(\bar{q}_n\, 
\gamma_\mu\, P_Y \, q_j) \,,
\label{eq_4qops}
\end{equation}
with $X,Y \in \{L,R\}$ and the indices $(m,i,n,j)$ denoting flavor. Then they derive a limit on their 
suppression scale $\Lambda$ by reinterpreting existing LHC dijet analyses.  We show the corresponding limits as brown areas in the figures.
\par
In our model, however, we are considering combined lepton and quark flavor 
violation. While our model also contains the effective operators 
\eqref{eq_4qops} 
we have additional operators of the type
\begin{equation}
 \C O_{ijkl}^{XY} = (\bar{q}_i \, \gamma^\mu \,P_{X} \,q_j)(\bar{\ell}_k 
\gamma_\mu \, P_{Y}\,\ell_l) \,.
\end{equation}
This type of operator can be generated from a $\zp$ exchange in the full theory 
and consequently a bound on it can be turned into a constraint on the 
corresponding $\zp$ couplings. In the following, we therefore want to 
reinterpret an existing ATLAS analysis of heavy neutral particles decaying to  
$e\mu$, $e\tau$ or $ \mu\tau$ \cite{Aad:2015pfa} in the light of our 
flavor-violating $\zp$ model.\\

\subsection{Reinterpreting collider searches}

The model we consider in this paper induces $\Delta F=2$ flavor-violating 
processes of the type  $qq' \rightarrow \zp \rightarrow \ell \ell'$. In order 
to constrain the relevant couplings $g_{qq'}$ and $g_{\ell\ell'}$ we first need 
an expression for the corresponding cross section within our model. Introducing 
the non-chiral reduced coupling 
\begin{equation}
 \bar{g} = \sqrt{\frac{g_L^2+g_R^2}{2}} \; ,
\end{equation}
we can derive an approximate expression for the cross section scaling as
\begin{equation}
  \sigma(s) \approx \frac{1}{3}\ \frac{s}{M^4_\zp}\ \frac{\bar{g}^2_{qq'}\, 
\bar{g}^2_{\ell\ell'}}{3\,\bar{g}^2_{qq'} + \bar{g}^2_{\ell\ell'}} \,.
\label{eq_partxsec}
\end{equation}
This expression gives a valid estimate for the cross section at parton level. 
However, we cannot access this cross section at the LHC directly. As we are 
dealing with a hadron collider we have to take into account parton distribution function effects and 
hadronization. Moreover, the observable cross section will also be affected by 
a number of detector effects like finite resolution, mistags, acceptance etc. 
\par
Our approach to incorporate all these effects is quite straight forward. We 
simulate the total cross section 
$\sigma_{_\mathrm{MC}}$ for the process 
$pp\rightarrow\zp\rightarrow \ell 
\ell^\prime$ in our model for a given combination $\{qq',\ell\ell'\}$ of flavor-violating interactions. The ratio of the simulated cross section
$\sigma_{_\mathrm{MC}}$ to the ATLAS limit on the cross section 
$\sigma_\mathrm{lim}$ allows us to derive an 
approximate limit on the off-diagonal quark coupling $\bar{g}_{qq'}$ as a 
function of the lepton coupling $\bar{g}_{\ell\ell'}$ 
according to
 \begin{align}
  |\bar{g}_{qq'}| \leq \left[\frac{\sigma_{_\mathrm{MC}}}{\sigma_\text{lim}}\ 
\frac{3+ 
r^2}{r^2\,\bar{g}^2_{qq',_\mathrm{MC}}} 
-\frac{3}{\bar{g}^2_{\ell\ell^\prime}}\right]^{-\frac{1}{2}} 
\label{eq_glimatl}\,.
 \end{align}
Here $\bar{g}_{qq',_\mathrm{MC}}$ and $\bar{g}_{\ell\ell^\prime,_\mathrm{MC}}$ 
denote the values of the reduced couplings used for the simulation and 
\begin{equation}
 r = \frac{ \bar{g}_{\ell\ell^\prime,_\mathrm{MC}} 
}{\bar{g}_{qq',_\mathrm{MC}}} \;,
\end{equation}
denotes the ratio of them. We immediately notice that
\cref{eq_glimatl} has a pole at
\begin{equation}
 \bar{g}^2_{\ell\ell^\prime}= \frac{\sigma_\text{lim}}{\sigma_{_\mathrm{MC}}}\ 
\frac{3\, r^2\,\bar{g}^2_{qq',_\mathrm{MC}}}{3+ 
r^2} \,.
\end{equation}
This simply indicates that for values of 
$\bar{g}_{\ell\ell^\prime}$ close to this pole the lepton coupling is so weak 
that even for a very large value of the quark coupling 
$\bar{g}_{qq^\prime}$ the signal cannot be distinguished from background, i.e. 
the process is unobservable at the LHC. 
In other words, given the observed cross section limit 
$\sigma_\mathrm{lim}$, we are not able anymore to set a limit on the quark 
coupling $\bar{g}_{qq^\prime}$ for such low values of 
$\bar{g}_{\ell\ell^\prime}$.
\par 
We still have to determine the simulated cross section 
$\sigma_\mathrm{MC}$. First, we calculate  the leading order (LO) cross section 
$\sigma_\mathrm{LO}$  with \textsc{madgraph v2.3.3} \cite{Alwall:2014hca} for 
the processes under consideration.
Next, we determine a mass-dependent $K$-factor to take into account next-to-next-to leading order (NNLO) 
effects. In the auxiliary material of Ref.~\cite{Aad:2015pfa} the NNLO 
cross sections of the Sequential Standard Model (SSM) $\zp$ are provided. In 
order to determine the values of the $K$-factor, we calculate the LO cross 
section for the SSM $\zp$ in \textsc{pythia v8.215} \cite{Sjostrand:2007gs} 
and compare to the provided NNLO results. The $K$-factor is then found 
from 
the ratio
\begin{align}
 K(M_\zp) = \frac{\sigma_\mathrm{NNLO}(M_\zp)}{\sigma_\mathrm{LO}(M_\zp)} \,.
\end{align}
The values we have determined in our analysis are given in 
Appendix~\ref{sec_appMC}  
in \cref{tab_kfacs}. Finally, we determine an effective mass-dependent 
acceptance times efficiency $A\times\epsilon$ from the ratio of 
the number of events that survived the detector plus analysis cuts to the 
expected total number of events at NNLO 
\begin{align}
 [A\times\epsilon](M_\zp) = 
\frac{N_\mathrm{survive}(M_\zp)}{N_\mathrm{NNLO}(M_\zp)} \,.
\end{align}
The number of events at NNLO $N_\mathrm{NNLO} = \sigma_\mathrm{NNLO} 
\times \int \mathrm{d}t\, \mathcal{L}$ is simply obtained from multiplying 
the cross section by the integrated luminosity. The numerical values of 
the determined acceptance times efficiency we used in our calculations can be 
found in \cref{sec_appMC} in \cref{tab_effs}.
Putting everything together we obtain the full simulated cross section as
\begin{align}
 \sigma_\mathrm{MC} = \sigma_\mathrm{LO} \times K \times [A\times\epsilon] \,.
\end{align}

\subsubsection{Numerical evaluation}

\begin{figure}[t!]
 \includegraphics[width=\textwidth]{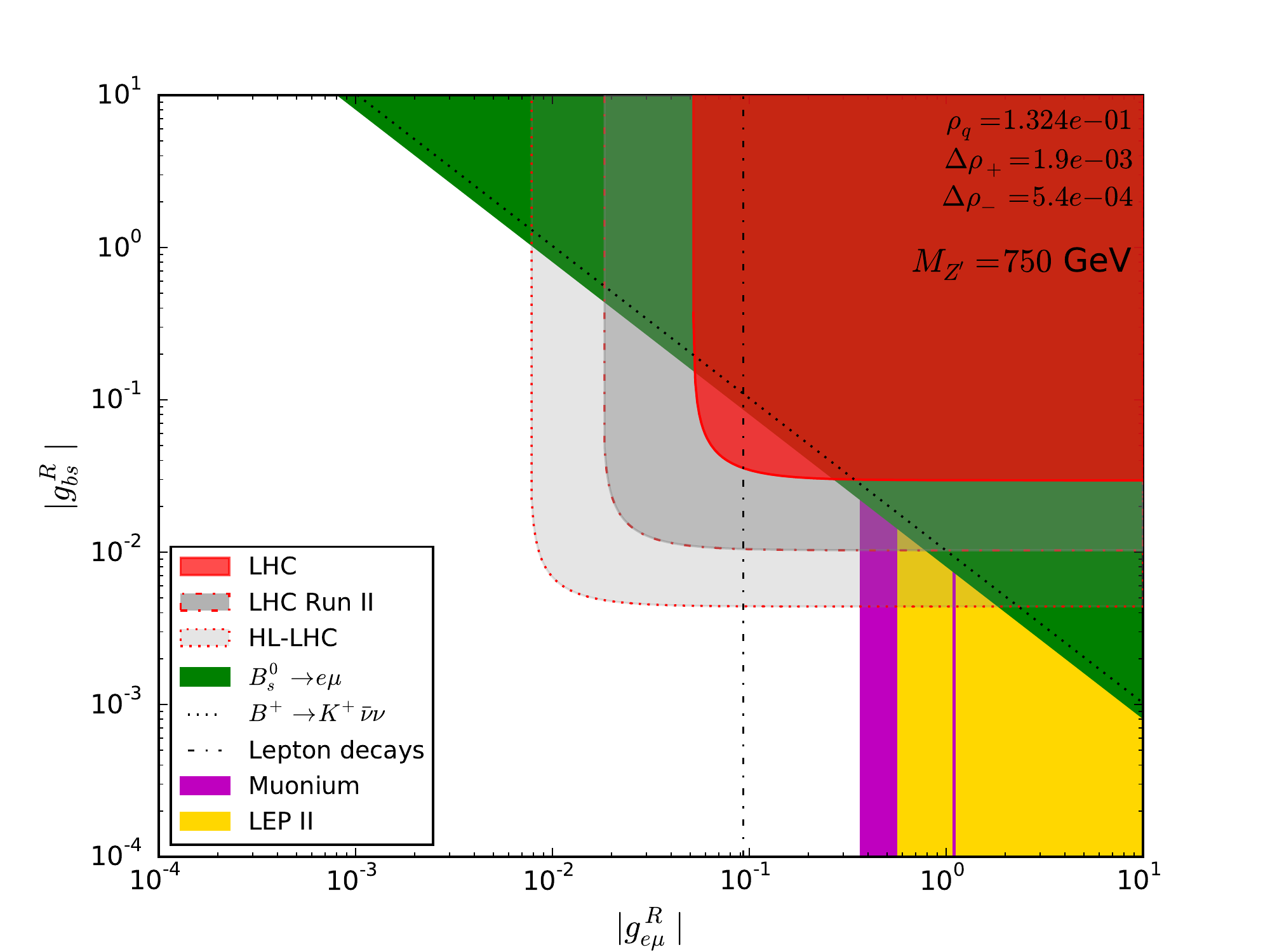} 
 \caption{The flavor-violating couplings $g^R_{bs}$ and $g^R_{e\mu}$ for a $\zp$ boson of $M_\zp=750$ GeV and a coupling ratio of $\rho_q= g^L_{bs}/g^R_{bs} = 0.1324$ (for an explanation of $\Delta\rho$ see \cref{sec_cancel}). The red area indicates the limit from the ATLAS analysis of the process $pp\rightarrow e\mu$ at $\sqrt{s}=8$ TeV. The red dash-dotted and dashed lines are projections to the LHC Run II and HL-LHC. The green area is the limit coming from the meson decay $B_s^0\rightarrow e\mu$. The gold and magenta areas are purely leptonic limits coming from LEP and muonium oscillation constraints. Finally, the black dotted line is the exclusion limit from $B^+\rightarrow K^+\bar{\nu}\nu$ and the black dash-dotted line the limit from lepton decays. These last two limits, however, apply only if we consider $g_{e\mu}^L$ instead of $g_{e\mu}^R$.}
 \label{fig_bsem}
\end{figure}

\begin{figure}[t!]
 \includegraphics[width=\textwidth]{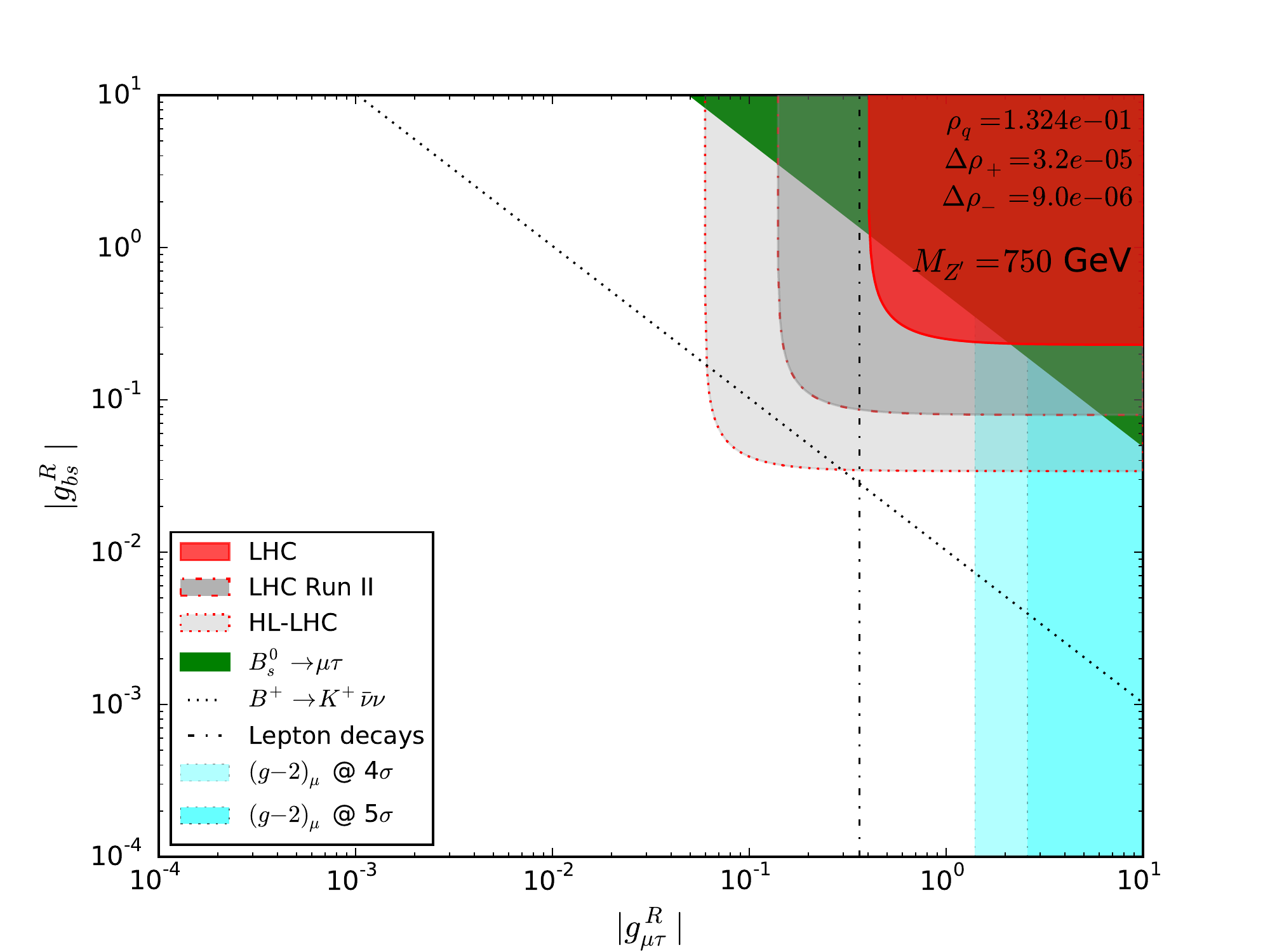}
 \caption{The flavor-violating couplings $g^R_{bs}$ and $g^R_{\mu\tau}$ for a $\zp$ boson of $M_\zp=750$ GeV and a coupling ratio of $\rho_q = g^L_{bs}/g^R_{bs}= 0.1324$ (for an explanation of $\Delta\rho$ see \cref{sec_cancel}). The red area indicates the limit from the ATLAS analysis of the process $pp\rightarrow \mu\tau$ at $\sqrt{s}=8$ TeV. The red dash-dotted and dashed lines are projections to the LHC Run II and HL-LHC. The green area represents the excluded region from the decay $B_s^0\rightarrow\mu\tau$. The black dotted line is the limit from $B^+\rightarrow K^+\bar{\nu}\nu$ and the black dash-dotted line the limit from lepton decays. These last two limits, however, apply only for left-handed lepton couplings. The light and dark cyan areas depict the 4 and $5\,\sigma$ exclusion bands from $\Delta a_\mu$.}
 \label{fig_bsmt}
\end{figure}

\begin{figure}[t!]
 \includegraphics[width=\textwidth]{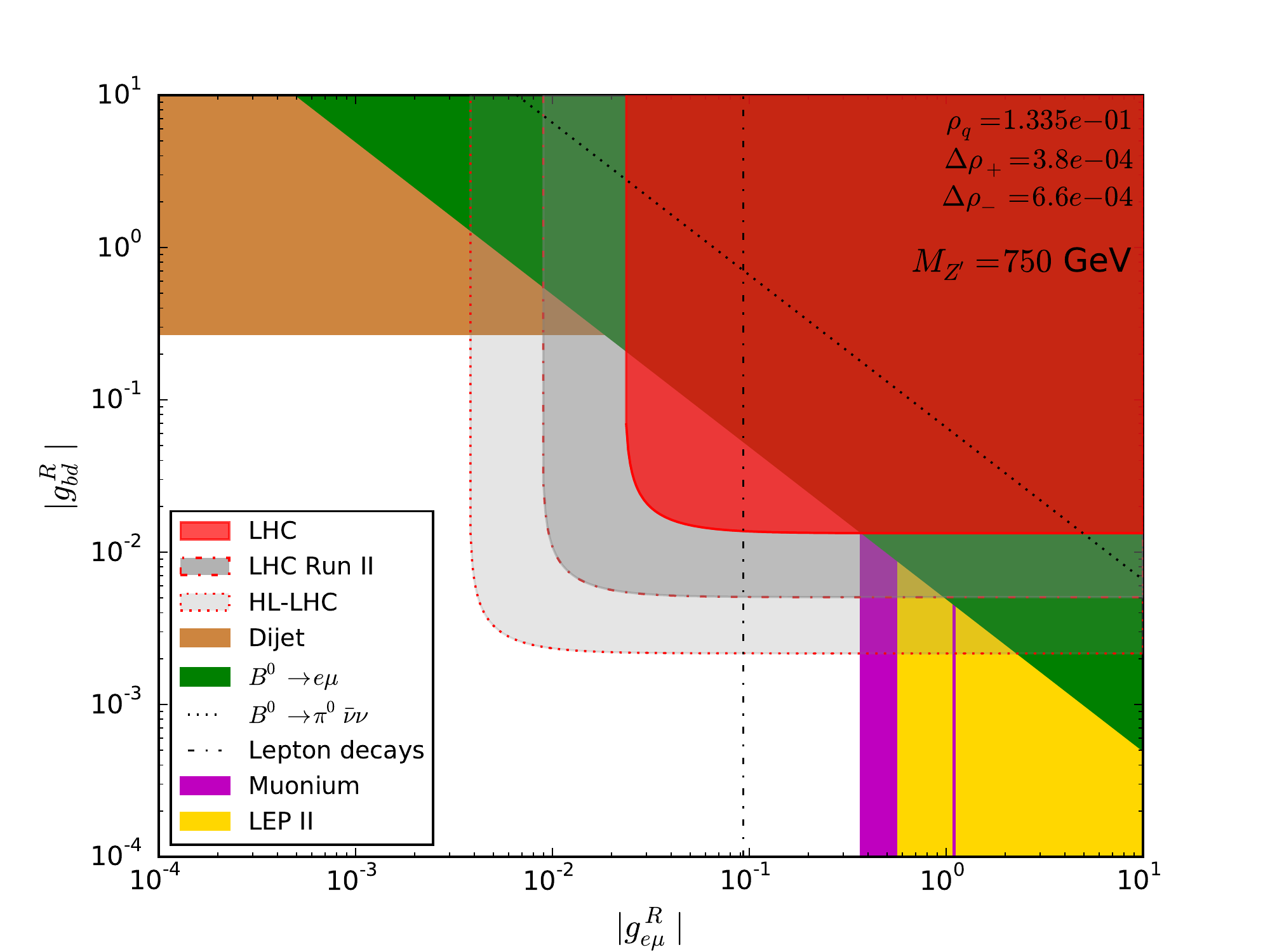}
 \caption{The flavor-violating couplings $g^R_{bd}$ and $g^R_{e\mu}$ for a $\zp$ boson of $M_\zp=750$ GeV and a coupling ratio of $\rho_q = g^L_{bd}/g^R_{bd}= 0.1335$ (for an explanation of $\Delta\rho$ see \cref{sec_cancel}). The red area indicates the limit from the ATLAS analysis of the process $pp\rightarrow e\mu$ at $\sqrt{s}=8$ TeV. The red dash-dotted and dashed lines are projections to the LHC Run II and HL-LHC. The green area is the limit coming from the meson decay $B^0\rightarrow e\mu$. The gold and magenta areas are purely leptonic limits coming from LEP and muonium oscillation constraints. The black dotted line is the exclusion limit from $B^0\rightarrow \pi^0\bar{\nu}\nu$  and the black dash-dotted line the limit from lepton decays. These last two limits, however, apply only for left-handed lepton couplings. The brown area depicts the limit from the four-quark contact interaction \cite{Davidson:2013fxa}.}
 \label{fig_bdem}
\end{figure}

In this section, we present as an example the exclusion limits on 
the couplings $g^R_{qq'}$ and $g^R_{\ell\ell'}$ for the combinations 
$\{qq',\ell\ell'\} = \{g_{bs}, g_{e\mu}\}, \{g_{bs}, 
g_{\mu\tau}\},\{g_{bd},g_{e\mu}\}$.  First, we define the ratios of left- 
to right-handed couplings
\begin{align}
 \rho_q \equiv \frac{g^L_{qq^\prime}}{g^R_{qq^\prime}} \, , &&  \rho_\ell 
\equiv 
\frac{g^L_{\ell\ell^\prime}}{g^R_{\ell\ell^\prime}} \,.
\label{eq_rhodef}
\end{align}
We present the derived bounds on the quark limits 
always as a limit in terms of the right-handed coupling $g^R$ and the 
corresponding ratio $\rho_q$ to the left-handed coupling. Hence, the limit on the
quark couplings derived in the last section reads
\begin{align}
  g^R_{qq'} (M_\zp,\rho_q, \bar{g}_{\ell\ell'}) \lesssim 
\sqrt{\frac{2}{1+\rho_q^2}}\times 
\left[\frac{\sigma_\mathrm{sim}(M_\zp)}{\sigma_\text{lim}(M_\zp)}\ \frac{3+ 
r^2}{r^2\,\bar{g}^2} -\frac{3}{\bar{g}^2_{\ell\ell'}}\right]^{-\frac{1}{2}} \,.
\label{eq_gratl}
 \end{align}
Examples of the limit derived from reinterpreting the ATLAS analysis  
\cite{Aad:2015pfa} as described above are shown  for the case of 
nonzero coupling pairs $\{g_{bs}, g_{e\mu}\}, \{g_{bs}, 
g_{\mu\tau}\},\{g_{bd},g_{e\mu}\}$
in \cref{fig_bsem,fig_bsmt,fig_bdem} as red areas. Additionally, several 
other constraints are depicted that will be explained in detail in 
\cref{sec_constraints}. 
We show plots for these three particular combinations of couplings as 
they illustrate all the  main features and relevant limits. First, LHC 
limits are 
generally strongest for a nonzero coupling in the $e\mu$ sector and weakest for 
a coupling in the $\mu\tau$ sector. The limits from meson decay into charged leptons, too, are in 
general stronger in the $e\mu$ sector than in the $e\tau$ and $\mu\tau$ 
sector. In addition, the $e\mu$ sector has further strong leptonic constraints, 
namely the one from LEP (which is also present in 
the $e\tau$ sector) and the one from muonium oscillations. In contrast to all 
other quark combinations the $bs$ sector is not tested by the dijet limits of 
Ref.~\cite{Davidson:2013fxa}. This can be seen by comparing 
\cref{fig_bsem,fig_bsmt} with \cref{fig_bdem}. The brown dijet region is 
present in the latter, but not in the former two. Lastly, only the $\mu\tau$ 
sector receives sizable constraints from the (muon) anomalous magnetic moment as 
shown in \cref{fig_bsmt}. In the plots we have indicated also limits from 
decays of mesons and leptons into neutrinos (black dotted  and dash-dotted lines). They are only applicable if we 
replace the right-handed lepton coupling by a left-handed one. Furthermore, we 
note that the limits from meson decays are absent in the  $cu$ sector (see plots in 
Appendix~\ref{sec_app_plots}).\\
\begin{figure}
\begin{subfigure}{.5\textwidth}
  \centering
   \includegraphics[width=\textwidth]{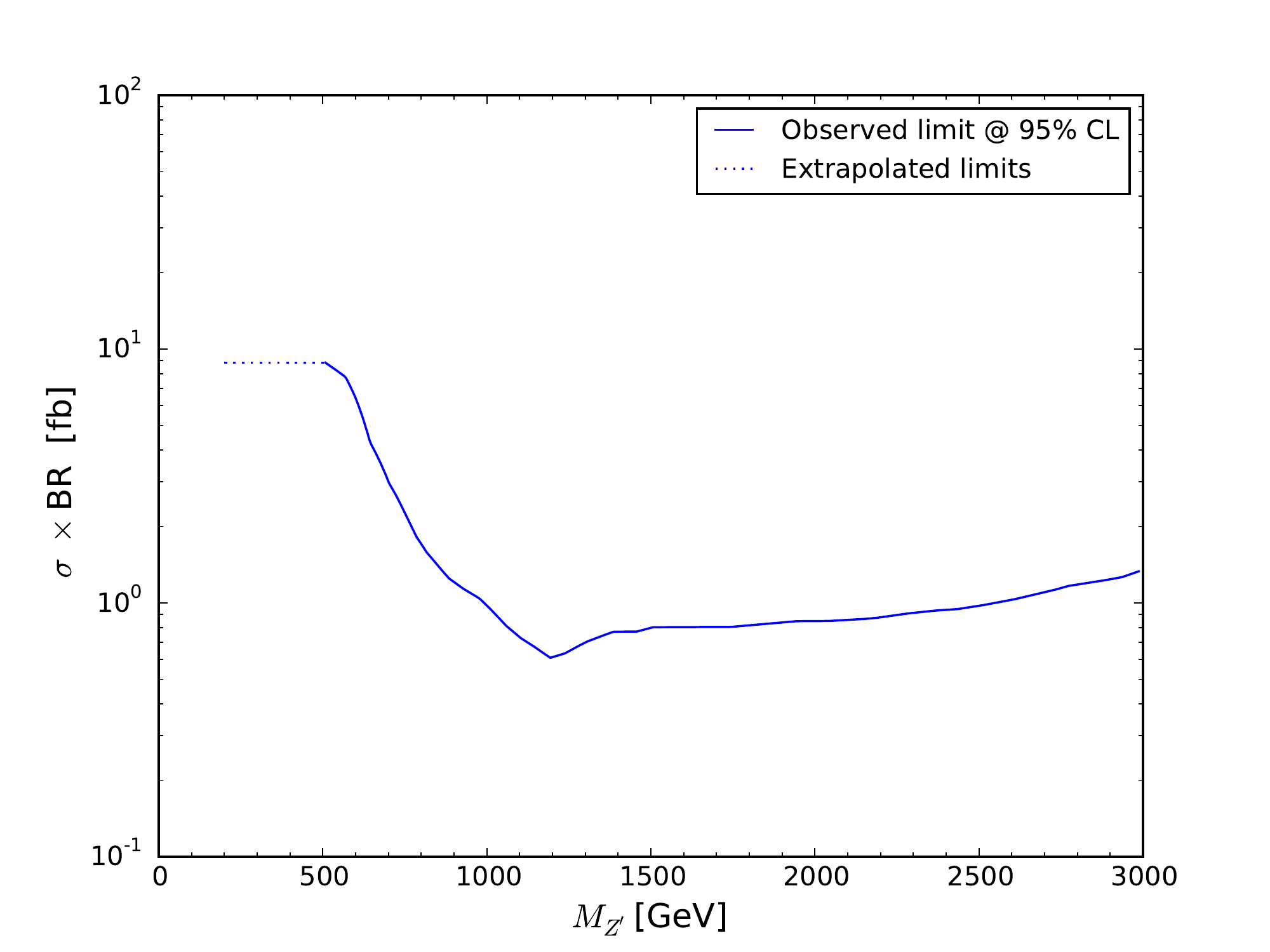}
\end{subfigure}                                
\begin{subfigure}{.5\textwidth}               
  \centering                                   
   \includegraphics[width=\textwidth]{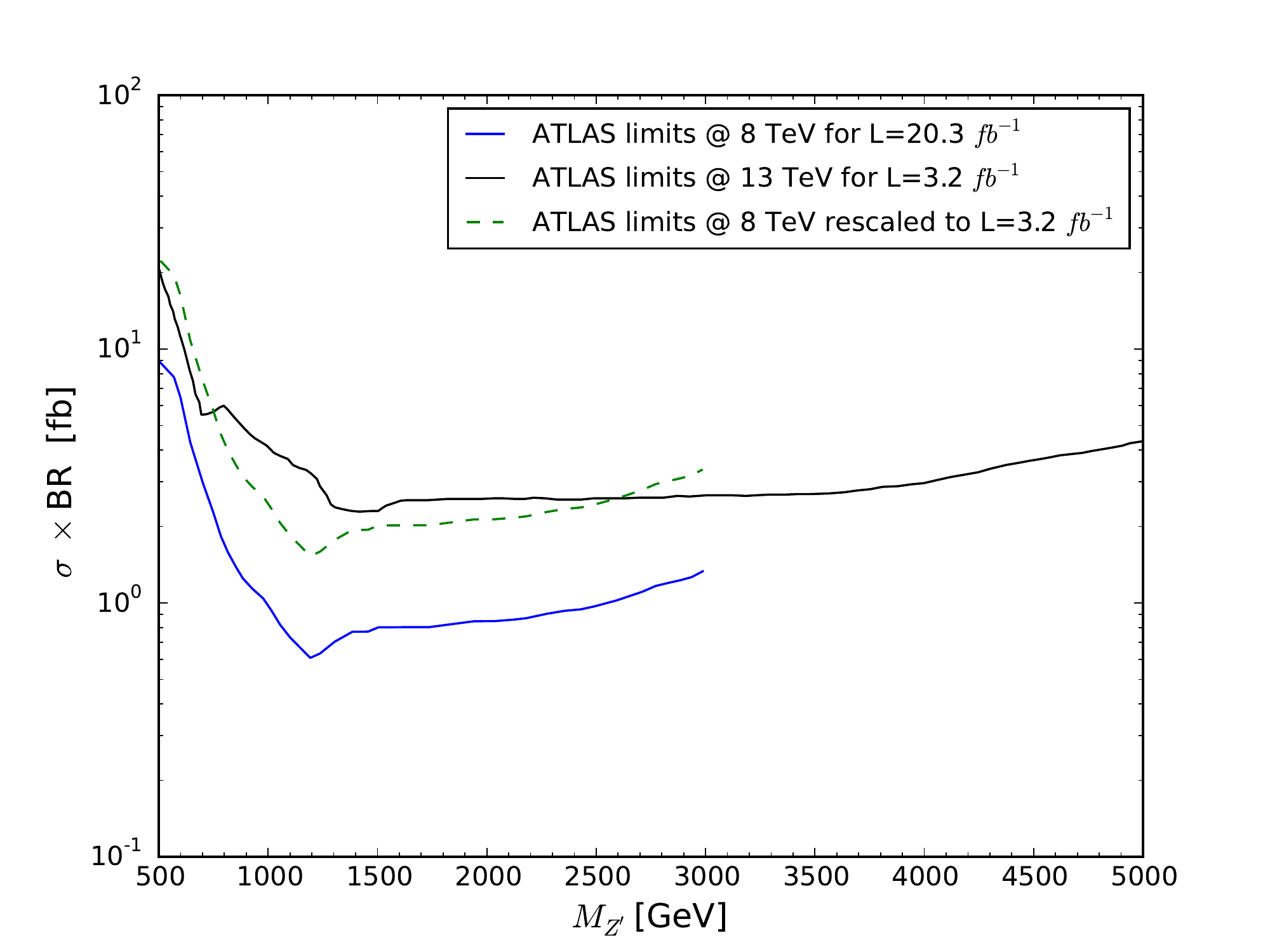}
\end{subfigure} 
\caption{(Left) Observed exclusion limits on the branching fraction times cross 
section in the $e\mu$ channel \cite{Aad:2015pfa}. The limits have 
been extrapolated down to masses of 200 GeV by applying a constant 
continuation.  (Right) The blue curve shows the exclusion limits on the 
branching fraction times cross section at $\sqrt{s}=8$ TeV given in 
Ref.\cite{Aad:2015pfa}. The black curve shows the preliminary limit 
at $\sqrt{s}=13$ TeV given in Ref.\cite{ATLAS-CONF-2015-072}. The green curve 
depicts the  8 TeV limit rescaled to the 13 TeV luminosity. The order of 
magnitude agreement between this curve and the 13 TeV limit in this region 
indicate that our luminosity rescaling is sensible.}
\label{fig_zplim}
\end{figure}
The limits shown in \cref{fig_bsem,fig_bsmt,fig_bdem} were derived for a $\zp$ 
mediator of $M_\zp = 750$ GeV. In general, we have derived these limits for 
various masses in the range of 200 GeV to 3 TeV for the  flavor combinations 
$qq'\in\{sd,bs,bd,cu\}$ and $\ell\ell'\in\{e\mu,e\tau,\mu\tau\}$. In 
Appendix~\ref{sec_app_plots} an example exclusion plot for each flavor 
combination is given for a $\zp$ boson with $M_\zp=1$ TeV. Under~\href{http://www.thphys.uni-heidelberg.de/~foldenauer/Zp_limits/}{\color{blue}{\texttt{http:$\slash\slash$www.thphys.uni-heidelberg.de$\slash\!\!\sim$foldenauer$\slash$Zp$_{-}$limits$\slash$}}}  and in the supplementary material to this paper the full set of 
exclusion plots can be found.
From these plots one can see that with increasing mass all limits weaken. Except for the case of direct production
at the LHC, limits typically scale as $g\sim M_{\zp}$.

We want to point out, however, that in order to obtain limits for
$\zp$ bosons with masses below $500$ GeV we have extrapolated the 
ATLAS limits rather optimistically. As shown in the left panel of 
\cref{fig_zplim}, we assume a constant scaling of limits down to masses of 
$M_\zp = 200$ GeV. Nevertheless, this seems to be justified as the ATLAS 
resonance search under consideration was designed for heavy mediators 
\cite{Aad:2015pfa}. Thus, in principle, a dedicated analysis in the 
low invariant mass range should yield better limits than those given in the 
analysis~\cite{Aad:2015pfa} we used.

Furthermore, we have projected the limits deduced from the ATLAS search at 
$\sqrt{s}=8$ TeV and $20.3\ \mathrm{fb}^{-1}$ of data to a Run 2 (LHC Run II) 
scenario with $\sqrt{s}=13$ TeV and $100\ \mathrm{fb}^{-1}$ and a high 
luminosity scenario (HL-LHC) with $\sqrt{s}=13$ TeV and $3000\ 
\mathrm{fb}^{-1}$. For this purpose we have rescaled the exclusion limits on 
the cross section by the respective luminosities
\begin{align}
 \sigma^{(13)}_\mathrm{lim} = \sqrt{\frac{\int \mathrm{d}t \ 
\mathcal{L}_8}{\int \mathrm{d}t \ \mathcal{L}_{13}}} \ 
\sigma^{(8)}_\mathrm{lim} \,,
\end{align}
assuming that scaling of the limits is only due to statistics.
As a cross-check for this prescription to work, we have compared luminosity 
rescaled limits from the ATLAS 8 TeV analysis to the preliminary limits from the 
ATLAS analysis at 13 TeV \cite{ATLAS-CONF-2015-072} in the $e\mu$-channel. 
As shown in the right panel of \cref{fig_zplim}, this method seems to give 
sufficiently accurate results for the purpose of a rough projection.
The projection to the LHC Run II scenario is illustrated by the red 
dash-dotted line in \cref{fig_bsem,fig_bsmt,fig_bdem}, the projection to the 
HL-LHC scenario by the red dotted line.

\section{Constraints}
\label{sec_constraints}

In this section we want to give a brief overview over the different 
constraints that already restrict our model. We will first consider pure 
quark-sector, 
then mixed quark- and lepton-sector and finally pure lepton-sector constraints.

\subsection{Meson mixing}
\label{sec_mixing}

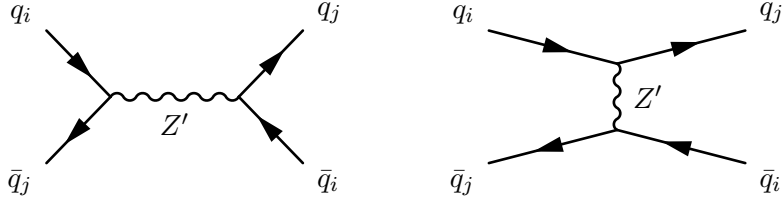
\begin{figure}[!t]
\begin{fmffile}{mesonmix_zp}
\begin{equation*}
\begin{gathered}
  \begin{fmfgraph*}(120,50)
    \fmfleft{i1,i2}
    \fmfright{o1,o2}
     
    \fmflabel{$\bar{q}_j$}{i1}
    \fmflabel{$q_i$}{i2}
    \fmflabel{$\bar{q}_i$}{o1}
    \fmflabel{$q_j$}{o2}
    
    \fmf{fermion}{i2,v1,i1}
    \fmf{fermion}{o1,v2,o2}
    
    \fmf{photon,label=$\zp$}{v1,v2}
  \end{fmfgraph*}
\end{gathered}
\qquad \qquad
\begin{gathered}
  \begin{fmfgraph*}(120,50)
    \fmfleft{i1,i2}
    \fmfright{o1,o2}
     
    \fmflabel{$\bar{q}_j$}{i1}
    \fmflabel{$q_i$}{i2}
    \fmflabel{$\bar{q}_i$}{o1}
    \fmflabel{$q_j$}{o2}
    
    \fmf{fermion}{o1,v1,i1}
    \fmf{fermion}{i2,v2,o2}
    
    \fmf{photon,label=$\zp$}{v1,v2}
  \end{fmfgraph*} 
\end{gathered}
\end{equation*}
\end{fmffile}
\caption{Tree-level meson mixing via $\zp$.}
\label{fig_mesonmix}
\end{figure}

One of the strongest probes of flavor violation in the quark sector is provided 
by meson mixing, where a meson $\C M$ oscillates into its conjugate state 
$\bar{\C M}$. In the Standard Model these processes are loop-suppressed as they 
require 
FCNCs and thus the matrix elements are rather small. Therefore, meson  mixing is 
very sensitive to new physics models that have 
flavor-changing couplings in the quark sector. This is the case for 
our model where meson mixing arises at tree level via $\zp$ exchange from the  
diagrams shown in \cref{fig_mesonmix}. \\
As the mass splitting $\Delta M_{\C M}$ of the conjugate 
meson states is directly proportional to the transition matrix element $M_{12}$,
\begin{align}
 \Delta M_{\C M} = 2 \, \Re(M_{12})  \propto g^2_{q_iq_j} \, ,
\end{align}
it is the appropriate observable for testing flavor violation. As meson mixing 
is a 
low-energy effect we follow Ref.~\cite{Buras:2012fs} and investigate the $\zp$ 
effects in an EFT approach, 
where we will integrate out the $\zp$ at the high scale $\mu_\mathrm{in} \sim 
M_\zp$. The resulting 
four-quark operators describing the low-energy phenomenology of the 
$\zp$-induced FCNCs are given by \cite{Buras:2012fs,Buras:2000if}
\begin{align}
\C O^\mathrm{VLL}_1 &= (\bar{q}_i\,\gamma^\mu P_{L}\, q_j)(\bar{q}_i\, 
\gamma^\mu 
P_{L}\, q_j)\,,  \label{eq_mixop1}\\
\C O^\mathrm{VRR}_1 &= (\bar{q}_i\,\gamma^\mu P_{R}\, q_j)(\bar{q}_i\, 
\gamma^\mu 
P_{R}\, q_j)\,, \label{eq_mixop2}\\
\C O^\mathrm{LR}_1 &= (\bar{q}_i\,\gamma^\mu P_{L}\, q_j)(\bar{q}_i\, 
\gamma^\mu 
P_{R}\, q_j)\,, \label{eq_mixop3}\\
\C O^\mathrm{LR}_2 &= (\bar{q}_i\,P_{L}\, q_j)(\bar{q}_i\,  
P_{R}\, q_j)\,. \label{eq_mixop4}
\end{align}
It should be noted that the operator $\C O^\mathrm{LR}_2$ is generated only 
through 
QCD-loop effects from operator mixing due to the running of operators from the
high to the low scale.

\begin{figure}[t!]
 \includegraphics[width=\textwidth]{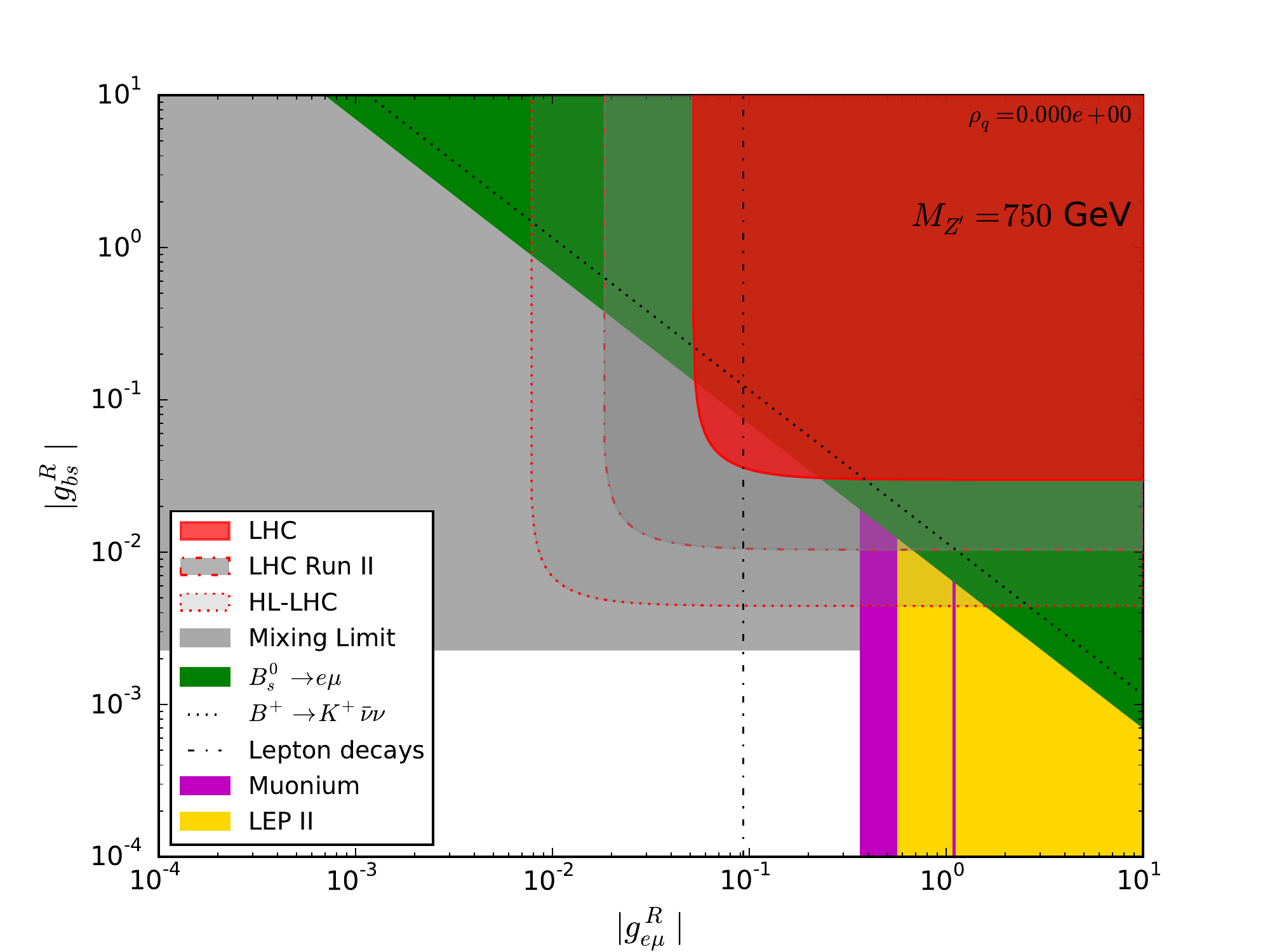}
 \caption{The flavor-violating couplings $g^R_{bs}$ and $g^R_{e\mu}$ for a 
$\zp$ boson of $M_\zp=750$ GeV with purely right-handed quark couplings 
($\rho_q=0$). The red area indicates the limit from the 
ATLAS 
analysis of the process $pp\rightarrow e\mu$ at $\sqrt{s}=8$ TeV. The red 
dash-dotted and dashed lines are projections to the LHC Run II and HL-LHC. The green area is the limit coming from the meson decay 
$B_s^0\rightarrow e\mu$. The gold and magenta areas are purely leptonic limits 
coming from LEP and muonium oscillation constraints. The black dotted 
line is the neutrino exclusion limit from $B^+\rightarrow K^+\bar{\nu}\nu$. The dash-dotted line originates from tests of lepton decays. However, 
these latter two limits only apply for left-handed lepton couplings. The gray area represents the 
$B_s-\bar{B}_s$ mixing limit.}
\label{fig_mixlim}
\end{figure}

After matching the operators of \cref{eq_mixop1,eq_mixop2,eq_mixop3,eq_mixop4} 
to the full 
theory we find with the off-diagonal matrix element $M^*_{12} = 
\bra\overline{\C M}|\mathcal{H}_\mathrm{eff}^{\Delta S = 2}|\C M 
\ket$ for the mass splitting \cite{Buras:2012fs}\footnote{We use $P^{{\rm VLL}}_{1}=P^{{\rm VRR}}_{1}$
to simplify the expression.}
\begin{align}
\Delta M^\mathrm{NP}_{\C M} = \frac{M_{\C M}\,f_{\C M}^2}{3} 
\,\frac{\left(g^R_{ij}\right)^2}{M^2_\zp}\, \Big[  
R^\mathrm{VLL}_1(\mu) \, P^\mathrm{VLL}_1\, (1+\rho_q^2) + 
\left(R^\mathrm{LR}_1(\mu) 
\, P^\mathrm{LR}_1 + R^\mathrm{LR}_2(\mu) \, 
P^\mathrm{LR}_2\right)\, \rho_q \Big] \,,  
\label{eq_masssplit}
\end{align}
where the $P_i$  denote the hadronic matrix elements corresponding to the 
operators $\C O_i$. We calculate the hadronic matrix elements for $K,B_d$ 
and $B_s$ mesons mainly from the relations given in 
Refs.~\cite{Buras:2012fs,Buras:2000if} and the lattice bag parameters from 
quenched QCD calculations given in \cite{Babich:2006bh,Becirevic:2001xt}. For 
$D$ mesons we rely on the relations given in Ref.~\cite{Golowich:2007ka}. The 
$R_i(\mu)$ are the renormalization group evolution coefficients  
encoding the running of the operators $\C O_i$ due to NLO QCD effects. They are 
normalized such that $R_i(\mu_\mathrm{in})=1$ at the scale $\mu_\mathrm{in}$ 
where the $\zp$ is integrated out. The coefficients are given by 
\cite{Buras:2012fs}
\begin{align}
 R^\mathrm{VLL}_1(\mu) = R^\mathrm{VRR}_1(\mu) &=  1 + \frac{\alpha_s}{4\pi} 
\left( \frac{11}{3} - 2 \, 
\log\frac{M^2_\zp}{\mu^2}\right) \,, \\
 R^\mathrm{LR}_1(\mu)  &=  1 - \frac{\alpha_s}{4\pi} \left( \frac{1}{6} + 
\log\frac{M^2_\zp}{\mu^2}\right)\, , \\
 R^\mathrm{LR}_2(\mu)  &= \quad  - \frac{\alpha_s}{4\pi} \left( 1 + 6 \, 
\log\frac{M^2_\zp}{\mu^2}\right)\, .
\end{align}
From the measurement of the 
meson mass splitting $\Delta M_{\C M}$ we can derive a limit on the 
coupling  
$g^R_{ij}$. In practice, we use the values provided by the {\textsc{UTfit}} collaboration~\cite{utfit} and more specifically the maximally allowed deviation between the 
measurement and the SM prediction.

In \cref{fig_mixlim} we show as an example again the limits for the 
combination $\{qq',\ell\ell'\} = \{bs,e\mu\}$. We consider purely right-handed 
quark couplings, i.e. $\rho_q=0$, and include the mixing limit as the gray 
area. It can be 
seen that the mixing limit in the quark sector, in general, is so strong that 
it excludes the whole region of parameter space that can be probed with 
multipurpose detectors at the LHC.

\subsubsection{Cancellation}
\label{sec_cancel}

We have just seen that for a general quark coupling configuration it seems 
hopeless to test flavor violation with ATLAS at the LHC. However, there is an 
important 
subtlety to these considerations that alters the picture just enough to serve 
as motivation for a search of $\zp$ induced FCNCs at the LHC. \par
The term in brackets in \cref{eq_masssplit} is a quadratic form in the 
parameter $\rho_q$. Thus, if the discriminant 
\begin{equation}
 \Delta = \Big(R^\mathrm{LR }_1(\mu) \, P^\mathrm{LR}_1 + R^\mathrm{LR}_2(\mu) 
\,P^\mathrm{LR}_2\Big)^2 - 4\ R^\mathrm{VLL}_1(\mu) \, P^\mathrm{VLL}_1 
\Big(R^\mathrm{LR}_1(\mu) \, P^\mathrm{LR}_1 + R^\mathrm{LR}_2(\mu) \, 
P^\mathrm{LR}_2\Big) \,,
\end{equation}
is greater than zero, we have two solutions $\rho_{0}$ for which the 
mass splitting due to $\zp$ exchange vanishes exactly. 
We will only consider the solution that has mostly right-handed couplings. The other solution is simply given by $1/\rho_{0}$ and has essentially the same behavior. \\
As we are interested in comparing flavor bounds on the $\zp$ couplings to collider 
bounds, we additionally define an upper and a lower tolerance  
$\Delta\rho_-$ and $\Delta\rho_+$ such that  the upper limit on the 
coupling $g^\mathrm{lim}_{ij}$ derived from mixing is less stringent than some 
reference limit 
$g^*$, i.e.
\begin{equation}
  \forall\ \rho_q \in I_0:= [\rho_0-\Delta\rho_-,\ \rho_0+\Delta\rho_+]:\  
g^\mathrm{lim}_{ij}(\rho_q) \geq g^* \,.
\end{equation}
With this definition we can find an interval $I_0$ around the 
exact cancellation point $\rho_0$, where the limit due to mixing is subdominant 
compared to the reference limit  $g^* = \min( g^\mathrm{col}_{ij})$, i.e. to 
the strongest bound we can set on $g^q_{ij}$ from the ATLAS search at a given 
mass $M_\zp$. For example, the exclusion plot in \cref{fig_bsem} shows the 
limits for a $\zp$ boson with  $M_\zp=750$ GeV coupling to ${bs}$ and ${e\mu}$. 
As can be seen in the plots the limits hold for $\rho_0=0.1324$ with 
$\Delta\rho_-=0.0005$ and $\Delta\rho_+=0.0019$, i.e. the tolerance interval in 
this case is $I_0=\[0.1319,0.1343\]$. 
\par
So far we have treated cancellation effects only at tree level. However, we 
have to make sure that these effects are persistent even at higher orders. 
Therefore, we investigate the impact of one-loop corrections on the 
cancellation solution $\rho_0$ and the associated tolerance interval $I_0$. The 
details can be found in Appendix~\ref{sec_app_cancel}. We find that the 
cancellation solution $\rho_0$ and the interval $I_0$ both receive an overall 
shift $\delta\rho$ at one-loop level. However, this shift is negligible in the 
sense that $\delta\rho < \rho_0$ in the region of parameter space that is 
probed by LHC bounds.

\subsection{Meson decays}

Another process in the meson sector that directly constrains our model is rare neutral meson decays ${\C M}^0\rightarrow\ell^+\ell^{'-}$, where ${\C 
M}^0$ can 
be the $K, D,B_d$ or $B_s$ meson. These decays involve two flavor-changing 
vertices and therefore are highly suppressed in the SM, whereas we can generate 
these processes on tree-level in our $\zp$ model. From the Lagrangian in 
\cref{eq_gen_lag} we can immediately construct the relevant four-fermion 
operators \cite{Golowich:2009ii} 
\begin{align}
 \C O_1 =  (\bar{\ell} \, \gamma_\mu \,P_L\, \ell')(\bar{q} \, \gamma^\mu \, 
P_L\, 
q') \,, && \C O_6 =\C O_1(L \leftrightarrow R)\,, \label{eq_mesops1}\\
 \C O_2 = (\bar{\ell} \, \gamma_\mu \,P_R\, \ell')(\bar{q} \, \gamma^\mu \, 
P_L\, 
q') \,, &&\C O_7 =\C O_2(L\leftrightarrow R) \,, \label{eq_mesops2}
\end{align}
with corresponding Wilson coefficients
\begin{align}
 C_1 = \frac{g^L_{\ell\ell'}\,g^L_{qq'} }{M^2_\zp}  \,, && C_2 = 
\frac{g^R_{\ell\ell'}\,g^L_{qq'} }{M^2_\zp} \,, && C_6 = 
\frac{g^R_{\ell\ell'}\,g^R_{qq'} }{M^2_\zp}  \,, && C_7 = 
\frac{g^L_{\ell\ell'}\,g^R_{qq'} }{M^2_\zp}  \,. \label{eq_mescoeffs}
\end{align}
With knowledge of the relevant operators and their associated Wilson 
coefficients we can calculate the branching ratio 
$\mathrm{BR}({\C M}^0\rightarrow\ell^+\ell^{'-})$ for the different mesons 
\cite{Golowich:2009ii,Golowich:2011cx}
\begin{align}
 \mathrm{BR}({\C M}^0 \rightarrow \ell^+\ell'^-) = & \frac{f^2_{{\C M}^0}\ 
M_{{\C M}^0}\ m^2_{\ell}}{ 32\,\pi\,\Gamma_{{\C M}^0}\,M^4_\zp}\ 
\left(1-\frac{m^2_\ell}{M^2_{{\C M}^0}}\right)^2 \, 
(g^L_{\ell\ell'}-g^R_{\ell\ell'})^2\left(g^L_{qq'}-g^R_{qq'}\right)^2 \,,
\label{eq_br_decay}
\end{align}
where $f_{{\C M}^0}, M_{{\C M}^0}$ and $\Gamma_{{\C M}^0}$ are the decay 
constant, the mass and the total width of the decaying meson. Furthermore, we 
have assumed that $\ell$ is the heavier of the two 
leptons and we have neglected the mass of the other one. Of course, these limits only exist when the mass of the relevant meson is bigger than the combined mass of the two leptons.\\
Finally, based on \cref{eq_br_decay} we can derive a limit on the $\zp$ coupling
 \begin{equation}
  |g^R_{qq'}| \leq \left[ \frac{32\,\pi\ \text{BR}({\C M}^0 \rightarrow \ell^+ 
\ell'^-) \, \Gamma_{{\C M}^0}\, M^4_\zp }{M_{{\C M}^0}\, 
f^2_{{\C M}^0}\, m^2_\ell \,\left(1-\frac{m^2_\ell}{M^2_{{\C M}^0}}\right)^2} 
\right]^\frac{1}{2}  \ 
\frac{1}{g^R_{\ell\ell'}\ |1-\rho_\ell|\ |1-\rho_q|} \,,
\label{eq_mesondec}
 \end{equation}
where we have made use of the relations in \cref{eq_rhodef}. The corresponding 
limits due to meson decays are depicted in \cref{fig_bsem,fig_bsmt,fig_bdem} 
by the green area. The power of the decay limits comes from the fact that it 
constrains the product $g_{qq'}g_{\ell\ell'}$. Therefore, meson decays can 
probe regions in parameter space where one of the two coupling is very small 
while the other one is big, a region hard to probe at the LHC. However, the LHC 
limits are generally more stringent in the direction of parameter space where 
both couplings become small but are of comparable size. Especially with Run II or HL-LHC data one can 
expect rather big gains along that direction.

\subsection{Neutrino limits}

It seems reasonable that an extension of the Standard Model should preserve the 
$\mathrm{SU}(2)_L$ gauge symmetry at high energies. In return this means for 
our effective model  that the $\zp$ gauge boson should couple to the quark and 
lepton doublets $Q_L$ and $L_L$, if left-handed couplings are present. In this 
scenario the $\zp$ couples  to neutrinos $\nu_i$ and $\nu_j$ with equal 
strength $g^L_{\ell_i\ell_j}$ as to charged leptons $\ell_i$ and $\ell_j$.  
 However, the coupling to neutrinos opens up a whole new class of constraints 
to our model. Especially, meson decays of the form ${\C M}_1^{0,\pm} 
\rightarrow{\C M}_2^{0,\pm} \, \bar{\nu} \,\nu$ can be a sensitive probe for 
the presence of left-handed $\zp$ couplings.\par
In this section we will now investigate such decays for the different neutral 
mesons that can have an impact on our model. In particular we consider decays of kaons and $B$-mesons. The corresponding measurements in the $D$-meson sector are not yet very restricting.

\subsubsection{Kaons}

\label{sec_kaonneutrinos}

\begin{center}
\begin{figure}[!t]
\begin{adjustwidth}{-\oddsidemargin-1in}{-\rightmargin-1in}
    \begin{fmffile}{kp2pip_sm}
    \begin{equation*}
    \begin{gathered}
    \begin{fmfgraph*}(100,90)
	\fmfstraight
	\fmfleft{i0,i1,i1b,i2,i3,i4,i5}
	\fmfright{o0,o1,o1b,o2,o3,o4,o5}
	
	\fmfv{l=$u$,l.a=-50}{i5}
	\fmfv{l=$u$,l.a=-140}{o5}
	\fmfv{l=$\bar{s}$,l.a=50}{i2}
	\fmfv{l=$\bar{d}$,l.a=140}{o2}

        \fmflabel{$\bar{\nu}_\ell$}{o1}
        \fmflabel{$\nu_\ell$}{o0}
       
	\fmf{fermion,tension=1}{i5,o5}
	\fmf{fermion,tension=1}{v1,i2}
	\fmf{fermion,tension=1}{o2,v2}
	\fmf{phantom}{v2,v1}
	\fmffreeze
	\fmf{phantom}{v2,v3,v1}
	\fmf{phantom}{i0,v3,o0}
	\fmf{fermion,left=0.35,label=$\bar{t}$}{v2,v3,v1}
	\fmf{photon,right,label=$W$}{v2,v1}
	\fmffreeze
	\fmf{fermion}{o1,v4}
	\fmf{photon,label=$Z$,label.side=left,tension=2}{v4,v3}
	\fmf{fermion}{v4,o0}
	\fmf{phantom,tension=1}{i0,v4}
	\fmfv{l=$K^+$,l.a=-135,l.d=.05w}{i4}
	\fmfv{l=$\pi^+$,l.a=-45,l.d=.05w}{o4}
    \end{fmfgraph*} 
    \end{gathered}   
    \qquad \qquad \qquad  
    \begin{gathered}
    \begin{fmfgraph*}(100,90)
	\fmfstraight
	\fmfleft{i0,i1,i1b,i2,i3,i4,i5}
	\fmfright{o0,o1,o1b,o2,o3,o4,o5}
	
	\fmfv{l=$u$,l.a=-50}{i5}
	\fmfv{l=$u$,l.a=-140}{o5}
	\fmfv{l=$\bar{s}$,l.a=50}{i2}
	\fmfv{l=$\bar{d}$,l.a=140}{o2}

        \fmflabel{$\bar{\nu}_\ell$}{o1}
        \fmflabel{$\nu_\ell$}{o0}
       
       \fmf{fermion,tension=1}{i5,o5}
	\fmf{fermion,tension=1}{v1,i2}
	\fmf{fermion,label=$\bar{t}$}{v2,v1}
	\fmf{fermion,tension=1}{o2,v2}
	\fmffreeze
	\fmf{fermion}{o1,v4}
	\fmf{fermion,label=$\ell^+$}{v4,v3}
	\fmf{phantom}{i1,v3}
	\fmffreeze
	\fmf{fermion}{v3,o0}
	\fmf{photon,label=$W$,label.side=left,tension=2}{v2,v4}
	\fmf{photon,label=$W$,tension=2}{v1,v3}
	\fmf{phantom}{i0,v3}
	\fmfv{l=$K^+$,l.a=-135,l.d=.05w}{i4}
	\fmfv{l=$\pi^+$,l.a=-45,l.d=.05w}{o4}
    \end{fmfgraph*} 
    \end{gathered} 
    \qquad \qquad \qquad  
    \begin{gathered}
    \begin{fmfgraph*}(100,90)
	\fmfstraight
	\fmfleft{i0,i1,i1b,i2,i3,i4,i5}
	\fmfright{o0,o1,o1b,o2,o3,o4,o5}
	
	\fmfv{l=$u$,l.a=-50}{i5}
	\fmfv{l=$u$,l.a=-140}{o5}
	\fmfv{l=$\bar{s}$,l.a=50}{i2}
	\fmfv{l=$\bar{d}$,l.a=140}{o2}

        \fmflabel{$\bar{\nu}_\ell$}{o1}
        \fmflabel{$\nu_\ell$}{o0}
       
       \fmf{fermion,tension=1}{i5,o5}
	\fmf{fermion,tension=1}{v1,i2}
	\fmf{fermion,label=$\bar{t}$}{v2,v1}
	\fmf{fermion,tension=1}{o2,v2}
	\fmffreeze
	\fmf{photon,label=$W$,label.side=left}{v2,v3}
	\fmf{photon,label=$W$}{v1,v3}
	\fmf{phantom}{i0,v3,o0}
	\fmffreeze
	\fmf{fermion}{o1,v4}
	\fmf{photon,label=$Z$,label.side=left,tesnion=2}{v4,v3}
	\fmf{fermion}{v4,o0}
	\fmf{phantom}{i0,v4}
	\fmfv{l=$K^+$,l.a=-135,l.d=.05w}{i4}
	\fmfv{l=$\pi^+$,l.a=-45,l.d=.05w}{o4}
    \end{fmfgraph*} 
    \end{gathered}       
    \end{equation*}
    \end{fmffile}
\hspace*{1cm}
\caption{Leading order SM contributions to the decay $K^+\rightarrow 
\pi^+\nu\bar{\nu}$.}
\label{fig_kp2pip_sm}
\end{adjustwidth}
\end{figure}
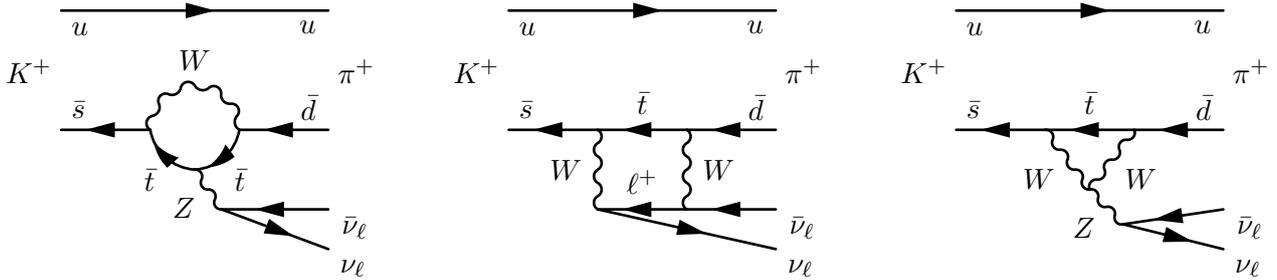
\end{center}

\begin{figure}[!t]
\centering
    \begin{fmffile}{kp2pip_zp}
    \begin{equation*} 
    \begin{gathered}
    \begin{fmfgraph*}(150,100)
	\fmfstraight
	\fmfleft{i0,i1,i1b,i2,i2b,i3,i4}
	\fmfright{o0,o1,o1b,o2,o2b,o3,o4}
	
        \fmfv{l=$u$,l.a=-50}{i3}
	\fmfv{l=$\bar{s}$,l.a=50}{i2}
	\fmfv{l=$u$,l.a=-140}{o3}
	\fmfv{l=$\bar{d}$,l.a=140}{o2}
	
        \fmflabel{$\bar{\nu}_i$}{o1}
        \fmflabel{$\nu_j$}{o0}
        
	\fmf{fermion}{i3,o3}
	\fmf{fermion,tension=1.5}{v2,i2}
	\fmf{fermion}{o2,v2}
	\fmffreeze
	\fmf{fermion}{o1,v3,o0}
	\fmf{photon,label=$\zp$,tension=2}{v2,v3}
	\fmf{phantom}{i0,v3}
	\fmfv{l=$g_{sd}$,l.a=90}{v2}
	\fmfv{l=$g_{\ell_i\ell_j}$,l.a=-90}{v3}
	\fmfv{l=$K^+$,l.a=180,l.d=.05w}{i2b}
	\fmfv{l=$\pi^+$,l.a=0,l.d=.05w}{o2b}
    \end{fmfgraph*} 
    \end{gathered}
    \end{equation*}
    \end{fmffile}
\hspace*{1cm}
\caption{Leading order $\zp$ contribution to the decay 
$K^+\rightarrow\pi^+\nu\bar{\nu}$.}
\label{fig_kp2pip_zp}
\end{figure}

\begin{figure}[!t]
\centering
    \begin{fmffile}{kp2pi0}
    \begin{equation*}
    \begin{gathered}
    \begin{fmfgraph*}(150,100)
	\fmfstraight
	\fmfleft{i0,i1,i1b,i2,i2b,i3,i4}
	\fmfright{o0,o1,o1b,o2,o2b,o3,o4}
	
        \fmfv{l=$u$,l.a=-50}{i3}
	\fmfv{l=$\bar{s}$,l.a=50}{i2}
	\fmfv{l=$u$,l.a=-140}{o3}
	\fmfv{l=$\bar{u}$,l.a=140}{o2}
	
        \fmflabel{$e^+$}{o1}
        \fmflabel{$\nu_e$}{o0}
        
	\fmf{fermion}{i3,o3}
	\fmf{fermion,tension=1.5}{v2,i2}
	\fmf{fermion}{o2,v2}
	\fmffreeze
	\fmf{fermion}{o1,v3,o0}
	\fmf{photon,label=$W$,tension=2}{v2,v3}
	\fmf{phantom}{i0,v3}
	\fmfv{l=$V^*_{us}$,l.a=90}{v2}
	\fmfv{l=$K^+$,l.a=180,l.d=.05w}{i2b}
	\fmfv{l=$\pi^0$,l.a=0,l.d=.05w}{o2b}
    \end{fmfgraph*} 
    \end{gathered}
    \end{equation*}
    \end{fmffile}
\hspace*{1cm}
\caption{Decay $K^+\rightarrow\pi^0\,e^+\,\nu_e$ in the SM.}
\label{fig_kp2pi0}
\end{figure}
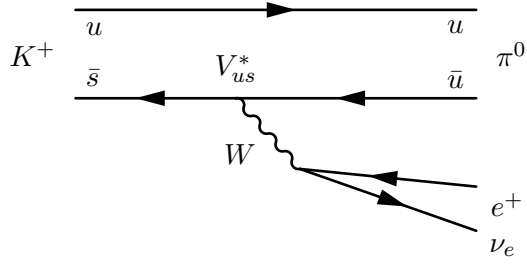

When we couple the $\zp$ to the first two quark generations we can constrain 
the left-handed lepton couplings from the kaon decay 
$K^+\rightarrow\pi^+\nu\bar{\nu}$. In order to extract a constraint on 
$g^L_{\ell_i\ell_j}$ we will calculate in the following the branching ratio 
$\mathrm{BR}(K^+\rightarrow\pi^+\nu\bar{\nu})$. For a detailed derivation of 
the branching fractions see Ref. \cite{Buras:1998raa}. \\
The relevant leading-order SM diagrams contributing to this decay are shown in 
\cref{fig_kp2pip_sm}. We can see that the leading SM contributions are 
already loop-suppressed. Thus, with the leading order $\zp$ contribution being 
a tree-level effect as shown in \cref{fig_kp2pip_zp}, one can expect that the 
$\zp$ can give a sizable contribution to the branching
fraction of this decay. Adopting the notation of 
Ref. \cite{Buras:1998raa}, where 
$(\bar{s}d)_{V\pm A}\equiv \bar{s}\gamma_\mu(1\pm\gamma_5)d$, the relevant 
operators for the low energy interaction read
\begin{align}
\C O_{K^0}^L = (\bar{s}d)_{ V-A }\ (\bar{\nu}_i\nu_j)_{V-A}\,, && \C O_{K^0}^R 
= (\bar{s}d)_{ V+A }\ (\bar{\nu}_i\nu_j)_{V-A} \,,
\end{align}
with corresponding Wilson coefficients
\begin{align}
 C_{K^0}^L = \frac{g^L_{sd} \,g^L_{\ell_i\ell_j}}{4\,M^2_\zp}\,, && 
C_{K^0}^R= \frac{g^R_{sd} \,g^L_{\ell_i\ell_j}}{4\,M^2_\zp}\,.
\end{align}
In order to calculate the branching ratio we will make use of 
isospin symmetry \cite{Buras:1998raa} to extract the hadronic matrix 
element for $(\bar{s}d)_{V-A}$ from the decay $K^+\rightarrow\pi^0\,e^+\,\nu_e$ 
shown in \cref{fig_kp2pi0}, 
\begin{equation}
 \bra\pi^+|(\bar{s}d)_{V-A}|K^+\ket = \sqrt{2} \, 
\bra\pi^0|(\bar{s}u)_{V\pm A}|K^+\ket \,.
\end{equation}
Additionally we take the hadronic matrix elements of the left- and right-handed 
currents to be the same \cite{Carrasco:2015pra} as the process of interest is 
purely 
governed by QCD and therefore should be independent of the underlying chirality 
structure.
The effective operator for the process in \cref{fig_kp2pi0} 
reads
\begin{equation}
 \C{O}_{K^+}^L = (\bar{s}u)_{V-A}\,
(\bar{\nu}_e e)_{V-A} \,,
\end{equation}
with coefficient
\begin{equation}
 C_{K^+}^L = \frac{G_F}{\sqrt{2}} V^*_{us}\,,
\end{equation}
where $V$ denotes the CKM matrix.
Thus, we can write for the branching ratios
\begin{align}
\frac{\mathrm{BR}(K^+\rightarrow\pi^+\nu\bar{\nu})}{\mathrm{BR}
(K^+\rightarrow\pi^0e^+\bar{\nu})} &=  \left( 
\frac{\sqrt{2}\,g^L_{\ell_i\ell_j}}{4\,G_F\,M^2_\zp} \right)^2 \ 
\frac{\left|g^L_{sd}\,\bra\pi^+|(\bar{s}d)_{V-A}|K^+\ket+g^R_{sd}\,
\bra\pi^+|(\bar{s} d)_{V+A}|K^+\ket\right|^2}{|V^*_{us}|^2\ 
|\bra\pi^0|(\bar{s}u)_{V-A} |K^+\ket |^2 } 
\\
&=\frac{g^{L^2}_{\ell_i\ell_j}\ g^{R^2}_{sd}\ 
|1+\rho_q|^2}{4\,|V^*_{us}|^2\,G^2_F\,M^4_\zp} 
\,.
\end{align}
To turn this into a limit on $g^{R}_{sd}$ we only consider the part of the 
branching fraction not explained by the SM 
\begin{equation}
 \mathrm{BR}(K^+\rightarrow\pi^+\nu\bar{\nu})^\mathrm{NP} = 
\mathrm{BR}(K^+\rightarrow\pi^+\nu\bar{\nu})^\mathrm{exp}  - 
\mathrm{BR}(K^+\rightarrow\pi^+\nu\bar{\nu})^\mathrm{SM} \,,
\end{equation}
where we used for our analysis
\begin{align}
 \mathrm{BR}(K^+\rightarrow\pi^+\nu\bar{\nu})^\mathrm{exp} 
= (1.73^{+1.15}_{-1.05})\times10^{-10} \,, &&  
(\text{cf. Ref. \cite{Artamonov:2008qb}}) \\
\mathrm{BR}(K^+\rightarrow\pi^+\nu\bar{\nu})^\mathrm{SM} = 
(9.1\pm0.7)\times 10^{-11} \,. &&  (\text{cf. Ref. \cite{Buras:2015qea}} )
\end{align}
Finally, we obtain the constraint on the leptonic coupling to be given by
\begin{equation}
 |g^R_{sd}| \leq \frac{2 \, |V_{us}|\,G_F\,M^2_\zp}{|g^{L}_{\ell_i\ell_j}|\, 
|1+\rho_q|} \, \left[ 
\frac{\mathrm{BR}(K^+\rightarrow\pi^+\nu\bar{\nu})^\mathrm{NP}}{\mathrm{BR}
(K^+\rightarrow\pi^0e^+\bar{\nu})}\right]^\frac{1}{2} \,.
\label{eq_limknunu}
\end{equation}
The resulting limits on the $\zp$ couplings are depicted by the black dotted 
line in the lower left panels in \cref{fig_plots1,fig_plots2,fig_plots3} in 
\cref{sec_app_plots}. The neutrino limits in the kaon sector are quite strong 
and like the limits from decays into charged leptons constrain the product 
$g_{qq'}g_{\ell\ell'}$. Especially in the $e\tau$ and $\mu\tau$ sector the 
kaon-neutrino limits  exclude all regions of parameter space that can 
be hoped to be tested at the LHC. However, as mentioned in the beginning of this section, 
the neutrino limits are only valid if we take the lepton couplings to 
be left-handed and can be fully circumvented by only considering 
right-handed lepton couplings.\newline

\subsubsection{B mesons}

\begin{figure}[!t]
\centering
    \begin{fmffile}{b0nunu}
    \begin{equation*}
        \begin{gathered}
    \begin{fmfgraph*}(150,100)
	\fmfstraight
	\fmfleft{i0,i1,i1b,i2,i2b,i3,i4}
	\fmfright{o0,o1,o1b,o2,o2b,o3,o4}
	
        \fmfv{l=$d$,l.a=-50}{i3}
	\fmfv{l=$\bar{b}$,l.a=50}{i2}
	\fmfv{l=$d$,l.a=-140}{o3}
	\fmfv{l=$\bar{d}$,l.a=140}{o2}
	
        \fmflabel{$\bar{\nu}_i$}{o1}
        \fmflabel{$\nu_j$}{o0}
        
	\fmf{fermion}{i3,o3}
	\fmf{fermion,tension=1.5}{v2,i2}
	\fmf{fermion}{o2,v2}
	\fmffreeze
	\fmf{fermion}{o1,v3,o0}
	\fmf{photon,label=$\zp$,tension=2}{v2,v3}
	\fmf{phantom}{i0,v3}
	\fmfv{l=$g_{bd}$,l.a=90}{v2}
	\fmfv{l=$g_{\ell_i\ell_j}$,l.a=-90}{v3}
	\fmfv{l=$B^0$,l.a=180,l.d=.05w}{i2b}
	\fmfv{l=$\pi^0$,l.a=0,l.d=.05w}{o2b}
    \end{fmfgraph*} 
    \end{gathered}
    \qquad \qquad \qquad 
    \begin{gathered}
    \begin{fmfgraph*}(150,100)
	\fmfstraight
	\fmfleft{i0,i1,i1b,i2,i2b,i3,i4}
	\fmfright{o0,o1,o1b,o2,o2b,o3,o4}
	
        \fmfv{l=$d$,l.a=-50}{i3}
	\fmfv{l=$\bar{b}$,l.a=50}{i2}
	\fmfv{l=$d$,l.a=-140}{o3}
	\fmfv{l=$\bar{u}$,l.a=140}{o2}
	
        \fmflabel{$e^+$}{o1}
        \fmflabel{$\nu_e$}{o0}
        
	\fmf{fermion}{i3,o3}
	\fmf{fermion,tension=1.5}{v2,i2}
	\fmf{fermion}{o2,v2}
	\fmffreeze
	\fmf{fermion}{o1,v3,o0}
	\fmf{photon,label=$W$,tension=2}{v2,v3}
	\fmf{phantom}{i0,v3}
	\fmfv{l=$V^*_{ub}$,l.a=90}{v2}
	\fmfv{l=$B^0$,l.a=180,l.d=.05w}{i2b}
	\fmfv{l=$\pi^-$,l.a=0,l.d=.05w}{o2b}
    \end{fmfgraph*} 
    \end{gathered}
    \end{equation*}
    \end{fmffile}
\hspace*{1cm}
\caption{(Left) Decay $B^0\rightarrow\pi^0\,\bar{\nu}_i\,\nu_j$ mediated by the $\zp$ boson. (Right) SM decay $B^0\rightarrow\pi^-\,e^+\,\nu_e$ mediated by the $W$ boson.}
\label{fig_b0nunu}
\end{figure}
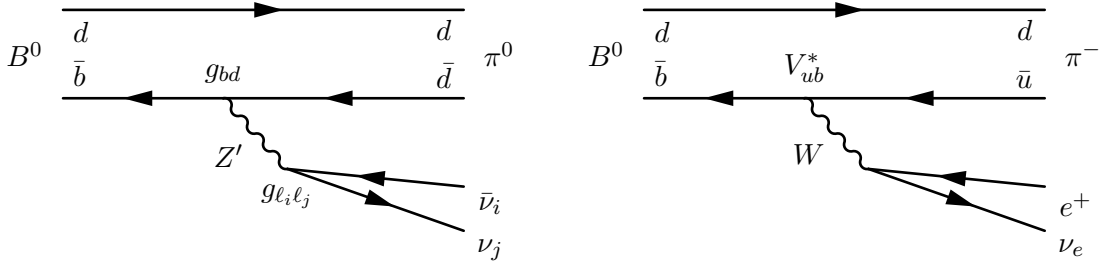

We can use the transition $b\rightarrow d\nu\bar\nu$ analogously to the case of the kaons. However, in this case the process that is induced through non-zero $g_{bd}$ and $g_{\ell_i\ell_j}$ couplings is $B^0\rightarrow\pi^0\,\bar{\nu}_i\,\nu_j$. This is shown in the left panel of \cref{fig_b0nunu}. As in the case for the kaons a limit on the coupling $g_{bd}$ can be derived from comparing the branching fraction of the latter $B^0$ decay to the one for the decay $B^0\rightarrow\pi^-\,e^+\,\nu_e$ (shown in the right panel of \cref{fig_b0nunu}). With the PDG values \cite{Olive:2016xmw} for the respective branching ratios 
\begin{align}
 \mathrm{BR}(B^0\rightarrow\pi^0\,\bar{\nu}\,\nu)^\mathrm{lim} & <  6.9\times10^{-5} \,, \\
\mathrm{BR}(B^0\rightarrow\pi^-\,e^+\,\nu_e) &= ( 1.45\pm0.05)\times10^{-4}\,,
\end{align}
we can then set a limit in analogy to \cref{eq_limknunu}\footnote{Since no detection of this process has been made and the resulting bounds are within regions that are already excluded we simply use the experimental limit on the branching ratio for our estimate.}
\begin{equation}
 |g^R_{bd}| \leq \frac{4 \, |V_{ub}|\,G_F\,M^2_\zp}{|g^{L}_{\ell_i\ell_j}|\, 
|1+\rho_q|} \, \left[ \frac{\mathrm{BR}(B^0\rightarrow\pi^0\,\bar{\nu}\,\nu)^\mathrm{lim} }{\mathrm{BR}(B^0\rightarrow\pi^-\,e^+\,\nu_e)}\right]^\frac{1}{2} \,.
\end{equation}

To constrain the transition $b\rightarrow s\nu\bar\nu$ we can use the decay 
$B^+\rightarrow K^+\nu\bar\nu$. To extract a limit on the the $\zp$ 
coupling $g_{bs}$ in the presence of left-handed lepton couplings we will 
again need the $\zp$ contribution to the branching ratio of this 
decay \cite{Buras:2014fpa}. First, we can 
parametrize any  contribution to this process in an EFT approach by 
the effective operators
\begin{equation}
 \C O^{L/R}_{B^0} = (\bar s \gamma_\mu 
P_{L/R}b)(\bar\nu\gamma^\mu (1-\gamma_5)\nu) \,.
\end{equation}
As for the kaons, the leading order SM contributions are 
coming from electroweak loop diagrams, which therefore are only involving 
left-handed fermions. The corresponding SM Wilson coefficient has been 
calculated \cite{Buras:2014fpa} and can be written as 
\begin{align}
 C_\mathrm{SM}^L = - \frac{e^2\,G_F}{\sqrt{2}\,\pi^2\,s_w^2}X_t \,, &&  X_t = 
1.469\pm0.017 \,, 
\end{align}
with $s_w=\sin\theta_w$ denoting the sine of the Weinberg angle.  
The leading-order contribution coming from the exchange of a $\zp$ as shown in 
\cref{fig_bp2kp} yields the Wilson coefficients
\begin{align}
 C_\zp^{L/R} = \frac{2\sqrt{2}\,\pi^2 \, g^{L/R}_{bs} \, 
g^L_{\ell_i\ell_j}}{e^2 
 \, V_{tb} V_{ts}^* \,  M^2_\zp} \,.
\end{align}
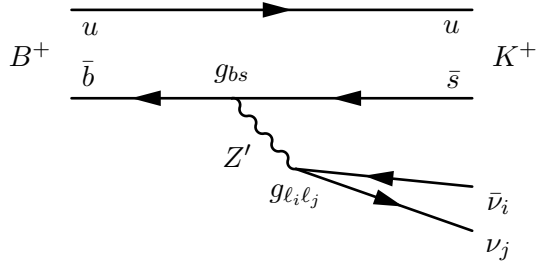
\begin{figure}[!t]
\centering
    \begin{fmffile}{bp2kp}
    \begin{equation*}   
    \begin{gathered}
    \begin{fmfgraph*}(150,100)
	\fmfstraight
	\fmfleft{i0,i1,i1b,i2,i2b,i3,i4}
	\fmfright{o0,o1,o1b,o2,o2b,o3,o4}
	
        \fmfv{l=$u$,l.a=-50}{i3}
	\fmfv{l=$\bar{b}$,l.a=50}{i2}
	\fmfv{l=$u$,l.a=-140}{o3}
	\fmfv{l=$\bar{s}$,l.a=140}{o2}
	
        \fmflabel{$\bar{\nu}_i$}{o1}
        \fmflabel{$\nu_j$}{o0}
        
	\fmf{fermion}{i3,o3}
	\fmf{fermion,tension=1.5}{v2,i2}
	\fmf{fermion}{o2,v2}
	\fmffreeze
	\fmf{fermion}{o1,v3,o0}
	\fmf{photon,label=$\zp$,tension=2}{v2,v3}
	\fmf{phantom}{i0,v3}
	\fmfv{l=$g_{bs}$,l.a=90}{v2}
	\fmfv{l=$g_{\ell_i\ell_j}$,l.a=-90}{v3}
	\fmfv{l=$B^+$,l.a=180,l.d=.05w}{i2b}
	\fmfv{l=$K^+$,l.a=0,l.d=.05w}{o2b}
    \end{fmfgraph*} 
    \end{gathered}
    \end{equation*}
    \end{fmffile}
\hspace*{1cm}
\caption{$\zp$ contribution to the decay 
$B^+\rightarrow K^+\nu\bar{\nu}$.}
\label{fig_bp2kp}
\end{figure}
Defining the differential branching fractions for the process as
\begin{equation}
 \frac{\mathrm{d\,BR}(B^+\rightarrow K^+\nu\bar\nu)}{\mathrm{d}q^2} \equiv 
\C{B}_K \,,
\end{equation} 
one finds for the ratio \cite{Buras:2014fpa}
\begin{align}
 \C{R}_K = \frac{\C{B}_K}{\C{B}^{SM}_K} = (1-2\eta)\epsilon^2 \,,
\end{align}
with the model-independent quantities 
\begin{align}
 \epsilon = \frac{\sqrt{|C^L|^2+|C^R|^2}}{|C^L_\mathrm{SM}|}\,, && \eta = 
\frac{- \mathrm{Re}(C^L\,C^{R^*})}{|C^L|^2+|C^R|^2} \,.
\end{align}
From the constraint that $\C{R}_K\leq4.3$ \cite{Buras:2014fpa} we can then 
derive a limit on the quark coupling 
\begin{equation}
 |g_{bs}^R| \leq \frac{\alpha\,|V_{tb} V_{ts}^*| 
\sqrt{2\,\C{R}_K}\,M^2_\zp}{\pi(1+\rho_q)\ |g^L_{\ell_i\ell_j}|} \ 
|C_\mathrm{SM}^L|\,.
\end{equation}
The limits for the $bd$ sector are depicted by the the black dotted line in the upper right panels of \cref{fig_plots1,fig_plots2,fig_plots3}. The limits for the $bs$ sector are depicted by the the black dotted line in 
\cref{fig_bsem,fig_bsmt,fig_bdem} and the upper left panels of \cref{fig_plots1,fig_plots2,fig_plots3}. Again these limits are only valid if we allow for left- instead of right-handed lepton couplings $g^L_{\ell_i\ell_j}$. The 
neutrino limits in the $B$-meson sector are not quite as strong as for the 
kaons. Nevertheless, for the $B_s$ meson (i.e. in the $bs$ sector) the neutrino 
limits in the $e\tau$ and $\mu\tau$ sector exclude close to all relevant 
regions in parameter space testable at ATLAS and CMS. In the $e\mu$ sector and 
generally for $B_d$ mesons (i.e. in the $bd$ sector) the neutrino limits are 
more comparable with those from meson decay into charged leptons. Indeed in 
the direction of both small quark and lepton couplings the ATLAS limits are more 
stringent.

\subsection{Lepton decays}

Another important leptonic constraint which involves neutrinos is due to charged  
lepton decays. Again this applies only to $\zp$ bosons with
couplings to left-handed leptons. As the process involves only leptons, the limits only depend on the leptonic couplings and will therefore correspond to vertical lines in the plots.

\begin{figure}[!t]
\centering
    \begin{fmffile}{lep_dec}
    \begin{equation*}
    \begin{gathered}
    \begin{fmfgraph*}(150,100)
        \fmfleft{i1,i2,i3,i4,i5}
        \fmfright{o1,o2,o3,o4}
        
        \fmflabel{$\ell_{iL}$}{i3}
        \fmflabel{$\ell_{jL}$}{o4}
        \fmflabel{$\nu_{iL}$}{o2}
        \fmflabel{$\bar{\nu}_{jL}$}{o1}
        
        \fmf{fermion,tension=2}{i3,v1}
        \fmf{fermion}{v1,o4}
        \fmf{phantom}{v1,o1}
        \fmffreeze
        \fmf{fermion}{o1,v2,o2}
        
        \fmf{photon,tension=2,label=$\zp$}{v1,v2}
    \end{fmfgraph*} 
    \end{gathered}
    \end{equation*}
    \end{fmffile}
\hspace*{1cm}
\caption{Diagrams of the lepton decay $\ell_i \rightarrow\ell_j \, \nu_i\, \bar{\nu}_j$. The same diagram also exists for right-handed charged leptons and for the lepton flavor-violating decays $\ell_i \rightarrow\ell_j \, \nu_j\, \bar{\nu}_i$, where the neutrino flavors are interchanged.}
\label{fig_lepdec}	
\end{figure}
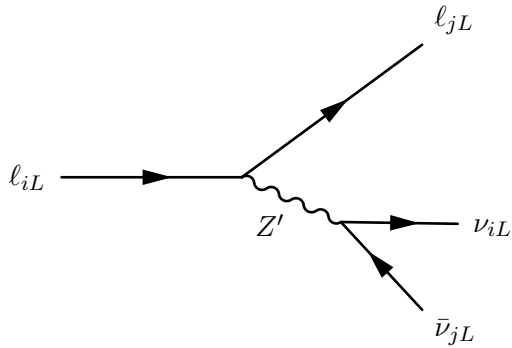

According to the diagram in Fig.~\ref{fig_lepdec} there is an additional decay contribution to the ordinary non flavor-violating lepton decay $\ell_{iL} \rightarrow\ell_{jL} \, \nu_{iL}\, \bar{\nu}_{jL}$. This contribution interferes with the SM contribution generated by a $W$ boson exchange. In addition, there are three new decay channels $\ell_{iL} \rightarrow\ell_{jL} \, \bar{\nu}_{iL}\, \nu_{jL}$, $\ell_{iR} \rightarrow\ell_{jR} \, \nu_{iL}\, \bar{\nu}_{jL} (\bar{\nu}_{iL}\, \nu_{jL})$ . These channels arise from diagrams as shown in Fig.~\ref{fig_lepdec} and the ones with neutrino flavors interchanged.
Together they modify the SM decay rate \cite{Altmannshofer:2016brv} into a given lepton plus neutrinos according to,
\begin{equation}
\Gamma_{\ell_{i}\to \ell_{j}+\bar{\nu}\nu}=\Gamma^\mathrm{SM}_{\ell_{i}\to \ell_{j}+\bar{\nu}\nu}\left[(1+x_{ij})^2+x_{ij}^2+2\,y_{ij}^2\right]\,,
\label{eq_gen_lepdec}
\end{equation}
where with the weak coupling constant denoted as $g$,
\begin{align}
x_{ij}=2\,\frac{(g^{L}_{\ell_{i}\ell_{j}})^2}{g^{2}}\frac{M^{2}_{W}}{M^{2}_{\zp}}, && y_{ij}=2\,\frac{g^{L}_{\ell_{i}\ell_{j}}\,g^{R}_{\ell_{i}\ell_{j}}}{g^{2}}\frac{M^{2}_{W}}{M^{2}_{\zp}}.
\end{align}
The first part is the contribution to the SM-like purely left-handed, non-flavor-violating channel. The second part are the non-SM-like chirality-flipped and/or flavor-violating channels.
Importantly we do not distinguish the neutrino species in the measurement of the final state. This is why all the contributions are summed. \par
As before, we will consider the case of purely left- or right-handed lepton couplings. In the case of purely right-handed couplings we do not get any contribution from the $\zp$ (as due to $g^{L}_{\ell_{i}\ell_{j}}=0$ also $x_{ij}, y_{ij}=0$). In the case of purely left-handed couplings the modification of the SM decay rate simplifies to 
\begin{equation}
\label{leptondecay}
\Gamma_{\ell_{i}\to \ell_{j}+\bar{\nu}\nu}=\Gamma^\mathrm{SM}_{\ell_{i}\to \ell_{j}+\bar{\nu}\nu}\left[(1+x_{ij})^2+x_{ij}^2\right]\,.
\end{equation}

\subsubsection{$\boldsymbol\mu$ decays}
Measurements of the $\mu$ lifetime are very precise with a relative uncertainty of the order of $10^{-6}$~\cite{Olive:2016xmw}. This suggests very tight constraints on $x_{\mu e}$. However, usually the decay of the $\mu$ is used to determine the Fermi constant $G_{F}$. Therefore this measurement cannot be used anymore to test new physics. To do so we need an additional measurement.
The $\beta$ decay of nucleons is possible only via a charged current and is therefore unaffected by our $\zp$.
However, it contains the CKM matrix element $V_{ud}$ which is usually extracted from those decays. The situation is similar for the decay of kaons which contain the matrix element $V_{us}$.

Nevertheless we can extract a limit from this comparison using the following argument. As can be seen from Eq.~\eqref{leptondecay} the $\zp$ contribution leads to an increase in the decay rate. Using the SM extraction of $V_{us}$ and $V_{ud}$ this would lead to smaller values of these CKM matrix elements.
Assuming unitarity for the CKM matrix we can then constrain $x_{\mu e}$ using the CKM matrix elements determined in the standard way~\cite{Olive:2016xmw},
\begin{eqnarray}
1-\frac{1}{(1+x_{\mu e})^2+x^2_{\mu e}}\approx 2\,x_{\mu e}
\!\!&\lesssim&\!\! 1-(|V_{ud}|^2+|V_{us}|^2+|V_{ub}|^2)
\\\nonumber
\!\!&\approx&\!\! 1-(|V_{ud}|^2+|V_{us}|^2)\approx 0.0005\pm 0.0005\lesssim 0.001.
\end{eqnarray}
On the right hand side we estimate the error by adding the errors for $V_{us}$ and $V_{ud}$ in quadrature and in the next step adding the small deviation from $1$.

The resulting limits are shown as dash-dotted lines in the figures and again only apply if we take the lepton coupling to be purely left- instead of right-handed.

In principle one could also derive limits from the angular dependence of the decay of polarized muons used to search for right-handed currents in Ref.~\cite{Jodidio:1986mz}. However, for these constraints to be effective requires the presence of both left- and right-handed couplings which we do not consider in the $e\mu$ sector.

\subsubsection{$\boldsymbol\tau$ decays}\label{taudecayssect}

In the following we want to discuss the impact of our model on various $\tau$ decay modes. We will treat the leptonic and hadronic  decay modes separately.

\subsubsection*{Leptonic mode}

Due to our choice of having only a single flavor-changing coupling in the lepton sector, either $e\tau$ or $\mu\tau$, a strong constraint can be obtained by comparing the branching ratios in these two channels~\cite{Altmannshofer:2016brv} (in~\cite{Foot:1994vd} a comparison with the SM branching ratio is used).\par
For definiteness, let us consider the case of a non-zero $\mu\tau$-coupling  first.  In the following we strongly rely on the derivation done in Ref.~\cite{Altmannshofer:2016brv}. A non-zero coupling $g_{\mu\tau}$  leads to an enhancement of the partial decay rate of the process $\tau\rightarrow\mu\bar{\nu}\nu$ according to \cref{eq_gen_lepdec}. Defining the ratio of the partial decay rates corresponding to $\tau\rightarrow\mu\bar{\nu}\nu$ and $\tau\rightarrow e\bar{\nu}\nu$,
\begin{equation}
 R_{\mu/e} \equiv \frac{\Gamma_{\tau\rightarrow\mu\bar{\nu}\nu}}{\Gamma_{\tau\rightarrow e\bar{\nu}\nu}}\,,
\end{equation}
we can rewrite \cref{eq_gen_lepdec} as
\begin{equation}
R_{\mu/ e}=R_{\mu/e}^\mathrm{SM}\left[(1+x_{ij})^2+x_{ij}^2+ 2 \, y_{ij}^2\right]\,. 
\label{eq_leptonratio}
\end{equation}
Within the SM the ratio $R_{\mu/ e}$ has been very accurately calculated \cite{Pich:2013lsa},
\begin{equation}
 R_{\mu/ e}^\mathrm{SM} = 0.972559\pm0.000005 \,.
\end{equation}
For the experimentally determined value we follow \cite{Altmannshofer:2016brv,Pich:2013lsa} and quote a precise measurement \footnote{We want to point out that previously also the ARGUS \cite{Albrecht:1992xa} and CLEO \cite{Anastassov:1996tc}  collaboration have determined the branching ratios entering $R_{\mu/e}$. These less precise measurement also enter the PDG world average \cite{Olive:2016xmw} $R_{\mu/ e}^\mathrm{PDG} = 0.976\pm0.04$.} by the \textsc{BaBar} collaboration \cite{Aubert:2009qj} yielding a value of 
\begin{equation}
 R_{\mu/ e} = 0.9796\pm0.0039 \,.
\end{equation}
Indeed the measured value is in slight disagreement with the SM prediction and the relative deviation amounts to $1.8 \,\sigma$ \cite{Pich:2013lsa} or 
\begin{equation}
 \Delta\mathcal{R}_{\mu/e} = \frac{R_{\mu/ e}}{R_{\mu/ e}^\mathrm{SM}} -1 = 0.0072 \pm 0.0040 \,.
 \label{eq_taureldev}
\end{equation}
Recalling \cref{eq_leptonratio} we can use this observed deviation as a constraint on our model. If we allow only for a left-handed coupling $g^L_{\mu\tau}$ (which is necessary for the decay into neutrinos to take place), we find the limit
\begin{equation}
  g^L_{\mu\tau} \leq \frac{g}{2}\ \frac{M_\zp}{M_W} \, \left[\left( 1+2\, \Delta\mathcal{R}_{\mu/e}\right)^\frac{1}{2} -1 \right]^\frac{1}{2}  \,.
\end{equation}
In the presence of both left- and  right-handed lepton couplings we can derive a limit as
\begin{equation}
 g^R_{\mu\tau} \leq \frac{g}{2\sqrt{1+\rho_\ell^2}}\,\frac{M_\zp}{M_W} \, \left[\left( 1+2\,(1+\rho_\ell^{-2})\, \Delta\mathcal{R}_{\mu/e}\right)^\frac{1}{2} -1 \right]^\frac{1}{2}  \,.
\end{equation}

\begin{figure}[t!]
\includegraphics[width=\textwidth]{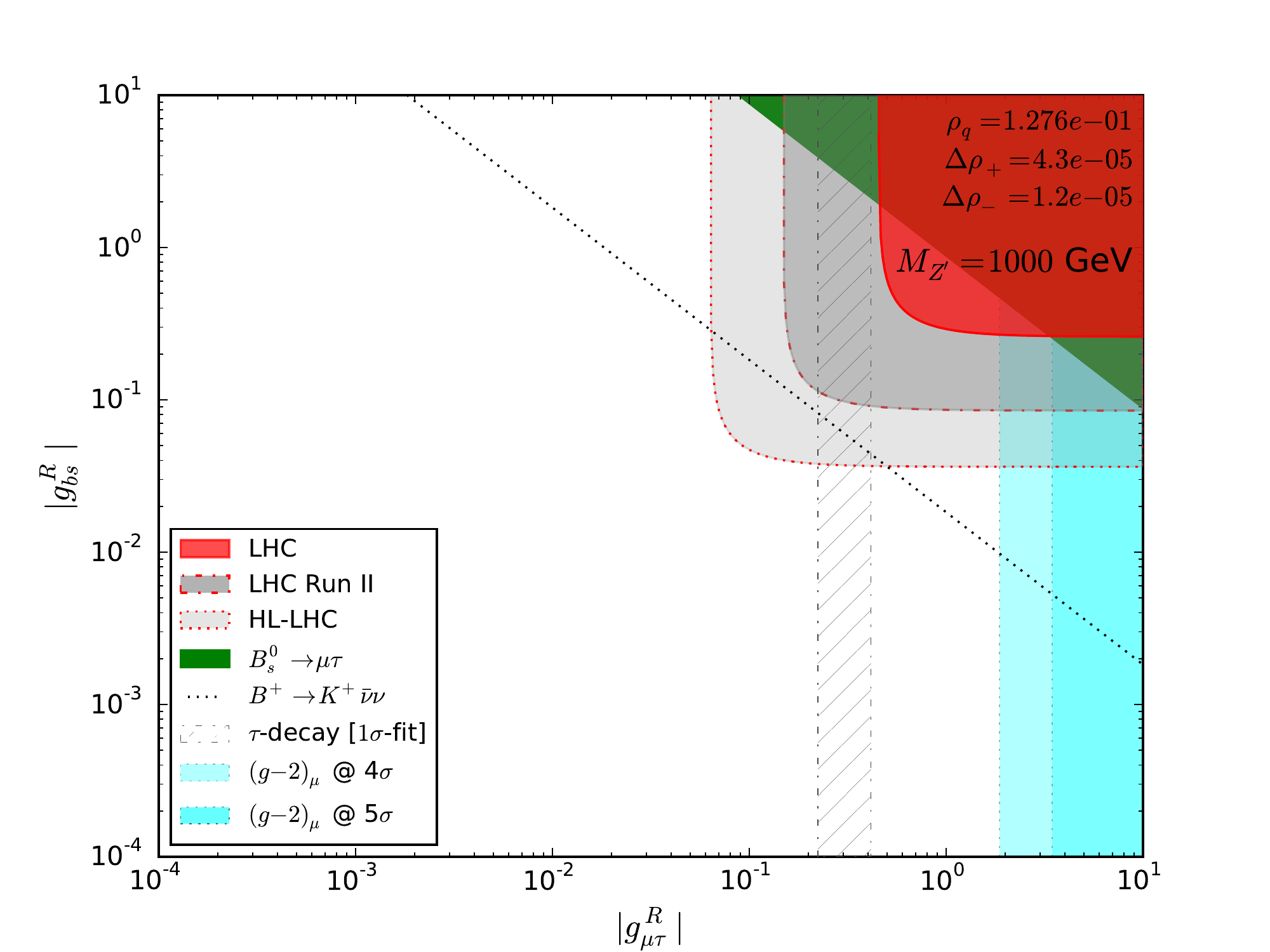}
 \caption{The flavor-violating couplings $g^R_{bs}$ and $g^R_{\mu\tau}$ for a $\zp$ boson of $M_\zp=1000$ GeV and a coupling ratio of $\rho_q = g^L_{bs}/g^R_{bs}= 0.1276$. The red area indicates the limit from the ATLAS analysis of the process $pp\rightarrow \mu\tau$ at $\sqrt{s}=8$ TeV. The red dash-dotted and dashed lines are projections to the LHC Run II and HL-LHC. The green area represents the excluded region from the decay $B_s^0\rightarrow\mu\tau$. The black dotted line is the limit from $B^+\rightarrow K^+\bar{\nu}\nu$ applying only for left-handed lepton couplings. The light and dark cyan areas depict the 4 and $5\,\sigma$ exclusion bands from $\Delta a_\mu$. The black and white hatched area is the preferred $1\,\sigma$ region of the observed deviation $\Delta\mathcal{R}_{\mu/e} = {R_{\mu/ e}}/{R_{\mu/ e}^\mathrm{SM}} -1$, also only applicable in the case of purely left-handed couplings.}
\label{fig_lfit_bsmt}
\end{figure}

Now we want to consider the case of non-zero $e\tau$-coupling. In this scenario the arguments are essentially analogous to the $\mu\tau$-case, however, using the inverted ratio
\begin{equation}
 R_{e/\mu} \equiv \frac{\Gamma_{\tau\rightarrow e\bar{\nu}\nu}}{\Gamma_{\tau\rightarrow\mu\bar{\nu}\nu}}=\frac{1}{R_{\mu/e}  }\,.
\end{equation}
It is worth noticing that in this case using the inverted ratio the relative deviation of the experimental value from the SM prediction as discussed in \cref{eq_taureldev} becomes negative
\begin{equation}
 \Delta\mathcal{R}_{e/\mu} = \frac{R_{e/\mu}}{R_{e/\mu}^\mathrm{SM}} -1 = -0.0072 \pm 0.0040 \,.
\end{equation}
From \cref{eq_gen_lepdec} we see that structurally the $\zp$ contribution from a non-vanishing $g_{e\tau}$ coupling will always lead to a positive shift in $R_{e/\mu}$. Therefore, the measured fluctuation leading to a negative shift will impose an anomalously stringent bound on $g_{e\tau}$ (cf. Fig.~\ref{fig_plots2} in the Appendix).

In both the case of non-zero $e\tau$- and $\mu\tau$-coupling we use the observed relative deviation $\Delta\mathcal{R}$ plus the $2\,\sigma$ uncertainty as exclusion bound. The corresponding limits are shown for example in \cref{fig_bsem,fig_bsmt,fig_bdem} as the vertical black dash-dotted line. It can be seen that lepton decay limits are by far the strongest purely leptonic limits and cut far into the region of parameter space testable with multipurpose experiments at the LHC. However, it has to be noted that these limits apply only if we consider purely left- instead of right-handed lepton couplings $g^{L}_{\ell_i\ell_j}$. In the right-handed case these limits are absent. \par

If we assume the previously  discussed $1.8\,\sigma$ relative  deviation  in the ratios of branching fractions \cref{eq_taureldev} not to be due to systematics or a fluctuation, we can speculate on a possible new physics origin. In order to justify such speculation, we checked with help of the accurate prediction of $R_{\mu/e}^\mathrm{SM}$ given in Ref.~\cite{Pich:2013lsa} and the measured $\tau$ lifetime that the excess $\Delta\mathcal{R}_{\mu/e}$ is indeed due to the observed value of $\Gamma_{\tau\rightarrow\mu\bar\nu\nu}$ being significantly higher than its SM prediction. Indeed, in previous work\cite{Crivellin:2015hha} it has been noticed that the relative deviation
\begin{equation}
 \frac{\Gamma_{\tau\rightarrow\mu\bar\nu\nu}}{\Gamma_{\tau\rightarrow\mu\bar\nu\nu}^\mathrm{SM}} -1 = (0.69\pm0.29)\% \,, 
\end{equation}
even amounts to $2.4\,\sigma$. The resulting increase in the total width $\Gamma_\mathrm{tot}$ is compatible with its observed value. Therefore, we additionally fitted the excess  within our model at the $1\,\sigma$ level. The preferred region is depicted by the black and white hatched area in \cref{fig_lfit_bsmt}. We can see that we can fit the observed deviation for a 1 TeV $\zp$ boson with moderate couplings $g_{\mu\tau}^L\sim \mathcal{O}(10^{-1})$. However, this only applies in the case of purely left-handed lepton couplings also implying that the limit from semi-leptonic meson decay into neutrinos applies (dotted line). In return, this meson decay limit excludes most of the region in parameter space testable with ATLAS or CMS at the LHC. 

\subsubsection*{Hadronic mode}

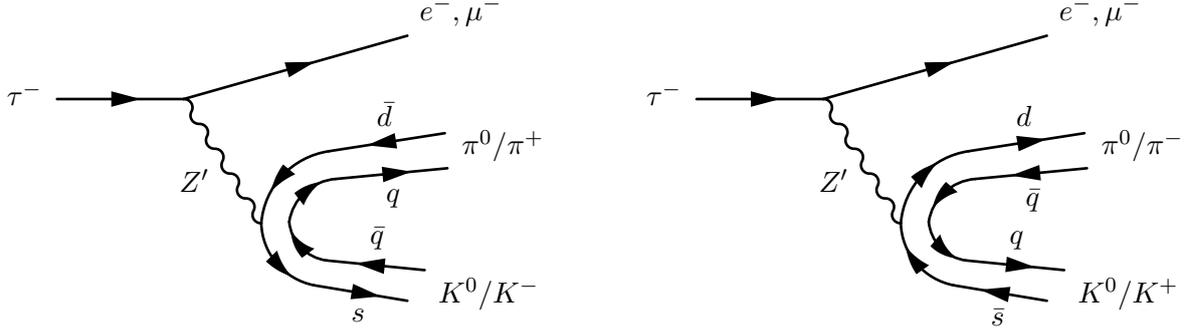
\begin{figure}[!t]
\centering
    \begin{fmffile}{taudec}
    \begin{equation*}
    \begin{gathered}
    \begin{fmfgraph*}(150,100)
        \fmfleft{i1,i2,i3,i3b,i4,i5,i6,i7,i8}
        \fmfright{o1,o2,o3,o3b,o4,o5,o6,o7,o8}
        
        \fmfv{l=$\tau^-$}{i6}
        \fmflabel{$e^-,\mu^-$}{o8}
        \fmfv{l=$\pi^0/\pi^+$,l.a=25}{o4}
        \fmfv{l=$K^0/K^-$}{o2}
        
        \fmf{fermion,tension=2}{i6,v}
        \fmf{phantom}{v,o6}
        \fmffreeze
        \fmf{fermion}{v,o8}
        
        \fmf{phantom}{i1,hd1}
        \fmf{fermion,tension=3,label=$s$,label.side=right}{hd1,o1}
        \fmf{phantom}{i2,hd2}
        \fmf{fermion,tension=3,label=$\bar{q}$,label.side=right}{o2,hd2}
        \fmf{phantom}{i3,h1}
        \fmf{phantom,tension=3}{h1,h2}
        \fmf{phantom}{i4,hu2}
        \fmf{fermion,tension=3,label=$q$,label.side=right}{hu2,o4}
        \fmf{phantom}{i5,hu1}
        \fmf{fermion,tension=3,label=$\bar d$,label.side=right}{o5,hu1}
       
        \fmf{fermion,right=0.35}{hu1,h1,hd1}
        \fmf{fermion,left=0.35}{hd2,h2,hu2}
        
        \fmffreeze
        
        \fmf{photon,tension=2,label=$\zp$,label.side=right}{v,h1}
    \end{fmfgraph*} 
    \end{gathered}
    \qquad \qquad \qquad \qquad
    \begin{gathered}
    \begin{fmfgraph*}(150,100)
        \fmfleft{i1,i2,i3,i3b,i4,i5,i6,i7,i8}
        \fmfright{o1,o2,o3,o3b,o4,o5,o6,o7,o8}
        
        \fmfv{l=$\tau^-$}{i6}
        \fmflabel{$e^-,\mu^-$}{o8}
        \fmfv{l=$\pi^0/\pi^-$,l.a=25}{o4}
        \fmfv{l=$K^0/K^+$}{o2}
        
        \fmf{fermion,tension=2}{i6,v}
        \fmf{phantom}{v,o6}
        \fmffreeze
        \fmf{fermion}{v,o8}
        
        \fmf{phantom}{i1,hd1}
        \fmf{fermion,tension=3,label=$\bar s$,label.side=left}{o1,hd1}
        \fmf{phantom}{i2,hd2}
        \fmf{fermion,tension=3,label=$q$,label.side=left}{hd2,o2}
        \fmf{phantom}{i3,h1}
        \fmf{phantom,tension=3}{h1,h2}
        \fmf{phantom}{i4,hu2}
        \fmf{fermion,tension=3,label=$\bar q$,label.side=left}{o4,hu2}
        \fmf{phantom}{i5,hu1}
        \fmf{fermion,tension=3,label=$d$,label.side=left}{hu1,o5}
       
        \fmf{fermion,left=0.35}{hd1,h1,hu1}
        \fmf{fermion,right=0.35}{hu2,h2,hd2}
        
        \fmffreeze
        
        \fmf{photon,tension=2,label=$\zp$,label.side=right}{v,h1}
    \end{fmfgraph*} 
    \end{gathered}
    \end{equation*}
    \end{fmffile}
\hspace*{1cm}
\caption{Diagrams of a possible signature of a doubly flavor-changing $\zp$ boson with non-zero couplings in the $sd$ and $e\tau/\mu\tau$ sector ($q$ denotes either a $d$- or a $u$-quark). The resulting decays are $\tau^\pm\rightarrow\ell^\pm\,(\pi^0\,K^0/\pi^\pm\,K^\mp)$, where $\ell \in \{e,\mu\}$. Only in the $sd$ sector such a semi-leptonic $\tau$ decay into a pion and a kaon is kinematically allowed. Especially for a non-zero quark coupling involving a $b$-quark such a decay is not possible.} 
\label{fig_taudec}	
\end{figure}

In the context of $\tau$ decays we want to mention the hadronic decay mode $\tau^\pm\rightarrow\ell^\pm\,(\pi^0\,K^0/\pi^\pm\,K^\mp)$ due to the diagrams shown in \cref{fig_taudec}, where $\ell \in \{e,\mu\}$. This mode is present in the case of non-vanishing quark couplings in the $sd$ sector\footnote{For a recent example of an explicit model with flavor-violating couplings in the quark sector as well as non-vanishing lepton couplings, motivated by observed anomalies in $B$ decays as well as $(g-2)_{\mu}$, see~\cite{ColuccioLeskow:2016dox}.}. As it is not present in the SM the detection of such a decay would be a smoking gun for a doubly flavor-changing $\zp$. It should be noted that such a decay is only possible into pions and kaons as $B$ and $D$ mesons are too heavy for the $\tau$ to decay into.

The decays into charged mesons $\tau^-\rightarrow\ell^-\,\pi^\pm\,K^\mp$  and $\tau^+\rightarrow\ell^+\,\pi^\pm\,K^\mp$ have been searched for at \textsc{BaBar} \cite{Aubert:2005tp} and \textsc{Belle} \cite{Miyazaki:2009wc}. The corresponding branching fraction limits can be turned into a limit on the $Z^\prime$ couplings $g_{\ell\tau}$ and $g_{sd}$. The relevant operators contributing to this process are again those of \cref{eq_mesops1,eq_mesops2} with corresponding Wilson coefficients given in \cref{eq_mescoeffs}. In order to calculate the branching fraction due to the $Z^\prime$ induced decay we need the hadronic matrix elements $\bra \pi^+K^-|(\bar{s}d)_{L/R}|0\ket$, where we have introduced the shorthand $(\bar{s}d)_{L/R} = \bar s\, \gamma_\mu P_{L/R} d$. Similar to \cref{sec_kaonneutrinos} we will use isospin symmetry to estimate the matrix element from the observed decay $\tau\rightarrow\nu_\tau\, K^-\pi^0$. The SM operator for this decay reads 
\begin{equation}
 \mathcal{O}_{\tau\nu} = (\bar{\nu}_\tau \gamma^\mu P_L\tau)(\bar s\, \gamma_\mu P_L u)\,,
\end{equation}
with corresponding Wilson coefficient $C_{\tau\nu} = 2\sqrt{2}\, G_F \, V_{us}$. 
In the following, we assume the electron and muon to be massless, which seems to be sensible as $m_e,m_\mu \ll m_\tau$. However, this implies that the outgoing leptons have definite handedness and consequently lead to distinguishable final states.
Therefore, the ratio of branching fractions reads
\begin{align}
 \frac{\Gamma_{\tau^-\rightarrow\ell^- \pi^+ K^-}}{\Gamma_{\tau^-\rightarrow\nu_\tau\, \pi^0 K^- }} =  \left(\frac{1}{2\sqrt{2}\, G_F \, V_{us}\, M_\zp^2} \right)^2 \ & \frac{\left| g^L_{\ell\tau}\bra\ell|(\bar{\ell} \tau)_L|\tau\ket \right|^2 + \left| g^R_{\ell\tau}\bra\ell|(\bar{\ell} \tau)_{R}|\tau\ket \right|^2}{\left| \bra{\nu}_\tau|(\bar{\nu}_\tau \tau)_L|\tau\ket  \right|^2} \notag \\
 \times & \frac{\left| g^L_{sd}\bra\pi^+K^-|(\bar{s} d)_L|0\ket + g^R_{sd}\bra\pi^+K^-|(\bar{s} d)_R|0\ket \right|^2}{\left| \bra\pi^0K^-|(\bar{s} u)_L|0\ket  \right|^2} \,. \label{eq_hadratio}
\end{align}
Treating the leptons as massless translates into $ \bra\ell|(\bar{\ell} \tau)_L|\tau\ket \simeq \bra{\nu}_\tau|(\bar{\nu}_\tau \tau)_L|\tau\ket $.
 Furthermore, we use isospin symmetry to relate the hadronic matrix elements
\begin{equation}
 \bra\pi^+K^-|(\bar{s} d)_L|0\ket \simeq \sqrt{2} \, \bra\pi^0K^-|(\bar{s} u)_L|0\ket \,.
\end{equation}
As QCD is a non-chiral theory, the same relation holds also with the operator $(\bar{s} d)_L$ replaced by $(\bar{s} d)_R$. Put together, we can thus estimate the branching fraction from \cref{eq_hadratio} to be
\begin{equation}
 \Gamma_{\tau^-\rightarrow\ell^- \pi^+ K^-} \simeq \left(\frac{g^R_{\ell\tau}\, g^R_{sd}}{2\, G_F \, M_\zp^2} \right)^2\, \frac{(1+\rho_\ell^2)\ |1+\rho_q|^2}{|V_{us}|^2} \ \Gamma_{\tau^-\rightarrow\nu_\tau\, \pi^0 K^- }^\mathrm{exp} \,. \label{eq_tauhadbr}
\end{equation}
This relation allows us to put a limit on $g_{\ell\tau}$ and $g_{sd}$. Using the \textsc{Belle} limits\cite{Miyazaki:2009wc}
\begin{align}
 \Gamma_{\tau^-\rightarrow e^- \pi^+ K^-} < 5.6\times10^{-8}\,, \\
 \Gamma_{\tau^-\rightarrow \mu^- \pi^+ K^-} < 16\times10^{-8} \,,
\end{align}
we have derived the corresponding bounds in the $\{sd,e\tau\}$ and $\{sd,\mu\tau\}$ sector shown in purple in the lower left panel of \cref{fig_plots2,fig_plots3}. It can be observed that the ATLAS bounds from the 8 TeV dataset are almost entirely lying within the purple areas. However, the LHC Run II and the HL-LHC scenario are expected to yield superior limits along the direction of small  $g_{\ell\tau}$ and $g_{sd}$.

\subsection{Muonium - Antimuonium oscillations}

\begin{figure}[!t]
\centering
    \begin{fmffile}{tree_mix}
    \begin{equation*}
    \begin{gathered}
    \begin{fmfgraph*}(100,60)
        \fmfleft{i1,i2}
        \fmfright{o1,o2}
        
        \fmflabel{$e^-$}{i1}
        \fmflabel{$\mu^+$}{i2}
        \fmflabel{$\mu^-$}{o1}
        \fmflabel{$e^+$}{o2}
        
        \fmf{fermion}{i1,v1}
        \fmf{fermion}{v1,i2}
        \fmf{fermion}{o2,v2}
        \fmf{fermion}{v2,o1}
        
        \fmf{photon,label=$\zp$}{v1,v2}
    \end{fmfgraph*}
    \end{gathered}
    \qquad \qquad \qquad
    \begin{gathered}
    \begin{fmfgraph*}(100,60)
        \fmfleft{i1,i2}
        \fmfright{o1,o2}
        
        \fmflabel{$e^-$}{i1}
        \fmflabel{$\mu^+$}{i2}
        \fmflabel{$\mu^-$}{o1}
        \fmflabel{$e^+$}{o2}
        
        \fmf{fermion}{i1,v1,o1}
        \fmf{fermion}{o2,v2,i2}
        
        \fmf{photon,label=$\zp$}{v1,v2}
    \end{fmfgraph*} 
    \end{gathered}
    \end{equation*}
    \end{fmffile}
\hspace*{1cm}
\caption{Relevant diagrams leading to muonium-antimuonium oscillations.}
\label{fig_muonium}	
\end{figure}
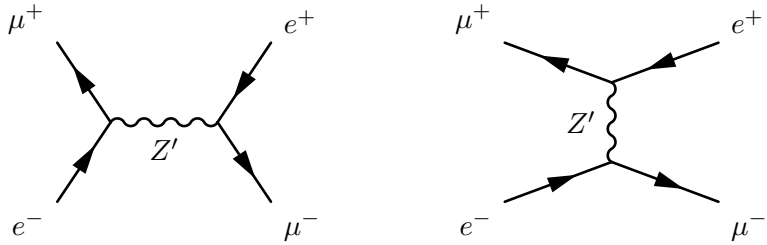

The two leptons $e^-$ and $\mu^+$ can form a hydrogen-like bound state called 
muonium $M$. In presence of flavor-changing processes this bound 
state can oscillate into its conjugate state consisting of $e^+\mu^-$ -- the 
antimuonium $\bar{M}$. The MACS 
experiment at the Paul Scherrer Institut in Villigen, Switzerland has searched for $M-\bar{M}$ transitions in a muon 
fixed target experiment. The non-observation of such transitions gives an upper bound for the probability of spontaneous 
muonium to antimuonium conversion $P_{M\bar{M}} \leq 8.2\times10^{-11}$ at 90 \% C.L.\cite{Willmann:1998gd}. \\
In our model we have tree-level contributions to $M-\bar{M}$  transitions from the diagrams depicted in 
\cref{fig_muonium}. These diagrams generate a transition matrix element $\C{M}_{M\bar{M}}$ resulting in a mass 
splitting of the two states \cite{hep-ph/0307264}
\begin{equation}
 \Delta M = 2\, |\C{M}_{M\bar{M}}| \;.
\end{equation}
We can calculate the matrix element and consequently the mass splitting in an 
effective operator approach. From our 
model Lagrangian \cref{eq_gen_lag} we obtain the corresponding low-energy 
interaction encoded in the operators
\begin{equation}
 \C{O}_\mathrm{e\mu}^{XY} = (\bar{\mu}\,\gamma^\nu\,P_X\,e) 
(\bar{\mu}\,\gamma_\nu\,P_Y\,e) \,,
\end{equation}
where $X,Y\in\{L,R\}$ and the Wilson coefficients
\begin{equation}
C^{XY} = 2\, \frac{g^X_{e\mu}\,g^Y_{e\mu}}{M^2_\zp}\,.
\end{equation}
In order to calculate the amplitude we will perform a non-relativistic field 
expansion and calculate the resulting 
effective potential from the Born approximation. Muonium is a non-relativistic 
Coulomb bound state and therefore the transitions $M-\bar{M}$ can be 
described  by 
a non-relativistic effective potential $V_\mathrm{eff}(\vec{x})$. Taking 
into 
account that the two fermions can either be in a spin 
singlet or triplet bound state we obtain for the potentials
\begin{align}
 V_\mathrm{singlet}(\vec{x}) &= \ \ 2 \, [C^{LL} - 2\,C^{LR} + C^{RR}] \ 
\delta^{(3)}(\vec{x})\;,\\
 V_\mathrm{triplet}(\vec{x}) &= -2 \, [C^{LL} + 2\,C^{LR} + C^{RR}] \ 
\delta^{(3)}(\vec{x})\;.
\end{align}
Assuming the muonium to be in its electronic ground state we can calculate 
the mass splitting \cite{hep-ph/0307264}
\begin{align}
 \Delta M \simeq 2\ \bra \bar{M} | (|\Re\, V_\mathrm{eff}(\vec{x}) | )|M 
\ket &= 2 \int \mathrm{d}^3x \  
\phi^*_{100}(\vec{x}) \, |\Re\, V_\mathrm{eff}(\vec{x})| 
\,\phi_{100}(\vec{x}) \\
& = \frac{4}{\pi \, a^3_{M\bar{M}}}\ [C^{LL} \mp 2\,C^{LR} + C^{RR}]\;,
\end{align}
with the Bohr radius of the muonimum $a_{M\bar{M}}=\frac{1}{\alpha \, 
m_\mathrm{red}}$ and the reduced mass 
$m_\mathrm{red} = \frac{m_e\,m_\mu}{m_e+m_\mu}$.  In order to get in contact 
with the experiment, we need to know the transition probability $P_{M\bar{M}} $ 
of an initially prepared muonium atom $M$ to oscillate into 
$\bar{M}$. Therefore, we need to know the time evolution of the two-state 
system that is generally obtained by solving the Schr\"{o}dinger equation 
\cite{Nierste:2009wg}
\begin{align}
i\, \frac{\mathrm{d}}{\mathrm{d}t}\left(
\begin{matrix}
 \ |M(t) \ket \ \\
 \ |\bar{M}(t) \ket \
\end{matrix}
\right) 
= 
\left(
\begin{matrix}
 M-i\frac{\Gamma}{2} && \frac{\Delta M}{2} \\
 \frac{\Delta M}{2} && M-i\frac{\Gamma}{2}
\end{matrix}
\right) 
\left(
\begin{matrix}
 \ |M(t) \ket \ \\
 \ |\bar{M}(t) \ket \
\end{matrix}
\right)  \,.
\label{eq_2schroed}
\end{align}
After diagonalizing the Hamiltonian of \cref{eq_2schroed} one finds for the 
time evolution of an initially pure muonimum state
\begin{equation}
 |M(t)\ket = \left( \cos\left(\frac{\Delta M}{2} t\right) |M\ket + i 
\sin\left(\frac{\Delta M}{2} t\right) |\bar{M}\ket \right) 
e^{-\frac{\Gamma}{2}t}\ e^{iMt} \,.
\end{equation}
If the 
bound state is mainly antimuonium it consists of $e^+$ and $\mu^-$. The muon 
will decay via $\mu^-\rightarrow e^- \,\bar\nu_e \, \nu_\mu$ with a highly 
energetic electron $e^-$ hitting the detector. For muonium it would instead be a 
highly energetic positron.
The measurement principle to detect that muonium has oscillated into antimuonium 
is 
to start with a pure muonium initial state and then look for the highly 
energetic electron resulting from the muon decay inside the antimuonium bound 
state.  In essence this means counting 
the number of outgoing energetic $e^-$. Their number can be determined by 
integrating the partial decay rate into electrons. This is given by the 
probability that the system is in an antimuonium state multiplied by the muon 
decay rate. Integrating over time we have~\cite{Harnik:2012pb},
\begin{equation}
 P_{M\bar{M}} = \int_0^\infty \mathrm{d}t \ \Gamma_\mu \, 
\sin^2\left(\frac{\Delta M}{2}\, t\right)\, e^{-\Gamma_\mu \,t} = 
\frac{1}{2\left(\frac{\Gamma^2_\mu \,}{\Delta M^2}+1\right)} \,.
\end{equation}
Assuming either left- or right-handed lepton couplings this can be translated into a 
rough limit on the off-diagonal lepton coupling of the $\zp$
\begin{equation}
 |g^{L/R}_{e\mu}| \leq \frac{1}{S_B}\[ \frac{\pi\, M^2_\zp \, \Gamma_\mu}{8\, 
\alpha^3\,m^3_\mathrm{red}}  \ \(\frac{2P_{M\bar{M}}}{1 - 
2P_{M\bar{M}}}\)^{\frac{1}{2}} \]^{\frac{1}{2}} \;,
\end{equation}
with $S_B = 0.35$ \cite{hep-ph/0307264} a correction factor for the muonium 
splitting 
in the magnetic field  coming from the $(V\pm A)\times(V\pm A)$ Lorentz 
structure of the interaction. The corresponding limits are shown for example in 
\cref{fig_bsem,fig_bdem} by the magenta band. As these limits are purely 
leptonic they result in a vertical exclusion line. In general, it is the 
strongest limit for the lepton sector only. In contrast, the LHC can 
probe $e\mu$ couplings significantly smaller than those excluded by muonium. 
However, this is only true in combination with same order of magnitude 
quark-sector couplings whereas the muonium limits are universal. Furthermore, 
as muonium is a bound state of $e^+$ and $\mu^-$ the $e\tau$ and $\mu\tau$ 
sectors are completely unaffected by this limit. The relevant $e\tau$ or $\mu\tau$ bound states in these cases would be difficult to access as the $\tau$ decays very rapidly.

\subsection{LEP limits}

\begin{figure}[t!]
\begin{subfigure}{.4\textwidth}
\centering
    \begin{fmffile}{tree_lep}
    \begin{equation*}
    \begin{gathered}
    \begin{fmfgraph*}(120,80)
        \fmfleft{i1,i2}
        \fmfright{o1,o2}
        
        \fmflabel{$e^-$}{i1}
        \fmflabel{$e^+$}{i2}
        \fmflabel{$\ell^-$}{o1}
        \fmflabel{$\ell^+$}{o2}
        
        \fmf{fermion}{i1,v1,o1}
        \fmf{fermion}{o2,v2,i2}
        
        \fmf{photon,label=$\zp$}{v1,v2}
    \end{fmfgraph*} 
    \end{gathered}
    \end{equation*}
    \end{fmffile}
\end{subfigure}                                
\begin{subfigure}{.6\textwidth}               
  \centering                                   
   \includegraphics[width=\textwidth]{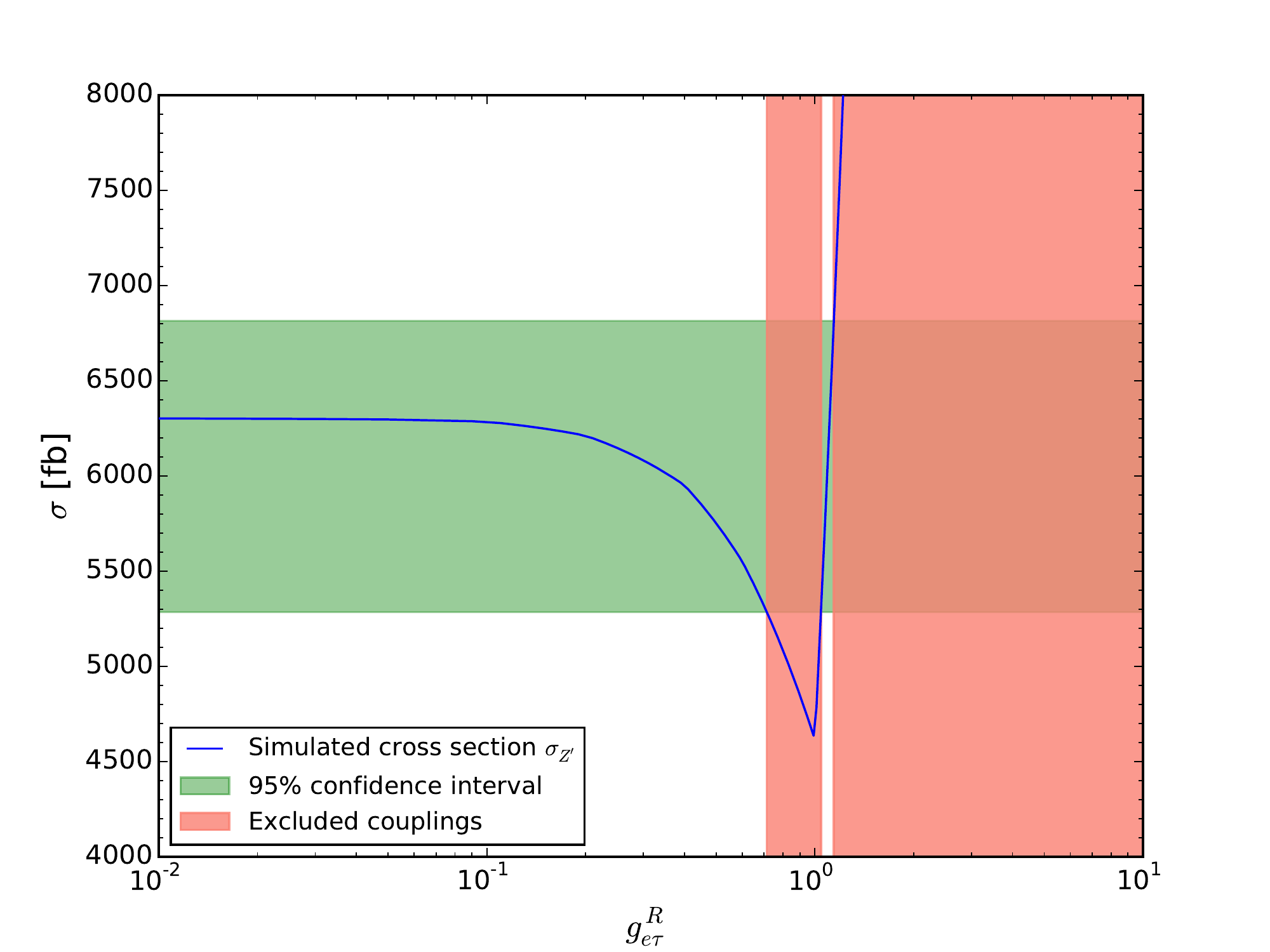}
\end{subfigure} 
\caption{(Left) Tree-level dilepton production via $\zp$ at LEP, where
$\ell$ is  either the $\mu$ or the $\tau$. (Right) The blue curve is the 
total simulated cross section $\sigma_\zp$ for the process 
$e^+e^-\rightarrow\tau^+\tau^-$ at $\sqrt{s}=207$ GeV. The green band 
depicts the 95\% confidence interval of the measured cross section 
$\sigma_\mathrm{exp}$. The red area shows the excluded couplings $g_{e\tau}^R$.
}
\label{fig_lep}	
\end{figure}

The LEP collider run at CERN from 1989 to 2000 produced a large amount of clean
$e^+e^-$ collisions. In particular the analyses of the processes 
$e^+e^-\rightarrow\mu^+\mu^-$ and $e^+e^-\rightarrow\tau^+\tau^-$  provide 
constraints on the $\zp$ couplings $g_{e\mu}$ and $g_{e\tau}$. \\
In order to derive constraints on these couplings in our model, we use the 
total inclusive cross sections $\sigma(e^+e^-\rightarrow\mu^+\mu^-)$ and 
$\sigma(e^+e^-\rightarrow\tau^+\tau^-)$  as measured by the ALEPH collaboration 
\cite{hep-ex/0609051}. Therefore, we have simulated the total inclusive cross 
section $\sigma_\zp$ of these two processes with \textsc{madgraph v2.3.3} 
\cite{Alwall:2014hca} at 
$\sqrt{s}=207$ GeV, including the $\zp$ diagram shown in the 
left panel of \cref{fig_lep}. We have simulated  $\sigma_\zp$ for a number of 
different values of the lepton couplings and the $\zp$ mass allowing for an 
additional hard photon  
in the final state. We then set a limit on $g_{e\mu}$ and $g_{e\tau}$ by 
using a two-sided hypothesis test. For this we assume that the measured cross 
section $\sigma_\mathrm{exp}$ (i.e the number of signal events) 
follows a Gaussian distribution. For the relative large total number of events 
$N_{\mu\mu}=683$ and $N_{\tau\tau}=402$ at $\sqrt{s}=207$ GeV \cite{hep-ex/0609051} this seems to be
justified. For a given $\zp$ mass we can then scan the simulated total 
inclusive cross 
section $\sigma_\zp$ for the lepton couplings $g_{e\mu}$ and $g_{e\tau}$ and 
exclude all coupling values that correspond to a cross section $\sigma \notin 
[\sigma_\mathrm{exp}-1.96\,\Delta\sigma, \sigma_\mathrm{exp}+1.96\,\Delta\sigma]$,\, 
which corresponds to a  two-sided 95\% confidence interval for a Gaussian 
distribution. \par 

The corresponding limits are found e.g. in \cref{fig_bsem,fig_bdem} and 
are depicted by the golden regions. As was the case for the muonium, these 
limits are purely leptonic (and therefore correspond to vertical bands in the 
$g_{qq'}-g_{\ell\ell'}$ plane). As LEP was an electron-positron collider, these 
limits do not concern the $\mu\tau$ sector. Generally, the LEP limits are 
weaker than the muonium limits and therefore are only of concern in the $e\tau$ 
sector (cf. Appendix~\ref{sec_app_plots} \cref{fig_plots2}) where the muonium 
limits are not present. A particular feature of the LEP limits is the gap in 
the excluded region of parameter space. The origin of this gap can be understood 
with help of the right panel of \cref{fig_lep}. For small couplings $g_{e\ell}$ 
the total cross section is mainly SM-like and agrees very well with the 
measurements (i.e. it lies within the 95\% confidence interval). For moderate 
couplings  $g_{e\ell} \lesssim 1$ the interference term, which is linear in 
$g_{e\ell}$ and has negative sign, starts to become important and eventually 
drives the cross section $\sigma_\zp$ below the confidence interval. This leads 
to the first exclusion band. With increasing couplings  $g_{e\ell}>1$ the pure 
$\zp$ contribution, which is quadratic in $g_{e\ell}$, starts to dominate and 
drives the cross section $\sigma_\zp$ well above the 95\% confidence interval. 
This leads to the second exclusion band. In between those two regimes we have a 
transition region where $\sigma_\zp$ lies within the 95\% confidence interval - 
the gap in the exclusion region.

\subsection{Magnetic dipole moments and the $\boldsymbol Z^{\boldsymbol\prime}$}
\label{sec_magmom}

A further possible constraint for the lepton sector of our model is coming from 
the measurements of $(g-2)$ of the electron and the muon. However, as 
$(g-2)_\mu$ exhibits a deviation between theory and experiment of about 
$3 \,\sigma$ or even more\cite{Bennett:2006fi,Jegerlehner:2009ry,Davier:2010nc,Hagiwara:2011af} 
it is tempting to speculate on a new physics origin. Hence, in this section we additionally want to 
explore the potential of the $\zp$ boson to play this role similar to earlier work~\cite{Altmannshofer:2014cfa,Altmannshofer:2014pba,Altmannshofer:2016brv,Heeck:2016xkh}.

\subsubsection{Experimental status}

We want to motivate our discussion by looking at one of the most 
precise measurements in the electroweak precision era: the 
determination  of the gyromagnetic ratio $g$ of the muon  at the E821 
experiment at the Brookhaven Alternating Gradient Synchrotron 
\cite{hep-ex/0208001}.
The naive SM tree-level calculation, i.e. the Dirac equation, yields a 
value for the gyromagnetic ratio of $g= 2$. Radiative corrections such as 
higher-order QED processes, electroweak loops or hadronic vacuum polarization 
lead to a shift of the gyromagnetic ratio, the so-called anomalous magnetic 
moment
\begin{equation}
a_\mu = \frac{(g-2)_\mu}{2}\,.
\end{equation}
Much interest has been triggered by the findings of the E821 experiment that 
point towards a mismatch between theory \cite{Davier:2010nc,Hagiwara:2011af} and experiment \cite{Bennett:2006fi} of up to $\sim3.6\,\sigma$ or 
\begin{equation}
 \Delta a_\mu = a^\mathrm{exp}_\mu - a^\mathrm{SM}_\mu = (2.87 \pm 0.80) \times 10^{-9}  \,.
 \label{eq_mumag_dev}
\end{equation}
For a suitable chirality structure\footnote{We thank Julian Heeck for pointing out to us that this requires a vector-like part.} the anomalous magnetic moment of the 
muon can get a positive shift due to radiative corrections from a $\zp$ loop 
through a nonzero $\mu\tau$-coupling. Therefore such a $\zp$ can potentially reconcile the 
experimental value with the theory prediction. In previous work\cite{Baek:2001kca,Altmannshofer:2014cfa} models have been studied, where the $\zp$ boson couples to the $L_\mu-L_\tau$ current,
\begin{equation}
 J^\alpha_\mathrm{lep} = Q_\ell\, (\bar{L}_2\,\gamma^\alpha\, L_2 
-\bar{L}_3\,\gamma^\alpha\, L_3 + \bar{\mu}_R\,\gamma^\alpha\, \mu_R - 
\bar{\tau}_R\,\gamma^\alpha\, \tau_R) \,,
\end{equation}
with $Q_\ell$ being the overall lepton charge, $L_2 = (\nu_\mu,\mu_L)$ and $L_2 = 
(\nu_\tau,\tau_L)$. In such models an explanation of the $(g-2)_\mu$ tension is ruled out for $\zp$ bosons with mass $M_\zp \gtrsim $ GeV by neutrino trident production $\nu\, N \rightarrow \nu\, 
\mu^+\mu^-N$ in the Coulomb field of a nucleus $N$ 
\cite{Altmannshofer:2014cfa,Altmannshofer:2014pba,Heeck:2016xkh,TheBABAR:2016rlg}. 
Nevertheless, neutrino trident production is not possible in our 
model as it requires diagonal couplings to $\mu$ and $\tau$ of the $\zp$ on 
tree level. Consequently neutrino trident constraints are not applicable because they are looking at two muon/tau signatures. Hence, in this case $(g-2)_{\mu}$ can be explained by a suitable pure $\mu\tau$ coupling~\cite{Altmannshofer:2016brv}.
In the future it may be possible to look for flavor changing trident signals with $e\mu$, $e\tau$ or $\mu\tau$ in the final state at the DUNE~\cite{Acciarri:2015uup} and SHiP~\cite{Anelli:2015pba,Alekhin:2015byh} facilities~\cite{Magill:2016hgc,Heeck2016}.
\par 
New physics can also be probed via the electron 
anomalous magnetic moment. Here the picture is different as 
with a deviation of only $\sim1.3\,\sigma$ between theory 
and experiment \cite{Giudice:2012ms}, the 
experimental result is in good agreement with the theoretical prediction and 
\begin{equation}
 \Delta a_e = a^\mathrm{exp}_e - a^\mathrm{SM}_e = (- 10.5 \pm 8.1)\times10^{-13}\,.
 \label{eq_emag_dev}
\end{equation}
The uncertainty in $\Delta a_e$ is expected to be reduced in the near 
future, enhancing its potential as a test of new physics. One 
subtlety that has to be taken into account is the fact that often the value of 
the fine structure constant $\alpha_{EM}$ is deduced from the electron 
magnetic moment measurement. To be sensitive to new physics in a consistent 
manner one should use in 
the calculation of the magnetic moment a value of $\alpha_{EM}$  determined by 
another independent measurement, as for example by interferometry of rubidium 
atoms \cite{0809.3177,Bouchendira:2010es}. In light of these prospects we 
will 
also investigate the shifts of the anomalous magnetic moment $a_e$ of the 
electron due to $\zp$ loops.

\subsubsection{$\boldsymbol Z^{\boldsymbol \prime}$ contribution to $\mathbf{\boldsymbol(g\boldsymbol-2\boldsymbol)}$}

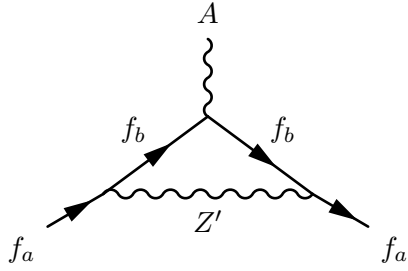
\begin{figure}[t!]
\begin{fmffile}{zp_triangle}
\begin{equation*}
\begin{gathered}
  \begin{fmfgraph*}(150,100)
    \fmfleft{i,i0,i1,i2}
    \fmfright{o,o0,o1,o2} 
    \fmflabel{$f_a$}{i}
    \fmflabel{$f_a$}{o}
    \fmflabel{$A$}{vu}

    \fmf{fermion,tension=4}{i,vcl}
    \fmf{fermion,label=$f_b$,label.side=left,tension=1}{vcl,vl}
    \fmf{fermion,label=$f_b$,label.side=left,tension=1}{vl,vcr}
    \fmf{fermion,tension=4}{vcr,o}
    \fmf{phantom,tension=1}{i2,vu,o2} 
    \fmf{phantom,tension=1}{i0,vcl,vcr,o0} 
    \fmf{photon,tension=2}{vu,vl}
    \fmf{photon,label=$\zp$,tension=0}{vcl,vcr}
  \end{fmfgraph*}
\end{gathered}
\end{equation*}
\end{fmffile} 
\caption{Anomalous magnetic moment of fermion $f_a$ due to $\zp$ exchange and 
$f_b$ running in the loop.}
\label{zp_triangle}
\end{figure}

We will now briefly recall the calculation of a general $\zp$ contribution to 
the 
anomalous magnetic moment $a$ in our model. The fermions that are coupled via 
flavor-changing interactions to the $\zp$ generically receive a contribution to 
their anomalous magnetic moment from the diagram of \cref{zp_triangle}. More 
specifically, we consider the $\zp$ interaction for two generic fermions $f_a$ 
and $f_b$
\begin{align}
 \mathcal{L} &= \bar{f}_a\, \gamma^\mu\, \left[g^L_{ab}\, P_L + g^R_{ab}\, 
P_R\right] f_b\, \zp_\mu + h.c.  \,. \label{eq_magint}
\end{align}
Using the notation of \cref{eq_magint}, we can derive  the $\zp$ 
contribution to the anomalous magnetic moment $a_{f_a}$ 
of the fermion $f_a$ with the fermion $f_b$ running in the loop 
\cite{hep-ph/0108081}. Introducing the mass ratios $x_a=\frac{m_a}{M_\zp}$ and 
$x_b=\frac{m_b}{M_\zp}$  and the vector and axial vector couplings $C_V=(g^R_{ab}+ g^L_{ab})/2$ and $C_A=(g^R_{ab}- g^L_{ab})/2$ the calculation \footnote{We want to thank Julian Heeck for kindly pointing out a missing overall factor of $-1$ in our calculation that appeared in an earlier version of this paper.} yields
 \begin{align}
  a_{f_a}(\zp) =   -\frac{x_a^2}{8\pi^2} \int^1_0\mathrm{d}u\,
  \Bigg[& u  (u-1) \Bigg(2\,(u-2) \,(C_V^2+ C_A^2) +  4\,\frac{x_a}{x_b}\ (C_V^2 - C_A^2) \Bigg) \notag \\
  & - u^2 \, \frac{x_a}{x_b}   \Bigg( (x_b-x_a)^2 \,C_V^2 - (x_b+x_a)^2\, C_A^2 \Bigg) \notag \\
  &  + u^2 \,   \Bigg( (x_b-x_a)^2 \,C_V^2 + (x_b+x_a)^2\, C_A^2 \Bigg) (u-1)\Bigg] \notag \\
  \times &\left[u\,\big((u-1)\,x_a^2+x_b^2\big)+(1-u)\right]^{-1}\,.
\label{eq_magmom}
 \end{align}
Assuming only right-handed couplings ($C_V=C_A=g^R_{ab}/2$), we can use this relation to turn the observed shift $\Delta a$ in the electron/muon magnetic moment into a limit on $g^R_{ab}$. \par
In order to get a better understanding of the $\zp$ contribution to the anomalous magnetic moment we derive an approximate formula. As we are mostly interested in $\zp$ bosons in the multi-GeV range we assume $M_\zp \gg m_a,m_b$. Therefore, we expand \cref{eq_magmom} for small ratios $x_a$ and $x_b$. Keeping only the 
leading powers yields the approximate formula 
\begin{equation}
 a_{f_a} \approx \frac{(g^R_{ab})^2}{4\,\pi^2}\,x_a \, \left[x_b\,\rho_\ell - \frac{x_a}{3}\, (1+\rho_\ell^2) \right] \,.
\label{eq_magapprox}
\end{equation}
As mentioned above we use the shift in the electron magnetic moment $\Delta a_e$ to constrain the 
off-diagonal couplings $g_{e\mu}$ and $g_{e\tau}$. From \cref{eq_magapprox} we can 
see that in the case of purely right-handed lepton couplings ($\rho_\ell =0$) the contribution to the electron magnetic moment $a_e$  is suppressed 
compared to the muon magnetic moment $a_\mu$  by a factor 
\begin{equation}
 \frac{x_e^2}{x_\mu^2} \approx \frac{1}{(200)^2} \sim \C O(10^{-5})\,.
\end{equation}
Comparing  \cref{eq_emag_dev} and \cref{eq_mumag_dev} we see that the precision of $\Delta a_e$  is only four orders of magnitudes smaller than the one of $\Delta a_\mu$. Hence, the constraints from $\Delta a_e$ are much weaker than those from $\Delta a_\mu$ and  play only a role for very light $\zp$ bosons\footnote{If we assume vector couplings ($\rho_\ell=1$), the leading term of the contribution to the electron magnetic moment $a_e$ relative to the muon magnetic moment $a_\mu$ is suppressed only by a factor ${x_e}/{x_\mu} \approx {1}/{200} \sim \C O(10^{-3})$. Furthermore, as the observed shift $\Delta a_e$ is negative whereas we obtain a positive shift, $a_e$ is a quite strong constraint for a vector coupling scenario.}. On the other hand, for light $\zp$ 
bosons we obtain rather strong limits from either LEP or muonium-antimuonium oscillation (cf. \cref{fig_plots1,fig_plots2}). Therefore, the constraint from $a_e$ proves to be practically irrelevant for our model. For the $\mu\tau$ sector the situation is more complicated. For purely right-handed lepton couplings the $\zp$ contribution goes in the wrong direction compared to the measurement (cf.~\cite{Altmannshofer:2016brv,Heeck:2016xkh}). As the current deviation between SM prediction and the measured value is greater than $3\,\sigma$ any contribution of such a $\zp$ is ruled out at this level. Therefore we show exclusions at the 4 and $5\,\sigma$ level e.g. in \cref{fig_bsmt,fig_lfit_bsmt,fig_plots3} as light and dark cyan bands. These limits are the only purely leptonic constraints in the $\mu\tau$ sector. 

\subsubsection{A hint for $\mathbf{\boldsymbol(g\boldsymbol-2\boldsymbol)_{\boldsymbol \mu}}$}

In view of the tension between the measured value and the theory prediction of $a_\mu$ we will now discuss the implications of a $\zp$ with non-zero $g_{\mu\tau}$ couplings on the muon magnetic moment. In order to explain the observed positive shift $\Delta a_\mu$ of \cref{eq_mumag_dev}, we also need a positive $\zp$ contribution to $a_\mu$. This is the case when the term in square brackets in \cref{eq_magapprox} is positive\footnote{This approximate relation holds only in the case of heavy $\zp$ bosons with $M_\zp \gg m_\tau$.}. This is a quadratic form in the coupling ratio $\rho_\ell$  and is positive only in between its two roots 
\begin{equation}
 \rho_{\ell,0} = \frac{3}{2} \left[ \frac{m_\tau}{m_\mu} \mp \sqrt{\left( \frac{m_\tau}{m_\mu}\right)^2 -\frac{4}{9}}\ \right] \,,
\end{equation}
i.e. for $ 0.02 \lesssim \rho_\ell \lesssim 50.75$. In this section we therefore consider a vector coupling scenario ($\rho_\ell =1$) and an optimized scenario where the limits from $\tau$ decay are weakest while the $\zp$ contribution to $a_\mu$ is still positive ($\rho_\ell=0.053$).

\subsubsection*{Vector coupling scenario}

\begin{center}
\begin{figure}[t!]
\begin{adjustwidth}{-\oddsidemargin-0.65in}{-\rightmargin}
\begin{subfigure}{.5\paperwidth}
  \centering
  \includegraphics[width=.45\paperwidth]{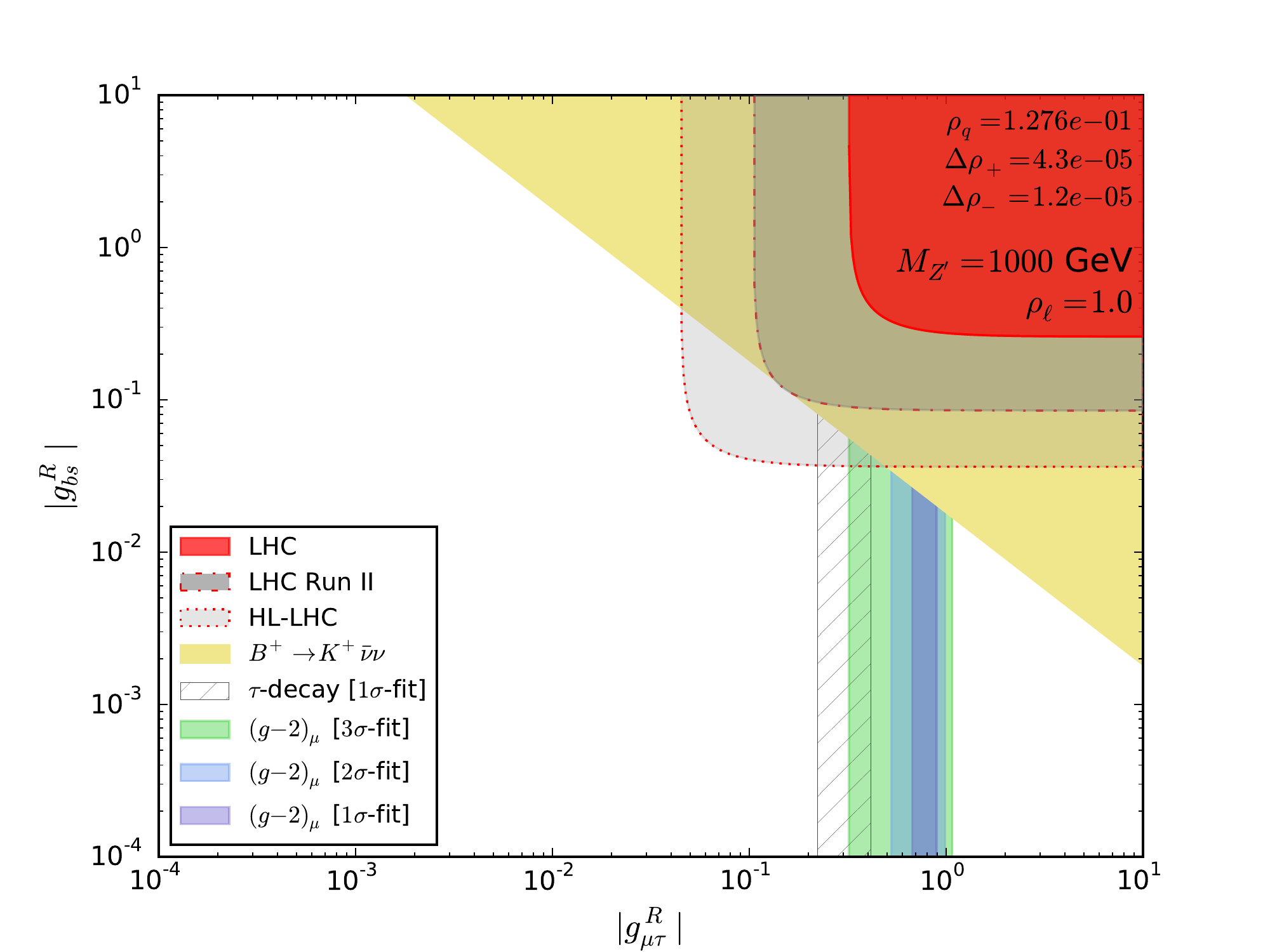}
\end{subfigure}\hspace*{-4.5em}
\begin{subfigure}{.5\paperwidth}
  \centering
  \includegraphics[width=.45\paperwidth]{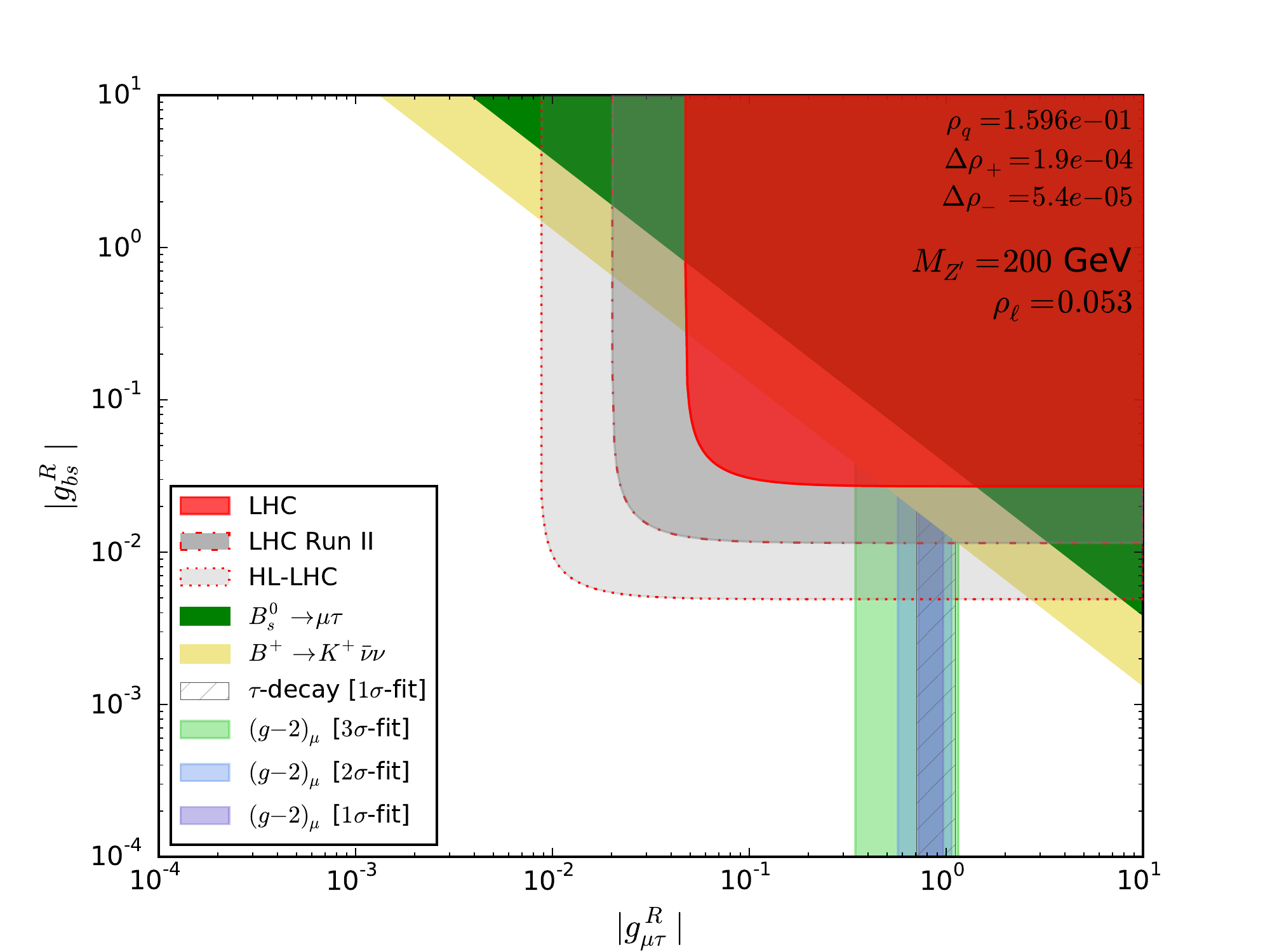}
\end{subfigure} 
\end{adjustwidth}
 \caption{The left (right) panel shows the $g^R_{bs} -g^R_{\mu\tau}$ plane for a $\zp$ boson with $M_\zp=$ 1000 (200) GeV and a lepton coupling ratio of $\rho_\ell=$ 1.0 (0.053). The red area indicates the limit from the ATLAS analysis of the process $pp\rightarrow \mu\tau$ at $\sqrt{s}=8$ TeV. The red dash-dotted and dashed lines are projections to the LHC Run II and HL-LHC. The green area represents the excluded region from the leptonic meson decay $B_s^0\rightarrow \mu\tau$. The yellow area is the limit from meson decay into neutrinos $B^+\rightarrow K^+ \bar\nu \nu$. The purple, blue and light green bands are the preferred 1, 2 and $3\,\sigma$ bands from $\Delta a_\mu$. The black and white hatched area depicts the preferred $1\,\sigma$ region of the observed deviation $\Delta\mathcal{R}_{\mu/e} = {R_{\mu/ e}}/{R_{\mu/ e}^\mathrm{SM}} -1$.}
\label{fig_propaganda}
\end{figure}
\end{center}

The left panel of \cref{fig_propaganda} shows the $g^R_{bs}-g^R_{\mu\tau}$ plane in a vector couplings scenario ($\rho_\ell=1$) for a $\zp$ with a mass of 1 TeV. First, we notice the absence of a limit from the leptonic meson decay $B^0_s\rightarrow\mu\tau$. The absence of such a limit is a peculiar feature of the vector coupling scenario in the lepton sector. This can be understood with help of \cref{eq_mesondec}, which features a term $|1-\rho_\ell|$ in the denominator and consequently diverges at $\rho_\ell=1$. Second, the limit from the meson decay $B^+\rightarrow K^+ \bar\nu \nu$ now becomes unavoidable due to the non-zero left-handed lepton coupling $g^L_{\mu\tau}$. This limit (shown in yellow) is much stronger than current 8 TeV ATLAS limits (shown in red) and possibly even stronger than limits from an LHC Run II scenario (red dash-dotted line). Even a future HL-LHC run could only slightly improve this limit along the direction of both small quark and lepton sector couplings.  As mentioned before we now get a positive $\zp$ contribution to $a_\mu$. Instead of using the observed deviation $\Delta a_\mu$ as a limit we can fit the excess. The purple, blue and green bands show the preferred 1, 2 and $3\,\sigma$ regions of $\Delta a_\mu$. It is worth noticing that for a 1 TeV $\zp$ the excess can naturally be accommodated  with $\mathcal{O}(1)$ lepton couplings $g_{\mu\tau}$. However, one has to be careful whether the limits from $\tau$ decay rule out such a $(g-2)_\mu$ explanation. On the $2\,\sigma$ level this is indeed the case~\cite{Altmannshofer:2016brv}. As done in \cref{fig_lfit_bsmt} we fit the observed deviation $\Delta \mathcal{R}_{\mu/e}$ at the $1\, \sigma$ level. This fit is shown by the black and white hatched area. The observed $\Delta a_\mu$ deviation is still compatible with the $\tau$ decay excess within $3\, \sigma$.

\subsubsection*{Optimized coupling scenario}

The right panel of \cref{fig_propaganda} shows the $g^R_{bs}-g^R_{\mu\tau}$ plane for a lepton coupling ratio of  $\rho_\ell=0.053$ for a $\zp$ with a mass of 200 GeV. This scenario is optimized such that for a positive $\zp$ contribution to $a_\mu$ the limit of $\tau$ decays is weakest. 
Previously, Altmannshofer et al. have shown explicitly in Ref.~\cite{Altmannshofer:2016brv} that for $\rho_\ell=0.1$ an explanation of $\Delta a_\mu$ is not ruled out by $\tau$ decay limits for $\zp$ masses greater than a few GeV.
Comparing to the vector coupling scenario we can see that we get a limit from the leptonic meson decay $B^0_s\rightarrow\mu\tau$. In addition, the relative strength of the limit from the meson decay $B^+\rightarrow K^+ \bar\nu \nu$ to the ATLAS limits is much weaker. This is due to the small fraction $\rho_\ell=0.053$ of left-handed lepton coupling, which drives the  decay into neutrinos. The most important point to notice is that now the $1\,\sigma$-fit of the $\tau$ decay excess  $\Delta \mathcal{R}_{\mu/e}$ lies on top of the $1\,\sigma$ band of the fit of $\Delta a_\mu$. Hence,  both excesses can be explained simultaneously with a coupling value of $0.7 \lesssim g^R_{\mu\tau} \lesssim 1.0 $  for a relatively low mass of $M_\zp=200$ GeV. This effect even persists for small perturbations around $\rho_\ell =0.053$ roughly in the region $0.03 \lesssim \rho_\ell \lesssim 0.35 $.  Furthermore, in the right panel \cref{fig_propaganda} we can see that  relevant parts of the parameter space of this scenario can be probed possibly already with LHC Run II data and definitely with a HL-LHC run.

\bigskip
Future more precise measurements of the branching fraction of the $\tau$ decays $\tau\rightarrow\mu\bar\nu\nu$ and $\tau\rightarrow e\bar\nu\nu$, e.g. at the \textsc{Belle-II} experiment~\cite{Aushev:2010bq} as well as the factor of four improvement in the precision of $(g-2)_{\mu}$ in the upcoming  E989 experiment at Fermilab~\cite{Grange:2015fou} can test this interpretation.
In addition to the purely leptonic tests the presence of $sd$ type quark couplings could present an opportunity to test this model in unusual $\tau$-decays, as discussed in Sect.~\ref{taudecayssect}. 
For example, assuming the maximally allowed quark coupling of $g^R_{sd} \approx 3\times10^{-3}$ for  the $200$ GeV $\zp$ discussed in this section yields a branching fraction of $\Gamma_{\tau^-\rightarrow\mu^-\pi^+K^-} \approx 9.0\times10^{-9}$. This could directly be searched for at \textsc{Belle-II}, which aims at a sensitivity of $1\times10^{-9}$ in branching fraction for $50 $ ab$^{-1}$ of data \cite{Aushev:2010bq}.

\section{Summary}
\label{sec_summary}

In this paper we have investigated simple test models of $\zp$ bosons with 
exclusively flavor-changing interactions, one in the lepton and one in the 
quark sector. For such models usually one would expect that precision tests of 
flavor-changing  neutral currents are far superior to direct production at the LHC. 
The latter could then be taken as just a nice confirmation of what we already 
know. For a generic chirality structure of the couplings this is indeed the case 
and LHC limits are eclipsed by limits on meson mixing as one can see for example 
from Fig.~\ref{fig_mixlim}. However, the latter limits depend on the relative 
strength of right- and left-handed couplings and there exist small regions where 
they can be evaded.  Here, the chirality independent LHC limits take over and 
become the best probe of new physics.
A similar situation arises with limits of mesons and leptons decaying into neutrinos (cf. e.g.~\cite{Buras:2014fpa,Buras:2015yca,Olive:2016xmw,Weinstein:1999de}). These 
limits are applicable for left-handed couplings, but can be evaded for purely 
right-handed ones.

Due to the coupling to leptons our $\zp$ boson also gives a contribution to 
$(g-2)$.  For couplings to $\mu\tau$ and a suitable chirality structure this allows for an explanation of the deviation in $(g-2)_{\mu}$ from the SM expectation (cf. also~\cite{Altmannshofer:2016brv}) as well as a small excess in the decay of $\tau$ leptons into muons and neutrinos. In  the future measurements of a $\zp$ decaying into $\mu$ and 
$\tau$ at ATLAS or CMS, or of $B$ decays at LHCb can probe into this 
parameter space.  In particular a dedicated ATLAS 
or CMS search at kinematics suitable for a relatively low mass resonance could 
be helpful.
As this can only test part of the interesting region it is worthwhile  to look 
for complementary probes. Here the study of $\tau$-decays, as it can be done, e.g. at \textsc{Belle-II}~\cite{Aushev:2010bq}, provides for interesting opportunities. For purely leptonic couplings in particular precision tests of lepton universality in these decays seem promising. In addition, flavor-violating trident production at high intensity experiments like DUNE~\cite{Acciarri:2015uup} or SHiP~\cite{Anelli:2015pba,Alekhin:2015byh} may allow to test this region~\cite{Magill:2016hgc,Heeck2016}.  Furthermore, additional couplings to relatively light quarks could provide for striking signals in unusual $\tau$-decays into $\mu$+hadrons.

\section*{Acknowledgments}
J.J. and P.F. would like to thank Florian Jetter, Tommaso Lari, Iacopo Vivarelli, Susanne Westhoff and especially Martin Bauer for stimulating and helpful discussions. We would also like to express our gratitude to Julian Heeck for noting an important factor of -1 and very helpful comments. 
P.F. is supported by the Graduiertenkolleg GRK 
1940 ``Particle Physics beyond the Standard Model''.
J.J. gratefully acknowledges support by the European Union's Horizon 2020 research and innovation programme under the Marie 
Sklodowska-Curie grant agreement Numbers 674896 and 690575.

\begin{appendix}

 \section{Higher order effects in cancellation}
 \label{sec_app_cancel}
 
 In \cref{sec_mixing} we have seen how meson mixing arises from four-quark 
operators that are generated in our model at tree level. Furthermore, we have 
argued that there are solutions of the coupling ratio $\rho_q$ for which the 
mixing exactly cancel. In the following we will investigate higher order 
effects contributing to mixing and its impact on the cancellation.
 
 \subsection{NLO effects}

\begin{figure}[!t]
\begin{fmffile}{loop_mix}
\begin{equation*}
\begin{gathered}
  \begin{fmfgraph*}(120,50)
    \fmfleft{i1,i2}    
    \fmfright{o1,o2}
     
    \fmflabel{$\bar{q}_j$}{i1}
    \fmflabel{$q_i$}{i2}
    \fmflabel{$\bar{q}_i$}{o1}
    \fmflabel{$q_j$}{o2}
    
    \fmf{fermion}{i2,v21}
    \fmf{fermion,tension=0}{v21,v11}
    \fmf{fermion}{v11,i1}
    \fmf{fermion}{o1,v12}
    \fmf{fermion,tension=0}{v12,v22}
    \fmf{fermion}{v22,o2}
    
    \fmf{photon,label=$\zp$,label.side=left,tension=1}{v21,v22}
    \fmf{photon,label=$Z$,tension=1}{v11,v12}
  \end{fmfgraph*}
\end{gathered}
\qquad \qquad
\begin{gathered}
  \begin{fmfgraph*}(120,50)
    \fmfleft{i1,i2}    
    \fmfright{o1,o2}
     
    \fmflabel{$\bar{q}_j$}{i1}
    \fmflabel{$q_i$}{i2}
    \fmflabel{$\bar{q}_i$}{o1}
    \fmflabel{$q_j$}{o2}
    
    \fmf{fermion}{v11,i1}
    \fmf{fermion}{v12,v11}
    \fmf{fermion}{o1,v12}
    \fmf{fermion}{v22,o2}
    \fmf{fermion}{v21,v22}
    \fmf{fermion}{i2,v21}
    
    \fmf{photon,label=$\zp$,label.side=left,tension=0}{v11,v21}
    \fmf{photon,label=$Z$,tension=0}{v12,v22}
  \end{fmfgraph*} 
\end{gathered}
\end{equation*}
\end{fmffile}
\caption{One-loop mixing diagram giving rise to higher order corrections of the 
four-quark operators \cref{eq_mixop1,eq_mixop2,eq_mixop3}. The same diagrams 
exist also with a Higgs boson instead of the additional $Z$. A third 
correction comes from the right diagram with the $Z$ exchanged by a 
second $\zp$.}
\label{gen_box}
\end{figure}
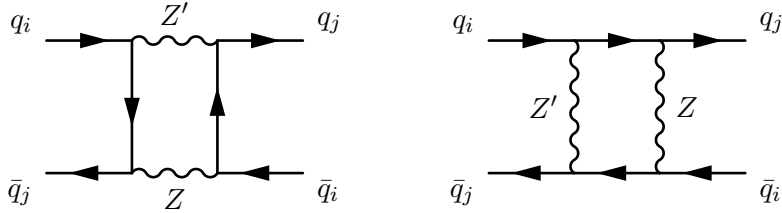

 Diagrams like the ones in \cref{gen_box} will give rise to 
corrections of the Wilson coefficients $C_i$ of the four-quark 
operators in \cref{eq_mixop1,eq_mixop2,eq_mixop3}. We define the corresponding 
Wilson coefficients as
\begin{align}
 C^\mathrm{VLL}_1 &= \frac{\left(g_{q_iq_j}^{L}\right)^{2}}{M^2_\zp}\, 
R^\mathrm{VLL}_1(\mu)\,,  \\
 C^\mathrm{VRR}_1 &= \frac{\left(g_{q_iq_j}^{R}\right)^{2}}{M^2_\zp}\, 
R^\mathrm{VRR}_1(\mu)\,, \\
 C^\mathrm{LR}_1 &= \frac{g_{q_iq_j}^{L} \, g_{q_iq_j}^{R}}{M^2_\zp}\, 
R^\mathrm{LR}_1(\mu) \,.
\end{align}
In the following we study the various contributions in more 
detail. Therefore, we consider diagrams of the type of 
\cref{gen_box}, where the additional boson in the loop can either be the SM $h, Z$ 
or a second $\zp$.
\subsubsection*{Higgs loop contribution}
First, we will consider a SM Higgs $h$ as the additional boson in the loop. 
The important feature of the Higgs is that it flips the chirality of the 
fermion at the vertex. Therefore, the Higgs introduces operator mixing in 
the sense that the Higgs correction to the Wilson coefficient of e.g.  
operator $\C O_1^\mathrm{VLL}$ will be proportional to 
$\C O_1^\mathrm{LR}$. A 
short calculation of the contributions from the one-loop diagrams in 
\cref{gen_box} with the $Z$ replaced by the Higgs leads to the estimate
\begin{align}
 {\delta C_i}^{(h)} \sim \frac{1}{8 \pi^2} \ \frac{m_q^2}{v^2}\ 
\log\left(\frac{M_h}{M_\zp}\right)  C_j \ \sim \ 10^{-7} \ C_j \,,
\end{align}
where we have used $m_q=1$ GeV and $M_\zp=1$ TeV to get the last relation. 
Hence, operator mixing due to NLO Higgs exchange is an effect roughly of the 
order 
of $10^{-7}$ and therefore much too small to be of any concern as will become 
clear in the following. 
\subsubsection*{$\boldsymbol Z$ loop contribution}
Next, we consider the diagrams as depicted in \cref{gen_box}, where a SM $Z$ 
boson plays the role of the additional boson running in the loop. Structurally 
the coupling of the $Z$ and $\zp$ are the same, so we do not introduce any 
 mixing amongst operators of the kind $\delta C_i \propto C_j$. Analogously, an 
order of magnitude 
calculation yields for the correction to the Wilson coefficients
\begin{align}
 {\delta C_i}^{(Z)} \sim \ \frac{1}{2\pi^2}\  (g^Z_q)^2 \ 
\log\left(\frac{M_Z}{M_\zp}\right) \ C_i \ \sim \ 10^{-3} \ C_i  \,,
\end{align}
where we have assumed for the coupling of the quark to the $Z$ a conservative 
value of $g^Z_q= 0.1$ and  $M_\zp=1$ TeV to get the last relation. 
First, we note that the correction coming from $Z$ contributions is much 
bigger than the one for the Higgs. Second, we note that the correction of the 
Wilson coefficient is proportional to the Wilson coefficient itself due to the 
absence of operator mixing. This implies that the correction will be universal 
to all Wilson coefficients and therefore only shift the cancellation solution 
$\rho_0$ and thereby not destroying it.
\subsubsection*{$\boldsymbol Z^{\boldsymbol\prime}$ loop contribution}
Finally, we consider the case of a pure $\zp$ induced loop diagram. One 
has to note that only the right diagram of \cref{gen_box} will contribute to 
generic meson mixing. The diagram on the left only contributes for mesons 
consisting of a quark-antiquark pair of same flavor. As before we can perform 
an estimate of the correction to the Wilson coefficients yielding
\begin{align}
 {\delta C_i}^{(\zp)} \sim \ \frac{1}{4\pi^2}\  (g^\zp_{qq'})^2  \ C_i \ \sim 
\ 10^{-2}\ (g^\zp_{qq'})^2 \ C_i  \,.
\end{align}
This contribution is in the same ballpark as the $Z$ contribution, but it 
introduces higher powers of the $\zp$ coupling. This changes the structure of 
\cref{eq_masssplit} fundamentally and therefore can possibly destroy the 
cancellation effect. 

\subsection{Numerical stability of cancellation}

We will now examine whether the correction due to a $\zp$ loop can spoil the 
cancellation solution. As the correction is of the order of ${\delta 
C_i}/{C_i} \sim 10^{-2}$  we will assume that the exact cancellation solution 
can be approximated by a perturbation series
\begin{align}
 \rho = \rho_0 + \delta \rho + \text{higher orders} \,. \label{eq_pertseries}
\end{align}
As we have just seen the Wilson coefficients at one-loop level due to $\zp$ 
corrections schematically read
\begin{align}
 C^{(\zp)}_{i}\ \sim \ \left(1 + \frac{1}{4\ \pi^2} \ g^2\right)\ 
C_{i} \ .
\end{align}
Hence the full cancellation equation at one-loop level becomes
\begin{align}
0  =& \left[ C^\mathrm{VLL}_1 \, P^\mathrm{VLL}_1\, (1+\rho_q^2) + 
\left(C^\mathrm{LR}_1 \, P^\mathrm{LR}_1 + C^\mathrm{LR}_2 \, 
P^\mathrm{LR}_2\right)\, \rho_q\right] \,g^2_{R}  &\leftarrow \text{tree-level 
relation} \notag \\
    & + \frac{1}{4 \pi^2}   \left[ C^\mathrm{VLL}_1 \, P^\mathrm{VLL}_1\, 
(1+\rho_q^4) +C^\mathrm{LR}_1 
\, P^\mathrm{LR}_1\, \rho_q^2 \right] 
\,g^4_{R} \,.   &\leftarrow \text{1-loop correction} \label{eq_pert_full}
\end{align}
If we then define the tree-level and one-loop terms as
\begin{align}
 f(\rho_q)  &=\left[ C^\mathrm{VLL}_1 \, P^\mathrm{VLL}_1\, (1+\rho_q^2) + 
\left(C^\mathrm{LR}_1 \, P^\mathrm{LR}_1 + C^\mathrm{LR}_2 \, 
P^\mathrm{LR}_2\right)\, \rho_q\right] \,g^2_{R} \,,  \\
    h(\rho_q) &=  \frac{1}{4 \pi^2}   \left[ C^\mathrm{VLL}_1 \, 
P^\mathrm{VLL}_1\, (1+\rho_q^4) +C^\mathrm{LR}_1 \, P^\mathrm{LR}_1\, \rho_q^2 
\right] \,g^4_{R} \,,
\end{align}
we find for the correction to the cancellation solution from perturbation theory
\begin{equation}
 \delta\rho = - \frac{h(\rho_0)}{f'(\rho_0) + h'(\rho_0)} \,.
\end{equation}

We checked these $\zp$ corrections for the $B_d$, $B_s$, $D$ 
and $K$ mesons. In all cases the corrections are reasonably small for  
reasonable values of the quark coupling $g^R_{qq'} \lesssim 1$. 
Especially the ratio $\delta \rho/\rho_0 < 1$ for all coupling combinations 
$\{qq',\ell\ell'\}$. We can calculate by the same method the correction on 
the tolerance $\delta(\Delta\rho)$. We find that this is generally 
much smaller than the tolerance itself $\delta(\Delta\rho)/\Delta\rho \ll 1$ 
and therefore negligible. Hence, we obtain a mere shift of the cancellation 
solution $\rho_0$ and its tolerance interval $I_0$. Therefore, the cancellation 
solution  $\rho_0$ is stable against higher order corrections and persists 
beyond tree-level. \\
However it should be noted that in the $\mu\tau$ sector 
and for high masses in the $e\tau$ sector the ratio of the shift to the 
tolerance $\delta\rho/\Delta \rho$ can be greater than 1. This is not a problem, 
as the cancellation solution still persists. It merely means that the shifted 
cancellation $\rho$ can lie outside of its original tolerance interval $I_0$ . 
This is an artefact of the extremely small tolerance interval in those 
channels.

\section{Monte Carlo simulation details}
\label{sec_appMC}

In this section we want to summarize important parameters we used for 
the determination of the simulated cross section $\sigma_\mathrm{MC}$ for the 
process $pp\rightarrow\zp\rightarrow \ell \ell^\prime$. \par
In \cref{tab_kfacs} we have summarized the values of the mass-dependent 
$K$-factors encoding NNLO contributions to the different final states.

\begin{table}[htb]
\centering
 \begin{tabular}{|c||c|c|c|c|c|c|c|c|c|c|}
  \hline
  $M_\zp$[GeV] & 500 & 550 & 600 & 650 & 700 & 750 & 800 & 1000 & 1200 & 2000 \\ 
 
  \hline \hline
  $K_{e\mu}$  &  1.449 & 1.447 & 1.414 & 1.436 & 1.427 & 1.423 & 1.480 & 1.494 
& 
1.501 & 1.602 \\
  \hline
  $K_{e\tau}$ & 1.355 & 1.313 & 1.379 & 1.428 & 1.351 & 1.391 & 1.424 & 1.455 & 
1.483 & 1.510 \\
  \hline
  $K_{\mu\tau}$ & 1.444 & 1.421 & 1.428 & 1.429 & 1.483 & 1.511 & 1.506 & 1.510 
& 1.525 & 1.661 \\
  \hline
 \end{tabular}
\caption{NNLO $K$-factors from SSM $\zp$}
\label{tab_kfacs}
\end{table}

In \cref{tab_effs} we find the efficiencies allowing to translate the Monte 
Carlo result into post-detector cross sections.

\begin{table}[htb]
\centering
 \begin{tabular}{|c||c|c|c|}
 \hline
$M_\zp$[GeV] &  $(A\times\epsilon)_{e\mu}$ & $(A\times\epsilon)_{e\tau}$ & 
$(A\times\epsilon)_{\mu\tau}$ \\
\hline
500    &   0.374   &     0.109  &    0.083   \\ 
550    &   0.380   &     0.116  &    0.086   \\ 
600    &   0.389   &     0.117  &    0.086   \\ 
650    &   0.401   &     0.122  &    0.090   \\ 
700    &   0.403   &     0.118  &    0.094   \\ 
750    &   0.410   &     0.123  &    0.091   \\ 
800    &   0.416   &     0.122  &    0.090   \\ 
900    &   0.428   &     0.116  &    0.098   \\ 
1000   &   0.440   &     0.115  &    0.095   \\ 
1100   &   0.441   &     0.117  &    0.103   \\ 
1200   &   0.441   &     0.118  &    0.098   \\ 
1400   &   0.449   &     0.119  &    0.096   \\ 
1600   &   0.445   &     0.119  &    0.099   \\ 
1800   &   0.431   &     0.114  &    0.096   \\ 
2000   &   0.415   &     0.109  &    0.089   \\ 
2200   &   0.386   &     0.104  &    0.082   \\ 
2500   &   0.358   &     0.093  &    0.071   \\ 
3000   &   0.283   &     0.069  &    0.053   \\
\hline
 \end{tabular}
\caption{Acceptance times efficiency from SSM $\zp$.}
\label{tab_effs}
\end{table}

\begin{figure}[t!]
 \begin{subfigure}{.5\paperwidth}
 {\section{Summary Plots} 
 \label{sec_app_plots}}
 \end{subfigure}
\begin{adjustwidth}{-\oddsidemargin-1in}{-\rightmargin}
\begin{subfigure}{.5\paperwidth}
  \centering
  \includegraphics[width=.45\paperwidth]{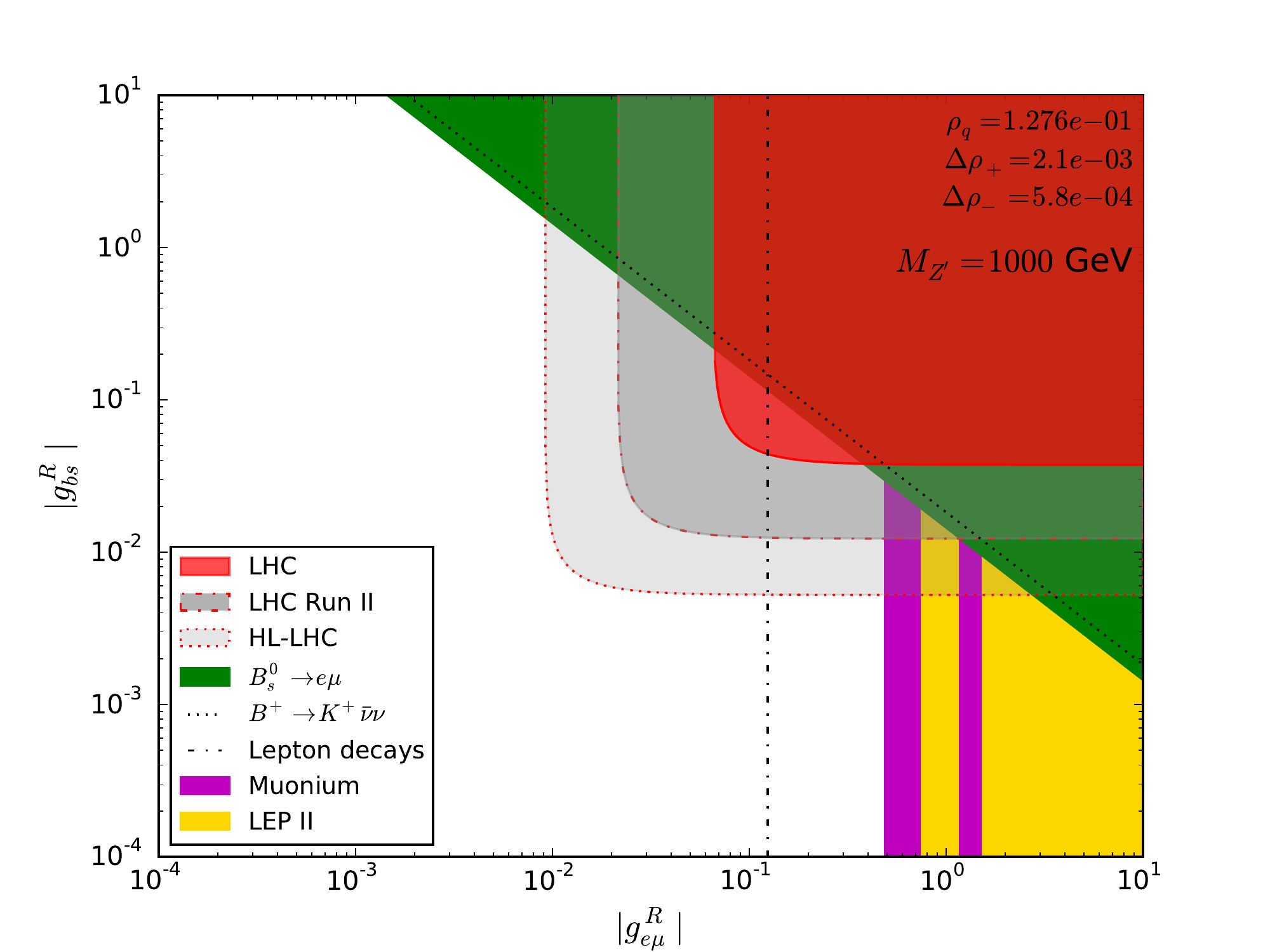}
  \caption{Flavor-violating couplings $g^R_{bs}$ and $g^R_{e\mu}$.}
\end{subfigure}
\begin{subfigure}{.5\paperwidth}
  \centering
  \includegraphics[width=.45\paperwidth]{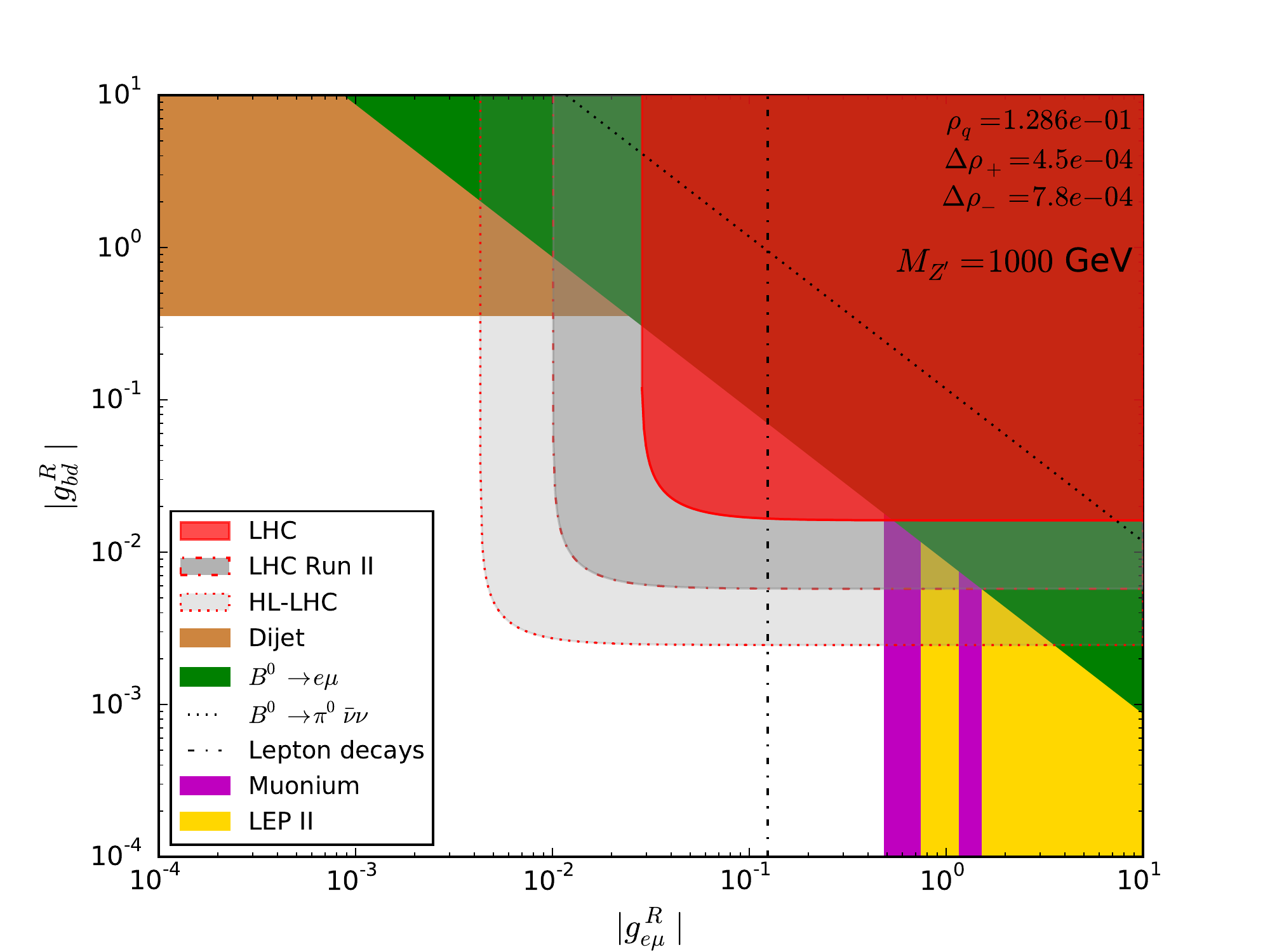}
  \caption{Flavor-violating couplings $g^R_{bd}$ and $g^R_{e\mu}$.}
\end{subfigure} \\
\begin{subfigure}{.5\paperwidth}
  \centering
  \includegraphics[width=.45\paperwidth]{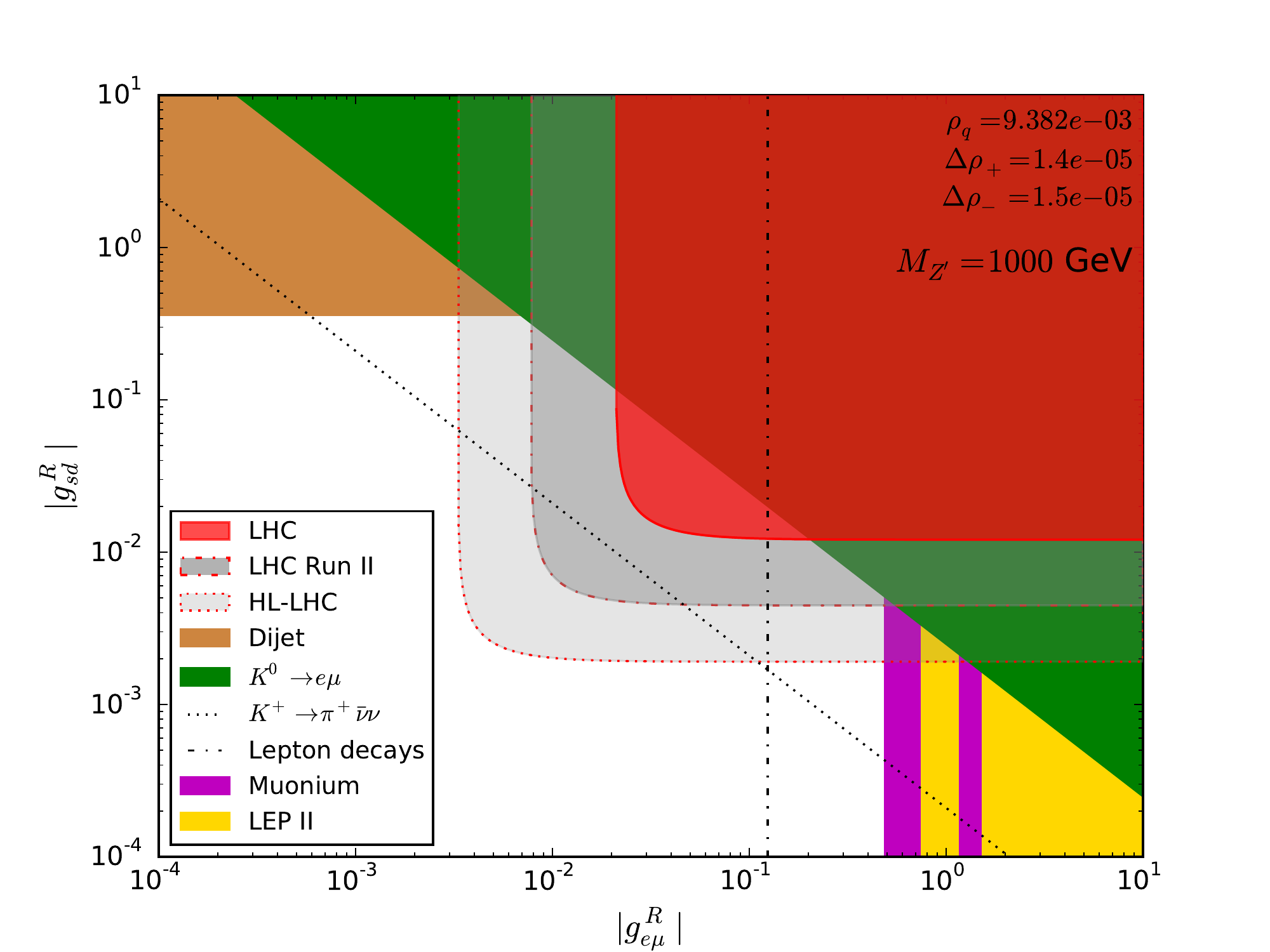}
  \caption{Flavor-violating couplings $g^R_{sd}$ and $g^R_{e\mu}$.}
\end{subfigure}
\begin{subfigure}{.5\paperwidth}
  \centering
  \includegraphics[width=.45\paperwidth]{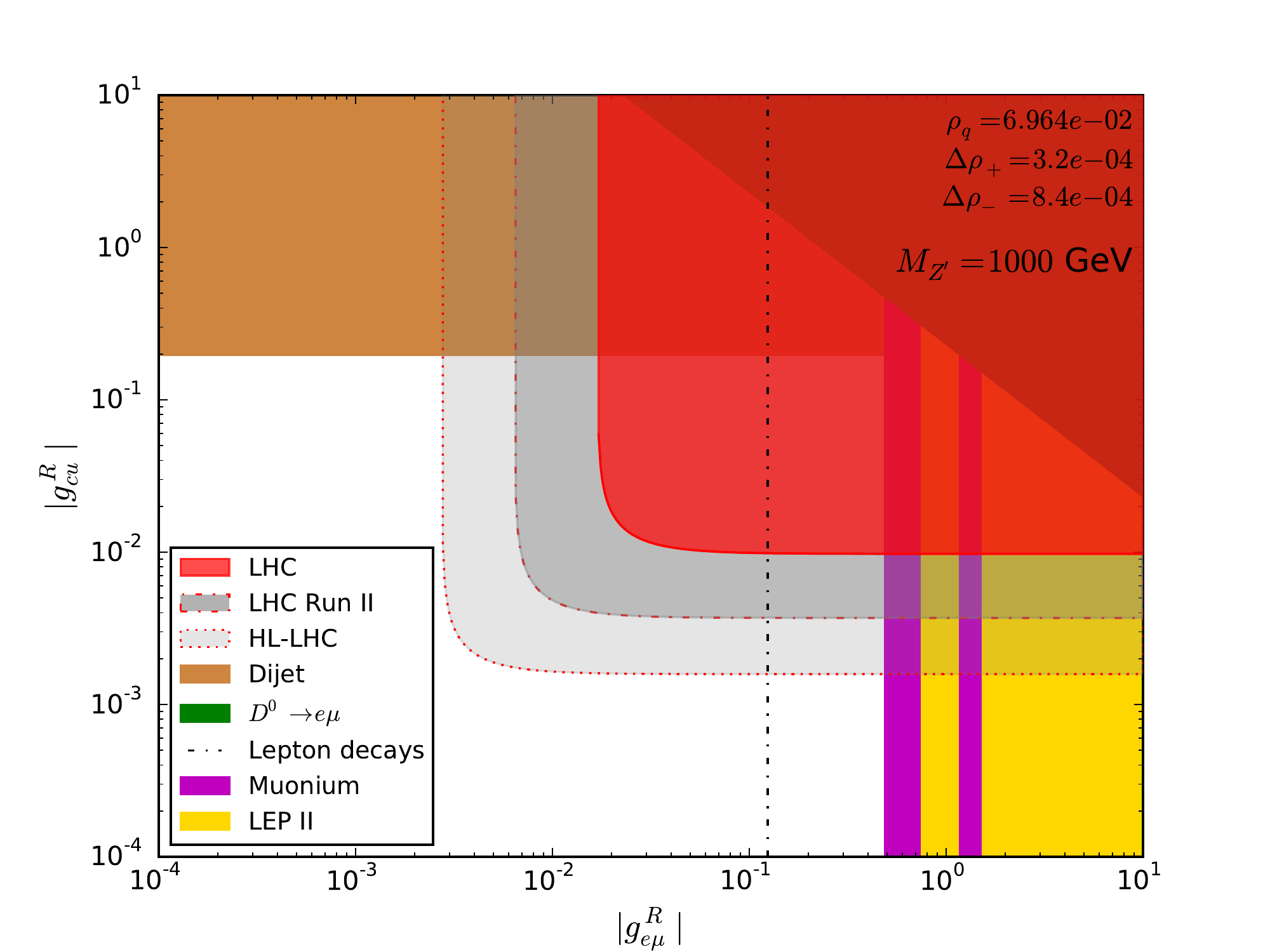}
  \caption{Flavor-violating couplings $g^R_{cu}$ and $g^R_{e\mu}$.}
\end{subfigure}
\end{adjustwidth}
 \caption{Flavor-violating couplings in the $e\mu$ sector for a $\zp$ boson of $M_\zp=1$ TeV.  The red areas depict the limits from the ATLAS analysis of the process $pp\rightarrow e\mu$ at $\sqrt{s}=8$ TeV. The red dash-dotted and dashed lines are projections to the LHC Run II and HL-LHC. The brown areas show four-quark contact interaction limits from LHC dijet analyses~\cite{Davidson:2013fxa}. The green areas are the limits coming from meson decays into charged leptons. The gold and magenta areas are purely leptonic limits coming from LEP and muonium oscillation constraints. The black dotted lines are meson decay limits into neutrinos and the black dash-dotted line the limits from lepton decays. These last two limits, however, apply only for left-handed lepton couplings  $g_{e\mu}^L$ instead of $g_{e\mu}^R$. The meson decay limits into neutrinos are absent in the $cu$ sector completely.}
\label{fig_plots1}
\end{figure}

\begin{figure}[ht!]
\begin{adjustwidth}{-\oddsidemargin-1in}{-\rightmargin}
\begin{subfigure}{.5\paperwidth}
  \centering
  \includegraphics[width=.45\paperwidth]{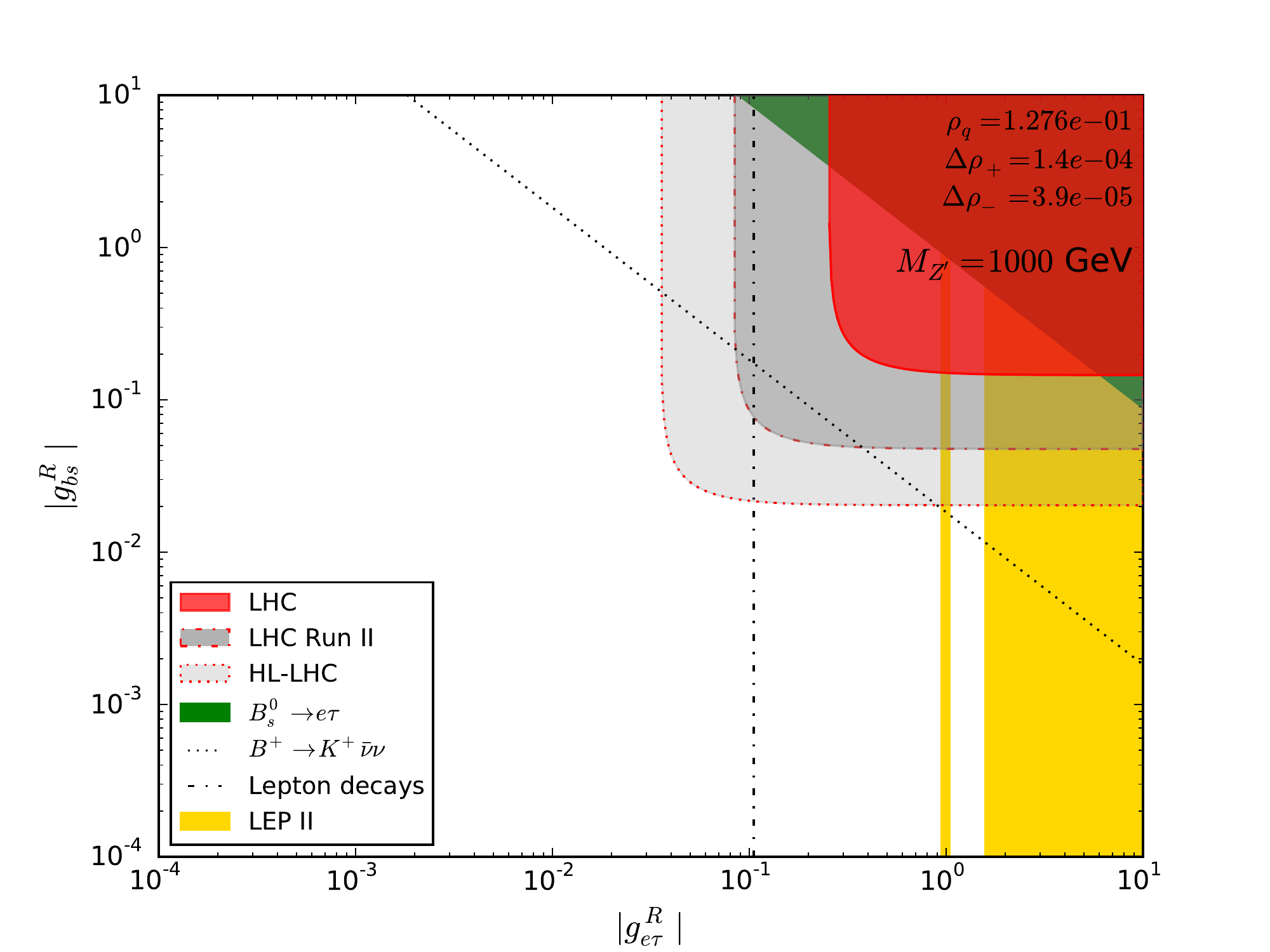}
  \caption{Flavor-violating couplings $g^R_{bs}$ and $g^R_{e\tau}$.}
\end{subfigure}
\begin{subfigure}{.5\paperwidth}
  \centering
  \includegraphics[width=.45\paperwidth]{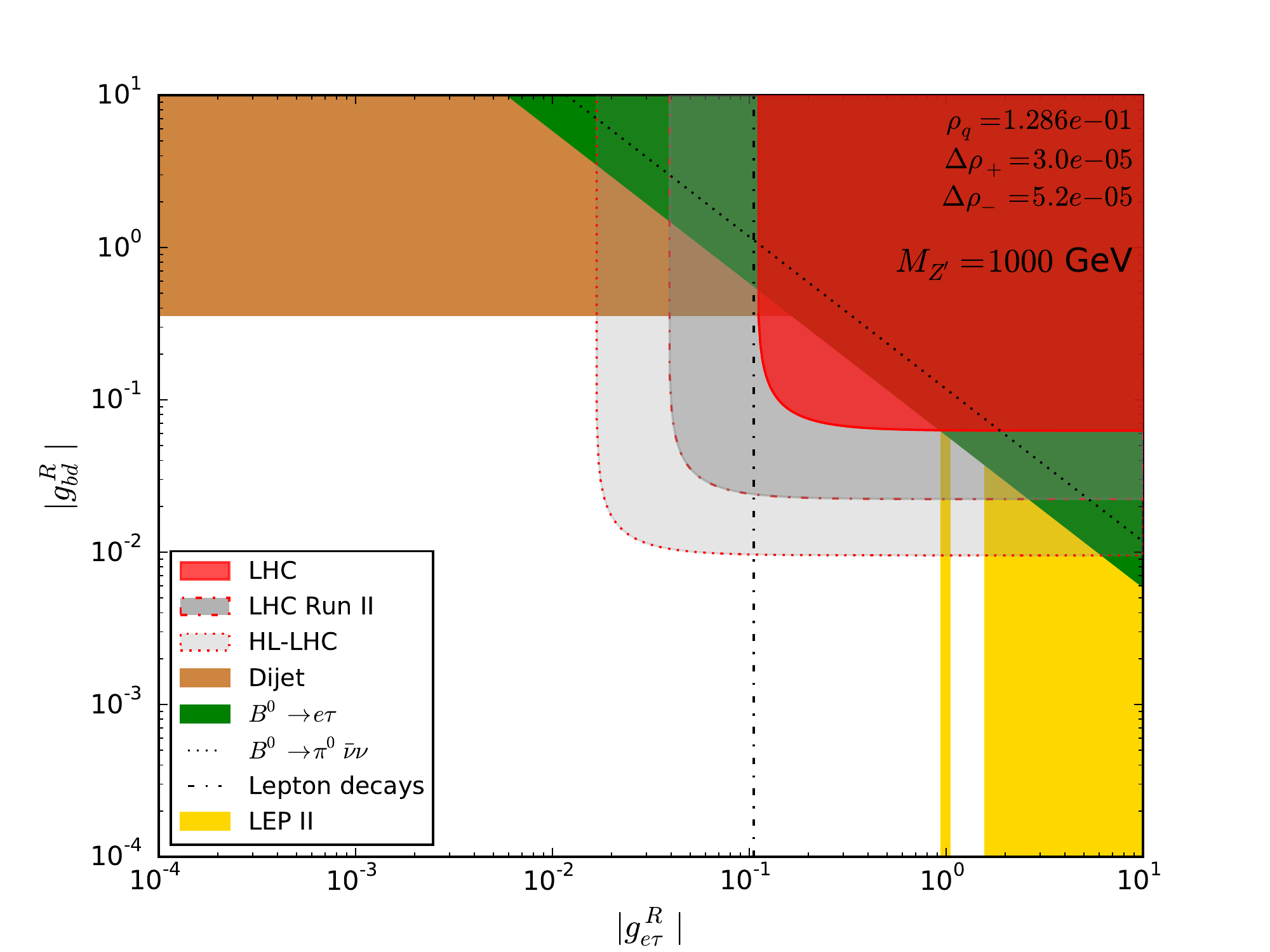}
  \caption{Flavor-violating couplings $g^R_{bd}$ and $g^R_{e\tau}$.}
\end{subfigure} \\
\begin{subfigure}{.5\paperwidth}
  \centering
  \includegraphics[width=.45\paperwidth]{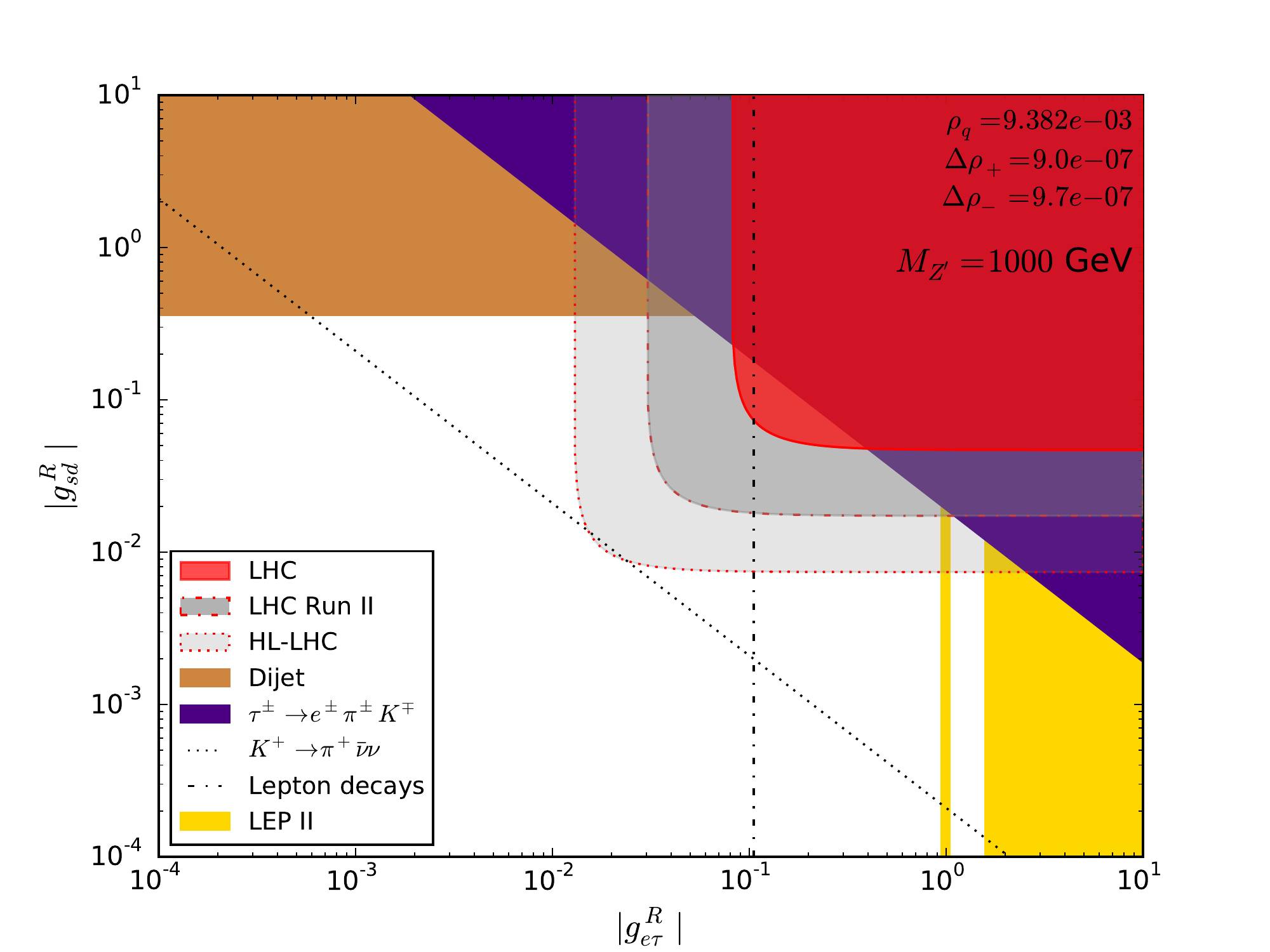}
  \caption{Flavor-violating couplings $g^R_{sd}$ and $g^R_{e\tau}$.}
\end{subfigure}
\begin{subfigure}{.5\paperwidth}
  \centering
  \includegraphics[width=.45\paperwidth]{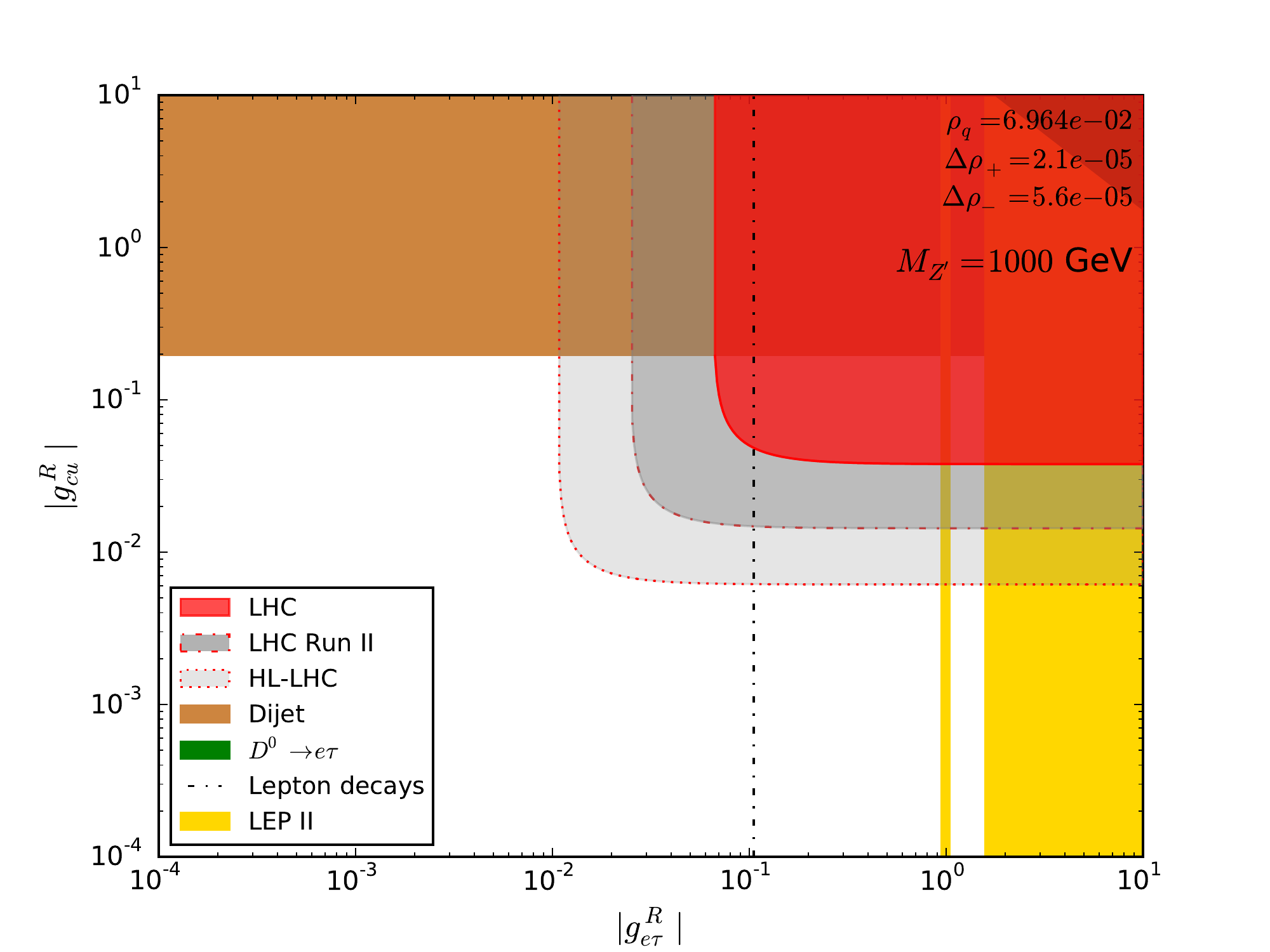}
  \caption{Flavor-violating couplings $g^R_{cu}$ and $g^R_{e\tau}$.}
\end{subfigure}
\end{adjustwidth}
 \caption{Flavor-violating couplings in the $e\tau$ sector for a $\zp$ boson of $M_\zp=1$ TeV.  The red areas depict the limits from the ATLAS analysis of the process $pp\rightarrow e\tau$ at $\sqrt{s}=8$ TeV. The red dash-dotted and dashed lines are projections to the LHC Run II and HL-LHC. The brown areas show four-quark contact interaction limits from LHC  dijet analyses~\cite{Davidson:2013fxa}. The green areas are the limits coming from meson decays into charged leptons. In purple we show the bounds from the rare decay $\tau^-\rightarrow e^- \pi^+ K^-$  only applicable in the $sd$ sector. The gold  areas are purely leptonic limits coming from LEP constraints. The black dotted lines are meson decay limits into neutrinos and the black dash-dotted line the limits from lepton decays. These last two limits, however, apply only for left-handed lepton couplings  $g_{e\tau}^L$ instead of $g_{e\tau}^R$. The meson decay limits into neutrinos are absent in the $cu$ sector completely.}
\label{fig_plots2}
\end{figure}

\begin{figure}[ht!]
\begin{adjustwidth}{-\oddsidemargin-1in}{-\rightmargin}
\begin{subfigure}{.5\paperwidth}
  \centering
  \includegraphics[width=.45\paperwidth]{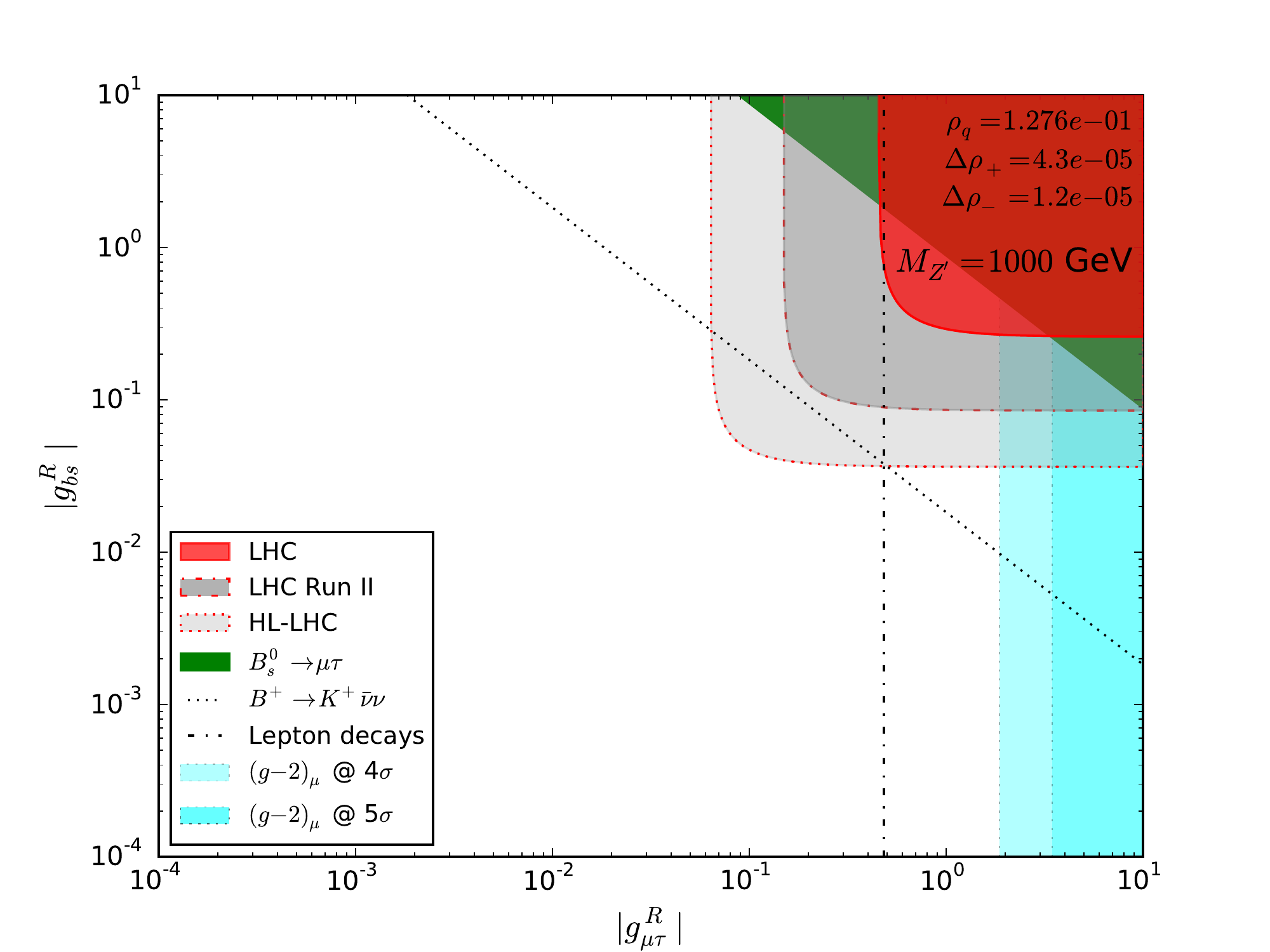}
  \caption{Flavor-violating couplings $g^R_{bs}$ and $g^R_{\mu\tau}$.}
\end{subfigure}
\begin{subfigure}{.5\paperwidth}
  \centering
  \includegraphics[width=.45\paperwidth]{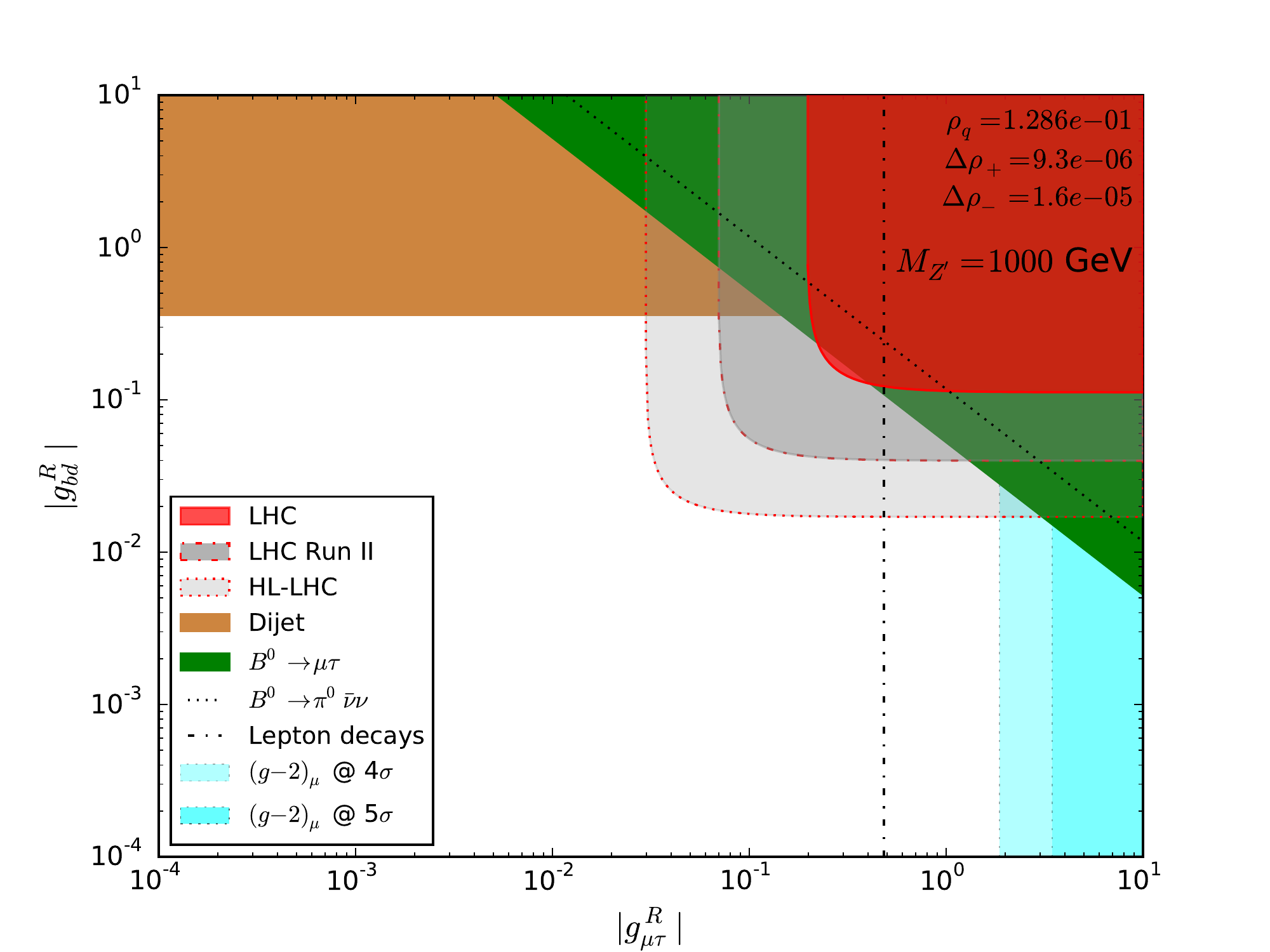}
  \caption{Flavor-violating couplings $g^R_{bd}$ and $g^R_{\mu\tau}$.}
\end{subfigure} \\
\begin{subfigure}{.5\paperwidth}
  \centering
  \includegraphics[width=.45\paperwidth]{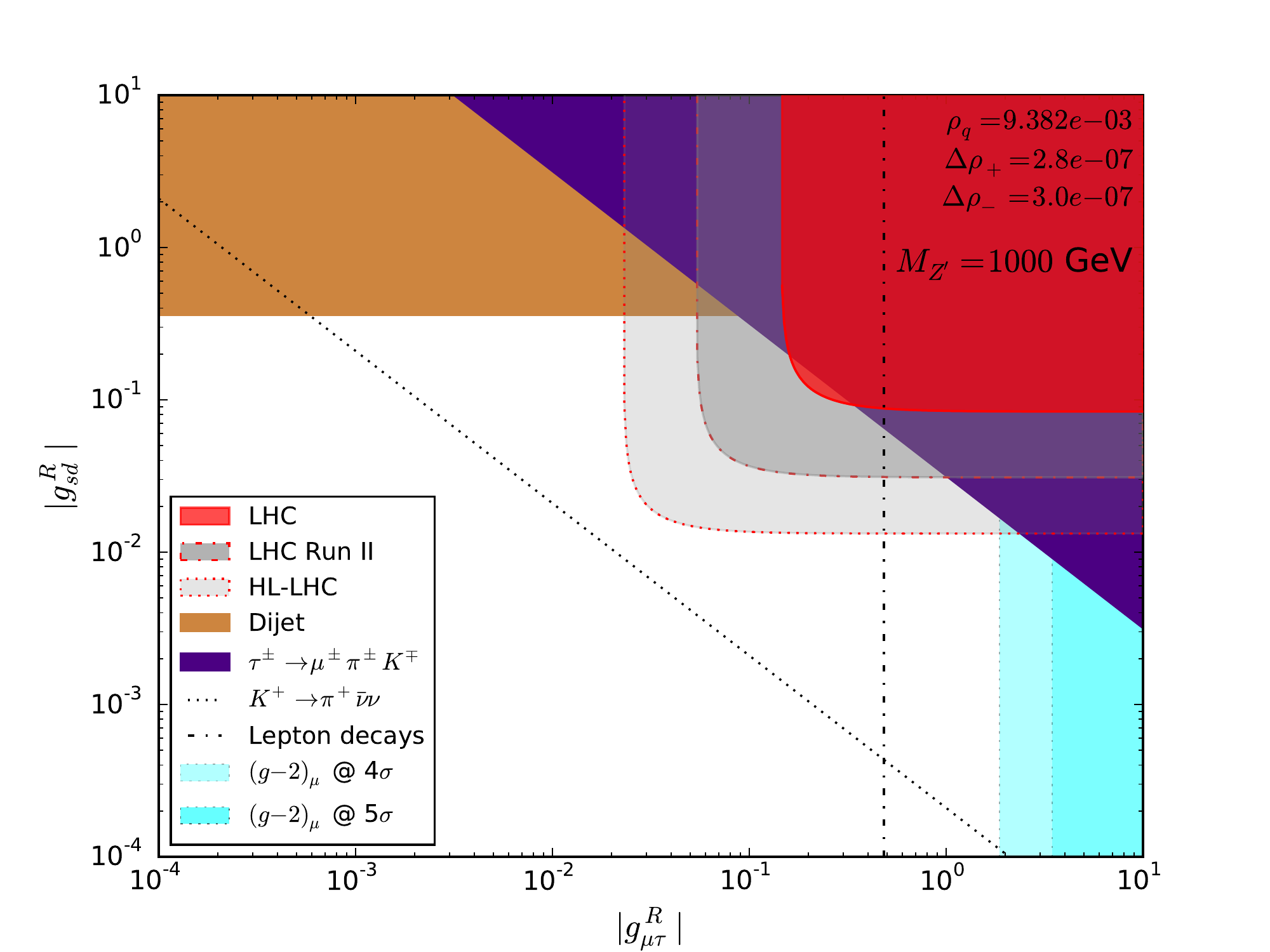}
  \caption{Flavor-violating couplings $g^R_{sd}$ and $g^R_{\mu\tau}$.}
\end{subfigure}
\begin{subfigure}{.5\paperwidth}
  \centering
  \includegraphics[width=.45\paperwidth]{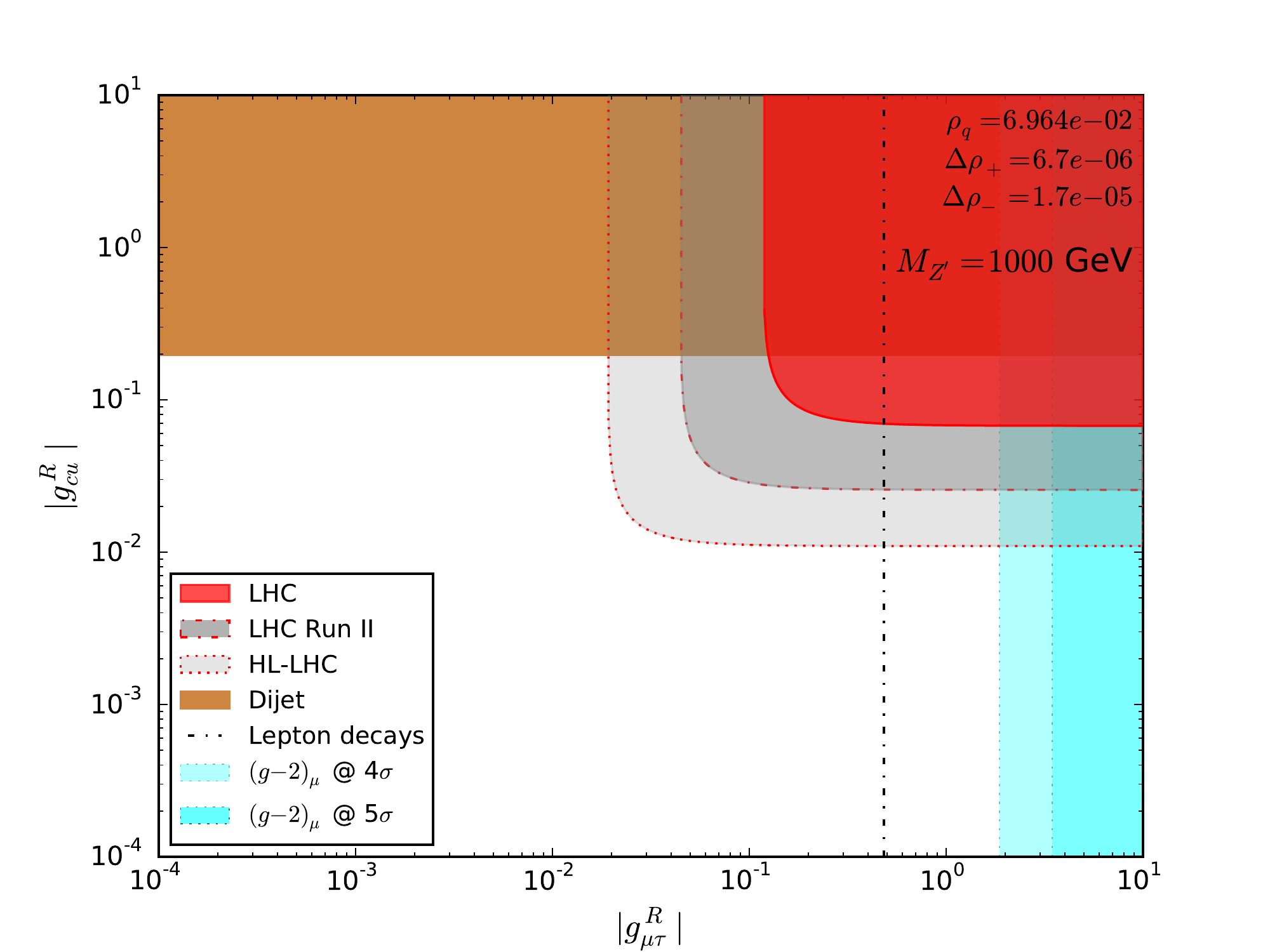}
  \caption{Flavor-violating couplings $g^R_{cu}$ and $g^R_{\mu\tau}$.}
\end{subfigure}
\end{adjustwidth}
 \caption{Flavor-violating couplings in the $\mu\tau$ sector for a $\zp$ boson of $M_\zp=1$ TeV.  The red areas depict the limits from the ATLAS analysis of the process $pp\rightarrow \mu\tau$ at $\sqrt{s}=8$ TeV. The red dash-dotted and dashed lines are projections to the LHC Run II and HL-LHC. The brown areas show four-quark contact interaction limits from LHC dijet analyses~\cite{Davidson:2013fxa}. The green areas are the limits coming from meson decays into charged leptons. In purple we show the bounds from the rare decay $\tau^-\rightarrow \mu^- \pi^+ K^-$ only applicable in the $sd$ sector. The light and dark cyan areas depict the 4 and $5\,\sigma$ exclusion bands from $\Delta a_\mu$. The black dotted lines are meson decay limits into neutrinos and the black dash-dotted line the limits from lepton decays. These last two limits, however, apply only for left-handed lepton couplings  $g_{\mu\tau}^L$ instead of $g_{\mu\tau}^R$. The meson decay limits into neutrinos are absent in the $cu$ sector completely.}
\label{fig_plots3}
\end{figure}

\end{appendix}


\clearpage
\begin{thebibliography}{10}


\bibitem{Barr:1978rv}
  S.~M.~Barr and A.~Zee,
  ``Calculating the Electron Mass in Terms of Measured Quantities,''
  Phys.\ Rev.\ D {\bf 17} (1978) 1854.


\bibitem{Wilczek:1978xi}
  F.~Wilczek and A.~Zee,
  ``Horizontal Interaction and Weak Mixing Angles,''
  Phys.\ Rev.\ Lett.\  {\bf 42} (1979) 421.


\bibitem{Davidson:1979wr}
  A.~Davidson, M.~Koca and K.~C.~Wali,
  ``U(1) as the Minimal Horizontal Gauge Symmetry,''
  Phys.\ Rev.\ Lett.\  {\bf 43} (1979) 92.


\bibitem{Ong:1978tq}
  C.~L.~Ong,
  ``Adding a Horizontal Gauge Symmetry to the {Weinberg-Salam} Model: An Eight Quark Model,''
  Phys.\ Rev.\ D {\bf 19} (1979) 2738.


\bibitem{Yanagida:1979gs}
  T.~Yanagida,
  ``Horizontal Symmetry and Mass of the Top Quark,''
  Phys.\ Rev.\ D {\bf 20} (1979) 2986.


\bibitem{Mohapatra:1974wk}
  R.~N.~Mohapatra,
  ``Gauge Model for Chiral Symmetry Breaking and Muon electron Mass Ratio,''
  Phys.\ Rev.\ D {\bf 9} (1974) 3461.


\bibitem{Langacker:2000ju}
  P.~Langacker and M.~Plumacher,
  ``Flavor changing effects in theories with a heavy $Z^\prime$ boson with family nonuniversal couplings,''
  Phys.\ Rev.\ D {\bf 62} (2000) 013006
  [hep-ph/0001204].


\bibitem{Guadagnoli:2011id}
  D.~Guadagnoli, R.~N.~Mohapatra and I.~Sung,
  ``Gauged Flavor Group with Left-Right Symmetry,''
  JHEP {\bf 1104} (2011) 093
  [arXiv:1103.4170 [hep-ph]].


\bibitem{Grinstein:2010ve}
  B.~Grinstein, M.~Redi and G.~Villadoro,
  ``Low Scale Flavor Gauge Symmetries,''
  JHEP {\bf 1011} (2010) 067
  [arXiv:1009.2049 [hep-ph]].


\bibitem{Leike:1998wr}
  A.~Leike,
  ``The Phenomenology of extra neutral gauge bosons,''
  Phys.\ Rept.\  {\bf 317} (1999) 143
  [hep-ph/9805494].


\bibitem{Langacker:2009im}
  P.~Langacker,
  ``The Physics of New U(1)-prime Gauge Bosons,''
  AIP Conf.\ Proc.\  {\bf 1200} (2010) 55
  [arXiv:0909.3260 [hep-ph]].


\bibitem{Aad:2015pfa}
  G.~Aad {\it et al.} [ATLAS Collaboration],
  ``Search for a Heavy Neutral Particle Decaying to $e\mu$, $e\tau$, or $\mu\tau$ in $pp$ Collisions at $\sqrt{s}=8$ TeV with the ATLAS Detector,''
  Phys.\ Rev.\ Lett.\  {\bf 115} (2015) no.3,  031801
  [arXiv:1503.04430 [hep-ex]].

  
\bibitem{Altmannshofer:2014cfa}
  W.~Altmannshofer, S.~Gori, M.~Pospelov and I.~Yavin,
  ``Quark flavor transitions in $L_\mu-L_\tau$ models,''
  Phys.\ Rev.\ D {\bf 89} (2014) 095033
  [arXiv:1403.1269 [hep-ph]].

\bibitem{Altmannshofer:2016brv}
  W.~Altmannshofer, C.~Y.~Chen, P.~S.~Bhupal Dev and A.~Soni,
  ``Lepton flavor violating $Z'$ explanation of the muon anomalous magnetic moment,''
  Phys.\ Lett.\ B {\bf 762} (2016) 389
  [arXiv:1607.06832 [hep-ph]].


\bibitem{Davidson:2013fxa}
  S.~Davidson and S.~Descotes-Genon,
  ``Constraining flavoured contact interactions at the LHC,''
  JHEP {\bf 1405} (2014) 066
  [arXiv:1311.5981 [hep-ph]].


\bibitem{Alwall:2014hca}
  J.~Alwall {\it et al.},
  ``The automated computation of tree-level and next-to-leading order differential cross sections, and their matching to parton shower simulations,''
  JHEP {\bf 1407} (2014) 079
  [arXiv:1405.0301 [hep-ph]].


\bibitem{Sjostrand:2007gs}
  T.~Sjostrand, S.~Mrenna and P.~Z.~Skands,
  ``A Brief Introduction to PYTHIA 8.1,''
  Comput.\ Phys.\ Commun.\  {\bf 178} (2008) 852
  [arXiv:0710.3820 [hep-ph]].


\bibitem{ATLAS-CONF-2015-072}
  The ATLAS collaboration,
  ``Search for beyond the Standard Model phenomena in $e\mu$ final states in pp collisions at $\sqrt{s}$ = 13 TeV with the ATLAS detector,''
  ATLAS-CONF-2015-072.


\bibitem{Buras:2012fs}
  A.~J.~Buras and J.~Girrbach,
  ``Complete NLO QCD Corrections for Tree Level Delta F = 2 FCNC Processes,''
  JHEP {\bf 1203} (2012) 052
  [arXiv:1201.1302 [hep-ph]].


\bibitem{Buras:2000if}
  A.~J.~Buras, M.~Misiak and J.~Urban,
  ``Two loop QCD anomalous dimensions of flavor changing four quark operators within and beyond the standard model,''
  Nucl.\ Phys.\ B {\bf 586} (2000) 397
  [hep-ph/0005183].


\bibitem{Babich:2006bh}
  R.~Babich, N.~Garron, C.~Hoelbling, J.~Howard, L.~Lellouch and C.~Rebbi,
  ``$K^0 - \bar{K}^0$ mixing beyond the standard model and CP-violating electroweak penguins in quenched QCD with exact chiral symmetry,''
  Phys.\ Rev.\ D {\bf 74} (2006) 073009
  [hep-lat/0605016].


\bibitem{Becirevic:2001xt}
  D.~Becirevic, V.~Gimenez, G.~Martinelli, M.~Papinutto and J.~Reyes,
  ``B parameters of the complete set of matrix elements of delta B = 2 operators from the lattice,''
  JHEP {\bf 0204} (2002) 025
  [hep-lat/0110091].


\bibitem{Golowich:2007ka}
  E.~Golowich, J.~Hewett, S.~Pakvasa and A.~A.~Petrov,
  ``Implications of $D^0$ - $\bar{D}^0$ Mixing for New Physics,''
  Phys.\ Rev.\ D {\bf 76} (2007) 095009
  [arXiv:0705.3650 [hep-ph]].


\bibitem{utfit}
The UTfit collaboration, http://www.utfit.org.



\bibitem{Golowich:2009ii}
  E.~Golowich, J.~Hewett, S.~Pakvasa and A.~A.~Petrov,
  ``Relating $D^0-\bar{D}^0$ Mixing and $D^0 \rightarrow \ell^+ \ell^-$ with New Physics,''
  Phys.\ Rev.\ D {\bf 79} (2009) 114030
  [arXiv:0903.2830 [hep-ph]].


\bibitem{Golowich:2011cx}
  E.~Golowich, J.~Hewett, S.~Pakvasa, A.~A.~Petrov and G.~K.~Yeghiyan,
  ``Relating $B_s$ Mixing and $B_s\to \mu^+\mu^-$ with New Physics,''
  Phys.\ Rev.\ D {\bf 83} (2011) 114017
  [arXiv:1102.0009 [hep-ph]].


\bibitem{Buras:1998raa}
  A.~J.~Buras,
  ``Weak Hamiltonian, CP violation and rare decays,''
  hep-ph/9806471.


\bibitem{Carrasco:2015pra}
  N.~Carrasco {\it et al.} [ETM Collaboration],
  ``$\Delta S=2$ and $\Delta C=2$ bag parameters in the standard model and beyond from N$_f$=2+1+1 twisted-mass lattice QCD,''
  Phys.\ Rev.\ D {\bf 92} (2015) no.3,  034516
  [arXiv:1505.06639 [hep-lat]].


\bibitem{Artamonov:2008qb}
  A.~V.~Artamonov {\it et al.} [E949 Collaboration],
  ``New measurement of the $K^{+} \to \pi^{+} \nu \bar{\nu}$ branching ratio,''
  Phys.\ Rev.\ Lett.\  {\bf 101} (2008) 191802
  [arXiv:0808.2459 [hep-ex]].


\bibitem{Buras:2015qea}
  A.~J.~Buras, D.~Buttazzo, J.~Girrbach-Noe and R.~Knegjens,
  ``$ {K}^{+}\to {\pi}^{+}\nu \overline{\nu} $ and $ {K}_L\to {\pi}^0\nu \overline{\nu} $ in the Standard Model: status and perspectives,''
  JHEP {\bf 1511} (2015) 033
  [arXiv:1503.02693 [hep-ph]].


\bibitem{Olive:2016xmw}
  C.~Patrignani {\it et al.} [Particle Data Group],
  ``Review of Particle Physics,''
  Chin.\ Phys.\ C {\bf 40} (2016) no.10,  100001.


\bibitem{Buras:2014fpa}
  A.~J.~Buras, J.~Girrbach-Noe, C.~Niehoff and D.~M.~Straub,
  ``$ B\to {K}^{\left(\ast \right)}\nu \overline{\nu} $ decays in the Standard Model and beyond,''
  JHEP {\bf 1502} (2015) 184
  [arXiv:1409.4557 [hep-ph]].


\bibitem{Jodidio:1986mz}
  A.~Jodidio {\it et al.},
  ``Search for Right-Handed Currents in Muon Decay,''
  Phys.\ Rev.\ D {\bf 34} (1986) 1967
   Erratum: [Phys.\ Rev.\ D {\bf 37} (1988) 237].

 
  
\bibitem{Foot:1994vd}
  R.~Foot, X.~G.~He, H.~Lew and R.~R.~Volkas,
  ``Model for a light $\zp$ boson,''
  Phys.\ Rev.\ D {\bf 50} (1994) 4571
  [hep-ph/9401250].


\bibitem{Pich:2013lsa}
  A.~Pich,
  ``Precision Tau Physics,''
  Prog.\ Part.\ Nucl.\ Phys.\  {\bf 75} (2014) 41
  [arXiv:1310.7922 [hep-ph]].

  
  

\bibitem{Albrecht:1992xa}
  H.~Albrecht {\it et al.} [ARGUS Collaboration],
  ``Evidence for the production of the charmed, doubly strange baryon $\Omega_c$ in $e^+ e^-$ annihilation,''
  Phys.\ Lett.\ B {\bf 288} (1992) 367.


\bibitem{Anastassov:1996tc}
  A.~Anastassov {\it et al.} [CLEO Collaboration],
  ``Experimental test of lepton universality in tau decay,''
  Phys.\ Rev.\ D {\bf 55} (1997) 2559
   Erratum: [Phys.\ Rev.\ D {\bf 58} (1998) 119904].


  
  
  
\bibitem{Aubert:2009qj}
  B.~Aubert {\it et al.} [BaBar Collaboration],
  ``Measurements of Charged Current Lepton Universality and $|V_{us}|$ using Tau Lepton Decays to $e^- \bar\nu_{e} \nu_{\tau}, \mu^- \bar \nu_{\mu} \nu_\tau, \pi^- \nu_\tau$ and $K^- \nu_\tau$,''
  Phys.\ Rev.\ Lett.\  {\bf 105} (2010) 051602
  [arXiv:0912.0242 [hep-ex]].
  
  
  
\bibitem{Crivellin:2015hha}
  A.~Crivellin, J.~Heeck and P.~Stoffer,
  ``A perturbed lepton-specific two-Higgs-doublet model facing experimental hints for physics beyond the Standard Model,''
  Phys.\ Rev.\ Lett.\  {\bf 116} (2016) no.8,  081801
  [arXiv:1507.07567 [hep-ph]].
  
 \bibitem{ColuccioLeskow:2016dox}
   E.~Coluccio Leskow, A.~Crivellin, G.~D'Ambrosio and D.~M\"{u}ller,
   ``$(g-2)_\mu$, Lepton Flavour Violation and $Z$ Decays with Leptoquarks: Correlations and Future Prospects,''
   arXiv:1612.06858 [hep-ph].
   
   
\bibitem{Aubert:2005tp}
  B.~Aubert {\it et al.} [BaBar Collaboration],
  ``Search for lepton-flavor and lepton-number violation in the decay $\tau^- \to \ell^\mp h^\pm h^{\prime -}$,''
  Phys.\ Rev.\ Lett.\  {\bf 95} (2005) 191801
  [hep-ex/0506066].
  
  
\bibitem{Miyazaki:2009wc}
  Y.~Miyazaki {\it et al.} [Belle Collaboration],
  ``Search for Lepton Flavor and Lepton Number Violating tau Decays into a Lepton and Two Charged Mesons,''
  Phys.\ Lett.\ B {\bf 682} (2010) 355
  [arXiv:0908.3156 [hep-ex]].
  
   
\bibitem{Willmann:1998gd}
  L.~Willmann {\it et al.},
  ``New bounds from searching for muonium to anti-muonium conversion,''
  Phys.\ Rev.\ Lett.\  {\bf 82} (1999) 49
  [hep-ex/9807011].


\bibitem{hep-ph/0307264}
  T.~E.~Clark and S.~T.~Love,
  ``Muonium - anti-muonium oscillations and massive Majorana neutrinos,''
  Mod.\ Phys.\ Lett.\ A {\bf 19} (2004) 297
  [hep-ph/0307264].


\bibitem{Nierste:2009wg}
  U.~Nierste,
  ``Three Lectures on Meson Mixing and CKM phenomenology,''
  arXiv:0904.1869 [hep-ph].


\bibitem{Harnik:2012pb}
  R.~Harnik, J.~Kopp and J.~Zupan,
  ``Flavor Violating Higgs Decays,''
  JHEP {\bf 1303} (2013) 026
  [arXiv:1209.1397 [hep-ph]].


\bibitem{hep-ex/0609051}
  S.~Schael {\it et al.} [ALEPH Collaboration],
  ``Fermion pair production in $e^{+} e^{-}$ collisions at 189-209-GeV and constraints on physics beyond the standard model,''
  Eur.\ Phys.\ J.\ C {\bf 49} (2007) 411
  [hep-ex/0609051].


\bibitem{Bennett:2006fi}
  G.~W.~Bennett {\it et al.} [Muon g-2 Collaboration],
  ``Final Report of the Muon E821 Anomalous Magnetic Moment Measurement at BNL,''
  Phys.\ Rev.\ D {\bf 73} (2006) 072003
  [hep-ex/0602035].


\bibitem{Jegerlehner:2009ry}
  F.~Jegerlehner and A.~Nyffeler,
  ``The Muon g-2,''
  Phys.\ Rept.\  {\bf 477} (2009) 1
  [arXiv:0902.3360 [hep-ph]].


\bibitem{Davier:2010nc}
  M.~Davier, A.~Hoecker, B.~Malaescu and Z.~Zhang,
  ``Reevaluation of the Hadronic Contributions to the Muon g-2 and to alpha(MZ),''
  Eur.\ Phys.\ J.\ C {\bf 71} (2011) 1515
   Erratum: [Eur.\ Phys.\ J.\ C {\bf 72} (2012) 1874]
  [arXiv:1010.4180 [hep-ph]].


\bibitem{Hagiwara:2011af}
  K.~Hagiwara, R.~Liao, A.~D.~Martin, D.~Nomura and T.~Teubner,
  ``$(g-2)_\mu$ and $\alpha(M_Z^2)$ re-evaluated using new precise data,''
  J.\ Phys.\ G {\bf 38} (2011) 085003
  [arXiv:1105.3149 [hep-ph]].




\bibitem{Altmannshofer:2014pba}
  W.~Altmannshofer, S.~Gori, M.~Pospelov and I.~Yavin,
  ``Neutrino Trident Production: A Powerful Probe of New Physics with Neutrino Beams,''
  Phys.\ Rev.\ Lett.\  {\bf 113} (2014) 091801
  [arXiv:1406.2332 [hep-ph]].


  
\bibitem{Heeck:2016xkh}
  J.~Heeck,
  ``Lepton flavor violation with light vector bosons,''
  Phys.\ Lett.\ B {\bf 758} (2016) 101
  [arXiv:1602.03810 [hep-ph]].
  

\bibitem{hep-ex/0208001}
  G.~W.~Bennett {\it et al.} [Muon g-2 Collaboration],
  ``Measurement of the positive muon anomalous magnetic moment to 0.7 ppm,''
  Phys.\ Rev.\ Lett.\  {\bf 89} (2002) 101804
   Erratum: [Phys.\ Rev.\ Lett.\  {\bf 89} (2002) 129903]
  [hep-ex/0208001].


  
\bibitem{Baek:2001kca}
  S.~Baek, N.~G.~Deshpande, X.~G.~He and P.~Ko,
  ``Muon anomalous g-2 and gauged $L_\mu - L_\tau$ models,''
  Phys.\ Rev.\ D {\bf 64} (2001) 055006
  [hep-ph/0104141].
  
  



\bibitem{TheBABAR:2016rlg}
  J.~P.~Lees {\it et al.} [BaBar Collaboration],
  ``Search for a muonic dark force at BABAR,''
  Phys.\ Rev.\ D {\bf 94} (2016) no.1,  011102
  [arXiv:1606.03501 [hep-ex]].


\bibitem{Acciarri:2015uup}
  R.~Acciarri {\it et al.} [DUNE Collaboration],
  ``Long-Baseline Neutrino Facility (LBNF) and Deep Underground Neutrino Experiment (DUNE) : Volume 2: The Physics Program for DUNE at LBNF,''
  arXiv:1512.06148 [physics.ins-det].
  
\bibitem{Anelli:2015pba}
  M.~Anelli {\it et al.} [SHiP Collaboration],
  ``A facility to Search for Hidden Particles (SHiP) at the CERN SPS,''
  arXiv:1504.04956 [physics.ins-det].

\bibitem{Alekhin:2015byh}
  S.~Alekhin {\it et al.},
  ``A facility to Search for Hidden Particles at the CERN SPS: the SHiP physics case,''
  Rept.\ Prog.\ Phys.\  {\bf 79} (2016) no.12,  124201
  [arXiv:1504.04855 [hep-ph]].
  
\bibitem{Magill:2016hgc}
  G.~Magill and R.~Plestid,
  ``Neutrino Trident Production at the Intensity Frontier,''
  arXiv:1612.05642 [hep-ph].
  
  \bibitem{Heeck2016}
  J.~Heeck, private communication.
  
  
\bibitem{Giudice:2012ms}
  G.~F.~Giudice, P.~Paradisi and M.~Passera,
  ``Testing new physics with the electron g-2,''
  JHEP {\bf 1211} (2012) 113
  [arXiv:1208.6583 [hep-ph]].


\bibitem{0809.3177}
M.~Cadoret, E.~De Mirandes, P.~Clad{\'e}, F.~Nez, L.~Julien, F.~Biraben, S.~Guellati-Kh{\'e}lifa,
``Atom interferometry based on light pulses: Application to the high precision measurement of the ratio h/m and the determination of the fine structure constant,''
The European Physical Journal Special Topics 2009, Vol.~172, No. 1, pp.121--136,


\bibitem{Bouchendira:2010es}
  R.~Bouchendira, P.~Clade, S.~Guellati-Khelifa, F.~Nez and F.~Biraben,
  ``New determination of the fine structure constant and test of the quantum electrodynamics,''
  Phys.\ Rev.\ Lett.\  {\bf 106} (2011) 080801
  [arXiv:1012.3627 [physics.atom-ph]].


\bibitem{hep-ph/0108081}
  K.~R.~Lynch,
  ``A Note on one loop electroweak contributions to g-2: A Companion to BUHEP-01-16,''
  hep-ph/0108081.


\bibitem{Aushev:2010bq}
  T.~Aushev {\it et al.},
  ``Physics at Super B Factory,''
  arXiv:1002.5012 [hep-ex].

  
  
\bibitem{Grange:2015fou}
  J.~Grange {\it et al.} [Muon g-2 Collaboration],
  ``Muon (g-2) Technical Design Report,''
  arXiv:1501.06858 [physics.ins-det].
  

\bibitem{Buras:2015yca}
  A.~J.~Buras, D.~Buttazzo and R.~Knegjens,
  ``$ K\to \pi \nu \overline{\nu} $ and $\varepsilon'/\varepsilon$ in simplified new physics models,''
  JHEP {\bf 1511} (2015) 166
  [arXiv:1507.08672 [hep-ph]].


\bibitem{Weinstein:1999de}
  A.~J.~Weinstein [CLEO Collaboration],
  ``Tau electroweak couplings,''
  PoS hf {\bf 8} (1999) 018
  [hep-ex/9911002].
  
  



\end{thebibliography}
\end{document}